\RequirePackage
		[
			abort,
			l2tabu,
			orthodox,
		]
		{nag}
\RequirePackage{import}

\makeatletter
\def\ece#1#2{\expandafter#1\csname#2\endcsname}%
\def\setproperty#1#2#3{\ece\protected@edef{#1@p#2}{\unexpanded{#3}}}%
\def\getproperty#1#2{%
  \expandafter\ifx\csname#1@p#2\endcsname\relax
  \else \csname#1@p#2\endcsname
  \fi
}%
\def\ifthenelsepropertydefined#1#2#3#4{%
  \expandafter\ifx\csname#1@p#2\endcsname\relax
  #4
  \else#3
  \fi
}%
\def\ifpropertydefined#1#2#3{%
    \ifthenelsepropertydefined{#1}{#2}{#3}{}
}
\def\ifpropertyundefined#1#2#3{%
    \ifthenelsepropertydefined{#1}{#2}{}{#3}
}
\def\raiseifpropertyundefined#1#2#3{%
    \ifpropertyundefined{#2}{#3}{\PackageError{#1}{Property #2 #3 needs to be defined. Put \@backslashchar setproperty{#2}{#3} to your settings file}{Grep for your property :)}}
}
\def\setpropertyifundefined#1#2#3{%
    \ifpropertyundefined{#1}{#2}{%
        \setproperty{#1}{#2}{#3}
    }{}
}
\def\setpropertyifundefinedwithusageinfo#1#2#3{%
    \setpropertyifundefined{#1}{#2}{TODO>> Put \backslash{}setproperty\{#1\}\{#2\}\{<your value here, e.g. ``#3``>\} into content/settings.tex <<}
}

\def\ifthenelseproperty#1#2#3#4{\providetoggle{#1@p#2}\settoggle{#1@p#2}{\getproperty{#1}{#2}}\iftoggle{#1@p#2}{#3}{#4}}
\def\ifthenelsepropertyequal#1#2#3#4#5{\ifthenelse{\equal{\getproperty{#1}{#2}}{#3}}{#4}{#5}}

\makeatother

\makeatletter
\newcommand\RequirePackageWithOption[2][]{%
    \@ifpackageloaded{#2}{%
        \PassOptionsToPackage{#1}{#2}
    }{%
        \RequirePackage[#1]{#2}
    }
}

\newcommand\ImportDefault[2][]{
    \IfFileExists{content/#1#2.tex}{
        \typeout{----- LOAD OWN #1#2 -----}
        \import{content/#1}{#2}
    }{
        \typeout{----- LOAD DEFAULT #1#2 -----}
        \ifthenelsepropertydefined{default}{#2}{%
            \IfFileExists{templates/\getproperty{default}{#2}/#1#2.tex}{
                \import{templates/\getproperty{default}{#2}/#1}{#2}
            }{
                \import{templates/Default/#1}{#2}
            }
        }{%
            \typeout{\getproperty{default}{#1}}
            \ifthenelsepropertydefined{default}{#1}{%
                \IfFileExists{templates/\getproperty{default}{#1}/#1#2.tex}{
                    \import{templates/\getproperty{default}{#1}/#1}{#2}
                }{
                    \import{templates/Default/#1}{#2}
                }
            }{
                \import{templates/Default/#1}{#2}
            }
        }
    }
}

\makeatother

\documentclass
    [
      a4paper,
      DIV=calc,
      BCOR=8mm,
      headinclude,
      bibliography=totoc,
      listof=totoc,
      index=totoc,
      british,
      10pt,
      captions=tableheading,
      chapterprefix,
      openright,
    ]
    {scrartcl}

    \newif\ifKOMA
    \KOMAtrue

    \setproperty{default}{titlepage}{Paper}
    \setproperty{default}{appendix}{Paper}
    \setproperty{default}{content}{Paper}

    \setproperty{compilation}{appendix}{true}
    \setproperty{compilation}{lof}{false}
    \setproperty{compilation}{lot}{false}
    \setproperty{compilation}{toc}{false}
    \setproperty{compilation}{complementglyphs}{false}
    \setproperty{compilation}{final}{true}
    \setproperty{compilation}{comment}{footnote}

    \usepackage{authblk}
    \setproperty{coauthors}{W7Xteaminclude}{True}
    \usepackage[]{algorithm2e}

    \usepackage{multirow}

    \usepackage[]{lineno}

    \usepackage[top=2.5cm, bottom=2.5cm, left=2.5cm, right=2.5cm]{geometry}

    \usepackage{xstring}
    \usepackage{catchfile}

\usepackage
		[
			ngerman,
			main=british
		]
		{babel}

\RequirePackageWithOption
		[
			log-declarations=false 
		]
		{xparse}

\usepackage{xpatch}
\usepackage{shellesc}

\setpropertyifundefined{color}{main}{black}
\setpropertyifundefined{color}{secondary}{black}

\setpropertyifundefinedwithusageinfo{author}{firstname}{J.R.R.}
\setpropertyifundefinedwithusageinfo{author}{familyname}{Tolkien}
\setpropertyifundefinedwithusageinfo{author}{acadtitle}{Prof.~Dr.~h.\,c.}
\setpropertyifundefinedwithusageinfo{author}{addressstreet}{Way of no return 77}
\setpropertyifundefinedwithusageinfo{author}{addresscity}{12345 Hobbingen}
\AtBeginDocument{%
    \setpropertyifundefined{author}{address}{\getproperty{author}{addressstreet}\\\getproperty{author}{addresscity}}
}
\setpropertyifundefinedwithusageinfo{author}{dateofbirth}{01.01.1970}
\setpropertyifundefinedwithusageinfo{author}{placeofbirth}{Helms Klamm, Rohan}
\setpropertyifundefinedwithusageinfo{author}{nationality}{german}
\setpropertyifundefinedwithusageinfo{author}{mobile}{+47 173 123456}
\setpropertyifundefinedwithusageinfo{author}{phone}{+53 001 123456}
\setpropertyifundefinedwithusageinfo{author}{faxnr}{+53 001 123456}
\setpropertyifundefinedwithusageinfo{author}{email}{minna@barnhelm.com}
\setpropertyifundefinedwithusageinfo{author}{orcid}{1234567}
\setpropertyifundefinedwithusageinfo{author}{signature}{\backslash{}igraph\{path to your signature here>\}}
\setpropertyifundefined{author}{affiliationindices}{1}
\setpropertyifundefined{affiliations}{1}{Max-Planck-Institute for Plasma Physics, 17491 Greifswald, Germany}

\setpropertyifundefined{document}{date}{\today}
\setpropertyifundefinedwithusageinfo{document}{title}{The Lord of the Rings}
\setpropertyifundefinedwithusageinfo{document}{titlehead}{There and back again}  
\setpropertyifundefinedwithusageinfo{document}{subject}{Novel}
\setpropertyifundefinedwithusageinfo{document}{keywords}{ring,mordor}
\setpropertyifundefinedwithusageinfo{document}{location}{Rivendale}
\setpropertyifundefinedwithusageinfo{document}{publishers}{Elrond Verlag}
\setpropertyifundefinedwithusageinfo{document}{dedication}{For my precious}
\setpropertyifundefinedwithusageinfo{document}{logos}{\backslash{}igraph\{<path to your logo here>\}}
\setpropertyifundefinedwithusageinfo{document}{type}{book}
\setpropertyifundefinedwithusageinfo{document}{doi}{145125.1234150}

\setpropertyifundefined{compilation}{externalfolder}{tikz/}
\setpropertyifundefined{compilation}{externalize}{false}
\setpropertyifundefined{compilation}{pdfa}{false}

\setpropertyifundefined{compilation}{final}{false}
\setpropertyifundefined{compilation}{comment}{footnote}

\AtEndPreamble{%
	\ifthenelseproperty{compilation}{pdfa}{\setpropertyifundefined{compilation}{externalize}{false}}{}
}

\setpropertyifundefined{compilation}{minionfonts}{false}
\setpropertyifundefined{compilation}{libertinusfonts}{false}
\setpropertyifundefined{compilation}{complementglyphs}{true}
\setpropertyifundefined{compilation}{fancychapterheadings}{false}
\setpropertyifundefined{compilation}{microtype}{false}
\setpropertyifundefined{compilation}{fontspec}{false}

\setpropertyifundefinedwithusageinfo{letter}{recipientname}{Sauron Mustermann}
\setpropertyifundefinedwithusageinfo{letter}{recipientaddress}{Ring Institute\\Tower 1\\12345 Mordor}
\setpropertyifundefinedwithusageinfo{letter}{subject}{Application for free Ork position}
\setpropertyifundefinedwithusageinfo{letter}{opening}{Dear Madam or Sir,}
\setpropertyifundefined{letter}{closing}{Sincerely yours,}

\setpropertyifundefined{thesis}{faculty}{FACULTY}
\setpropertyifundefined{thesis}{degree}{DEGREE}
\setpropertyifundefined{thesis}{defensedate}{DEFENSEDATE}
\setpropertyifundefined{thesis}{submissionyear}{YEAR OF SUBMISSION}

\makeatletter
\@ifundefined{ifKOMA}{%
    \newif\ifKOMA%
	\KOMAfalse%
}{}

\@ifclassloaded{book}{%
    \setpropertyifundefined{compilation}{titlepage}{true}
    \setpropertyifundefined{compilation}{abstract}{true}
    \setpropertyifundefined{compilation}{toc}{true}
    \setpropertyifundefined{compilation}{bibliography}{true}
    \setpropertyifundefined{compilation}{lof}{true}
    \setpropertyifundefined{compilation}{lot}{true}
    \setpropertyifundefined{compilation}{appendix}{true}
    \setpropertyifundefined{compilation}{affidavit}{true}
    \setpropertyifundefined{compilation}{clsdefineschapter}{true}
}{}
\@ifclassloaded{report}{%
    \setpropertyifundefined{compilation}{titlepage}{true}
    \setpropertyifundefined{compilation}{abstract}{true}
    \setpropertyifundefined{compilation}{toc}{true}
    \setpropertyifundefined{compilation}{bibliography}{true}
    \setpropertyifundefined{compilation}{lof}{true}
    \setpropertyifundefined{compilation}{lot}{true}
    \setpropertyifundefined{compilation}{appendix}{true}
    \setpropertyifundefined{compilation}{affidavit}{true}
    \setpropertyifundefined{compilation}{clsdefineschapter}{true}
}{}
\@ifclassloaded{article}{%
    \setpropertyifundefined{compilation}{titlepage}{true}
    \setpropertyifundefined{compilation}{abstract}{true}
    \setpropertyifundefined{compilation}{toc}{true}
    \setpropertyifundefined{compilation}{bibliography}{true}
}{}
\@ifclassloaded{scrbook}{%
    \setpropertyifundefined{compilation}{frontmatter}{true}
    \setpropertyifundefined{compilation}{titlepage}{true}
    \setpropertyifundefined{compilation}{abstract}{true}
    \setpropertyifundefined{compilation}{toc}{true}
    \setpropertyifundefined{compilation}{bibliography}{true}
    \setpropertyifundefined{compilation}{lof}{true}
    \setpropertyifundefined{compilation}{lot}{true}
    \setpropertyifundefined{compilation}{appendix}{true}
    \setpropertyifundefined{compilation}{affidavit}{true}
    \setpropertyifundefined{compilation}{clsdefineschapter}{true}
}{}
\@ifclassloaded{scrreprt}{%
    \setpropertyifundefined{compilation}{titlepage}{true}
    \setpropertyifundefined{compilation}{abstract}{true}
    \setpropertyifundefined{compilation}{toc}{true}
    \setpropertyifundefined{compilation}{bibliography}{true}
    \setpropertyifundefined{compilation}{lof}{true}
    \setpropertyifundefined{compilation}{lot}{true}
    \setpropertyifundefined{compilation}{appendix}{true}
    \setpropertyifundefined{compilation}{affidavit}{true}
    \setpropertyifundefined{compilation}{clsdefineschapter}{true}
}{}
\@ifclassloaded{scrartcl}{%
    \setpropertyifundefined{compilation}{titlepage}{true}
    \setpropertyifundefined{compilation}{abstract}{true}
    \setpropertyifundefined{compilation}{toc}{true}
    \setpropertyifundefined{compilation}{bibliography}{true}
}{}
\@ifclassloaded{moderncv}{%
    \setpropertyifundefined{compilation}{titlepage}{false}
    \setpropertyifundefined{compilation}{abstract}{false}
    \setpropertyifundefined{compilation}{toc}{false}
    \setpropertyifundefined{compilation}{bibliography}{false}
    \setpropertyifundefined{compilation}{glossaries}{false}
}{}
\setpropertyifundefined{compilation}{frontmatter}{false}
\setpropertyifundefined{compilation}{titlepage}{false}
\setpropertyifundefined{compilation}{abstract}{false}
\setpropertyifundefined{compilation}{toc}{false}
\setpropertyifundefined{compilation}{los}{false}
\setpropertyifundefined{compilation}{bibliography}{false}
\setpropertyifundefined{compilation}{lof}{false}
\setpropertyifundefined{compilation}{lot}{false}
\setpropertyifundefined{compilation}{lol}{false}
\setpropertyifundefined{compilation}{appendix}{false}
\setpropertyifundefined{compilation}{listofpublications}{false}
\setpropertyifundefined{compilation}{acknowledgement}{false}
\setpropertyifundefined{compilation}{affidavit}{false}
\setpropertyifundefined{compilation}{clsdefineschapter}{false}
\setpropertyifundefined{compilation}{acronyms}{false}
\setpropertyifundefined{compilation}{glossaries}{true}

\makeatother

\usepackage{ifluatex}

\ifluatex

	\newcommand{\FirstWord}[1]{\luaexec{tex.print(#1)}}
	
\else
\fi

\RequirePackageWithOption[
			fleqn 
		]
		{amsmath}

\usepackage{amsthm}
\usepackage{calc}
\ifluatex
\else
    \usepackage
		[
		]
		{amsfonts}

    \usepackage
		[
		]
		{amssymb}

\fi

\ifluatex
	\usepackage
		[
		]
		{fontspec}

\setproperty{compilation}{fontspec}{true}

	\usepackage
		[
			math-style=ISO,
			bold-style=ISO,
			nabla=upright,
		]
		{unicode-math}

	\AtEndPreamble{
        \ifthenelseproperty{compilation}{minionfonts}{
			\setmathfont{Cambria Math}
			\ifthenelseproperty{compilation}{complementglyphs}{%
				\setmathfont[range=\mathup/{num,latin,Latin,greek,Greek}]{Minion Pro}
				\setmathfont[range=\mathbfup/{num,latin,Latin,greek,Greek}]{MinionPro-Bold}
				\setmathfont[range=\mathit/{num,latin,Latin,greek,Greek}]{MinionPro-It}
				\setmathfont[range=\mathbfit/{num,latin,Latin,greek,Greek}]{MinionPro-BoldIt}
				\setmathfont[range={"0025,"002B,"002C,"002D,"002E,"2212,"2248,"007E,`±,`×,`·,`÷,`¬,`∂,`∆,`∕,`∞,`⌐,`<,`>,`=,`≥,`≤,`−}]{Minion Pro}
				\setmathfont[range={"2211,"221A}]{Latin Modern Math}
			}{}
			\renewcommand{\mathcal}[1]{{\textit{\addfontfeatures{Contextuals=Swash}#1}}}
			\newfontface{\MinionProSwash}{MinionPro-It}
					[
						Contextuals=Swash,
						Ligatures={TeX,Common,Rare,Historic,Contextual,Required}
					]
			\typeout{----- USING MINION FONTS -----}
		}{
		    \ifthenelseproperty{compilation}{libertinusfonts}{
		    	\setmainfont{Libertinus Serif}
		[
			BoldFont          = {* Bold},
  			ItalicFont        = {* Italic},
  			BoldItalicFont    = {* Bold Italic},
			SmallCapsFeatures = {Renderer=Basic},
			Ligatures         = {TeX,Common},
			Scale             = MatchLowercase,
			RawFeature=-lnum;+pnum;+onum;
		]

		    	\setsansfont{Libertinus Sans}
		[
			BoldFont          = {* Bold},
  			ItalicFont        = {* Italic},
  			BoldItalicFont    = {* -Bold Italic},
			SmallCapsFeatures = {Renderer=Basic},
			Ligatures         = {TeX,Common},
			Scale             = MatchLowercase,
			Numbers           = {Lining,Monospaced},
		]
		    	\setmonofont{Libertinus Mono}
		[
			BoldFont          = {* Bold},
  			ItalicFont        = {* Italic},
  			BoldItalicFont    = {* Bold Italic},
			SmallCapsFeatures = {Renderer=Basic},
			Ligatures         = {Common},
			Scale             = MatchLowercase,
			Numbers           = {Lining,Monospaced},
		]
		    	\setmathfont{Cambria Math}
				\ifthenelseproperty{compilation}{complementglyphs}{%
					\setmathfont[range=\mathup/{num,latin,Latin,greek,Greek}]{LibertinusSerif}
					\setmathfont[range=\mathbfup/{num,latin,Latin,greek,Greek}]{LibertinusSerif-Bold}
					\setmathfont[range=\mathit/{num,latin,Latin,greek,Greek}]{LibertinusSerif-Italic}
					\setmathfont[range=\mathbfit/{num,latin,Latin,greek,Greek}]{LibertinusSerif-BoldItalic}
					\setmathfont[range={"0025,"002B,"002C,"002D,"002E,"2212,"2248,"007E,"2208,"221D,"2A85,`±,`×,`·,`÷,`¬,`∂,`∆,`∕,`∞,`⌐,`<,`>,`=,`≥,`≤,`−,`;,}]{Libertinus Math}
					\setmathfont[range={"2211,"221A}]{Latin Modern Math}
					\setmathfont[range=\mathcal/{num,latin,Latin,greek,Greek},Contextuals=Swash]{Cambria Math}
				}{}
		    	\newfontface{\LibertinusSwash}{LibertinusSerif-Italic}
		    			[
		    				Contextuals=Swash,
		    				Ligatures={TeX,Common,Rare,Historic,Contextual,Required}
		    			]
		    	\newfontface{\LibertinusInitials}{LibertinusSerif-Initials}
		    	\typeout{----- USING LIBERTINUS FONTS -----}
		    }{
                \setmathfont{latinmodern-math.otf}
				\ifthenelseproperty{compilation}{complementglyphs}{%
                    \setmathfont[range=\setminus]{Asana Math}
				}{}
            }
        }
	}
\else
	\usepackage
		[
			T1
		]
		{fontenc}
	\usepackage
		[
			utf8
		]
		{inputenc}
	\usepackage{lmodern}
\fi

\usepackage{fontawesome5} 

\ifluatex
    \usepackage[a-3b,pdf17]{pdfx}
\usepackage{xmpincl}

\ifluatex
    \pdfvariable omitcidset=1
\else
\fi

\ifluatex
	\newcommand{\replaceNBSP}[1]{\luadirect{s,_ = string.gsub("\luatexluaescapestring{#1}", "\luatexluaescapestring{~}", " "); tex.print(-2,s)}}
\else
	\newcommand{\replaceNBSP}[1]{} 
\fi

\AtBeginDocument{
	\newwrite\metadatafile
	\immediate\openout\metadatafile=\jobname.xmpdata
	\immediate\write\metadatafile{\unexpanded{\Title}{\expanded{\replaceNBSP{\getproperty{document}{title}}}}}
	\immediate\write\metadatafile{\unexpanded{\Author}{\expanded{\getproperty{author}{firstname} \getproperty{author}{familyname}}}}
	\immediate\write\metadatafile{\unexpanded{\Subject}{\expanded{\getproperty{document}{subject}}}}
	\immediate\write\metadatafile{\unexpanded{\Keywords}{\getproperty{document}{keywords}}}
	\immediate\write\metadatafile{\unexpanded{\PublicationType}{\expanded{\getproperty{document}{type}}}}
	\immediate\write\metadatafile{\unexpanded{\Doi}{\expanded{\getproperty{document}{doi}}}}
	\immediate\closeout\metadatafile
}

\else
\fi

\usepackage
		[
			locale=UK,
			separate-uncertainty,
			retain-zero-exponent=true,
			binary-units, 
		]
		{siunitx}

\DeclareSIUnit{\arbitraryunit}{a.\,u.}

\DeclareSIUnit{\permille}{‰}

\DeclareSIUnit{\sample}{S}

\DeclareSIPrefix{\Femto}{f\kern0.1ex}{-15}

\DeclareSIUnit{\pb}{\pico\barn}
\DeclareSIUnit{\fb}{\femto\barn}
\DeclareSIUnit{\nb}{\nano\barn}

\usepackage{chemfig}

\usepackage{chemmacros}

\usechemmodule{all}

\renewtagform{reaction}[]{(}{)}

\AtBeginEnvironment{reactions}{}
\AtEndEnvironment{reactions}{}

\usepackage{chemformula}

\setchemformula
		{
			arrow-style={
							>=stealth',
							line cap=round,
							thick
						},
		}

\ifthenelseproperty{compilation}{minionfonts}{
	\setchemformula
			{
				arrow-style={
								>=stealth',
								line cap=round,
								thick
							},
				font-spec={[Numbers={Proportional,Lining}]MinionPro}
			}
}{}

\ifthenelseproperty{compilation}{libertinusfonts}{
	\setchemformula
			{
				arrow-style={
								>=stealth',
								line cap=round,
								thick
							},
				font-spec={[Numbers={Proportional,Lining}]LibertinusSerif}
			}
}{}

\RenewChemArrow{->}{\draw[chemarrow,->] (cf_arrow_start) -- (cf_arrow_end);}

\AtEndPreamble{
\usepackage{scrhack}
}

\AtEndPreamble{
	\ifthenelseproperty{compilation}{pdfa}{
	}{
		\usepackage{intopdf}
		
	}
}

\ifthenelseproperty{compilation}{final}{%
    \PassOptionsToPackage{disable}{todonotes}
}{}

\usepackage%
		[
			colorinlistoftodos, 
		]
		{todonotes}

\ifthenelseproperty{compilation}{final}{%
    \PassOptionsToPackage{disable}{todonotes}
}{}

\usepackage%
		[
			colorinlistoftodos, 
		]
		{todonotes}

\newcommand*{\TODO}[2][inline]{%
    \todo[#1]{#2}
}

\AtEndPreamble{%
    \RequirePackage{letltxmacro}

    \LetLtxMacro{\oldmissingfigure}{\missingfigure}
    \renewcommand{\missingfigure}[2][]{%
        \tikzexternaldisable\oldmissingfigure[{#1}]{#2}\tikzexternalenable%
    }

    \LetLtxMacro{\oldtodo}{\todo}
    \renewcommand{\todo}[2][]{%
        \tikzexternaldisable\oldtodo[#1]{#2}\tikzexternalenable%
    }
}

\ifthenelseproperty{compilation}{final}{%
    \PassOptionsToPackage{final}{changes}
}{%
    \ifpropertydefined{compilation}{comment}{%
        \PassOptionsToPackage{commentmarkup=\getproperty{compilation}{comment}}{changes}
    }
}

\usepackage%
		[
		]
		{changes}

\usepackage{seqsplit}

\usepackage{booktabs}

\ifthenelseproperty{compilation}{fontspec}{%
	\AtBeginEnvironment{tabular}{\addfontfeatures{Numbers={Monospaced}}}
    }{}

\usepackage{xcolor} 

\PassOptionsToPackage{dvipsnames,svgnames,x11names,rgb}{xcolor}
\definecolor{darkgreen}{RGB}{0,80,0} 
\definecolor{darkred}{RGB}{80,0,0} 

\usepackage{graphicx}

\usepackage{pgf}  

\SendSettingsToPgf

\usepackage
		[
			format=plain, 
			indention=1em, 
			textfont={color={black},small}, 
			width=0.925\linewidth, 
			hypcap=true, 
		]
		{caption}

\AtBeginDocument
			{
				\captionsetup
						{
                            labelfont={color=\getproperty{color}{main},small,sf,bf}, 
						}
			}

\usepackage{wrapfig}

\usepackage
		[
			hypcap=true, 
			textfont={color={black},small}, 
			width=0.925\textwidth, 
			indention=1em, 
		]
		{subcaption}

\usepackage{tikz}

\usetikzlibrary
			{%
				arrows,
				arrows.meta,
				bending,
				backgrounds,
				calc,
				fit,
				fadings,
				graphs,
				intersections,
				shadings,
				shapes,
				shapes.misc,
				shapes.geometric,
 				patterns,
 				plotmarks,
				chains,  
 				matrix,
 				through,
 				positioning,  
 				scopes,
				spy,
				decorations.shapes,
				decorations.text,
				decorations.pathmorphing,
				decorations.pathreplacing,
				decorations.markings,
				intersections,
			}
\ifluatex
    \usetikzlibrary
			{%
				graphdrawing,
			}
\fi

\makeatletter
\pgfkeys{/pgf/.cd,
  rectangle corner radius/.initial=3pt
}
\newif\ifpgf@rectanglewrc@donecorner@
\def\pgf@rectanglewithroundedcorners@docorner#1#2#3#4{%
  \edef\pgf@marshal{%
    \noexpand\pgfintersectionofpaths
      {%
        \noexpand\pgfpathmoveto{\noexpand\pgfpoint{\the\pgf@xa}{\the\pgf@ya}}%
        \noexpand\pgfpathlineto{\noexpand\pgfpoint{\the\pgf@x}{\the\pgf@y}}%
      }%
      {%
        \noexpand\pgfpathmoveto{\noexpand\pgfpointadd
          {\noexpand\pgfpoint{\the\pgf@xc}{\the\pgf@yc}}%
          {\noexpand\pgfpoint{#1}{#2}}}%
        \noexpand\pgfpatharc{#3}{#4}{\cornerradius}%
      }%
    }%
  \pgf@process{\pgf@marshal\pgfpointintersectionsolution{1}}%
  \pgf@process{\pgftransforminvert\pgfpointtransformed{}}%
  \pgf@rectanglewrc@donecorner@true
}
\pgfdeclareshape{rectangle with rounded corners}
{
  \inheritsavedanchors[from=rectangle] 
  \inheritanchor[from=rectangle]{north}
  \inheritanchor[from=rectangle]{north west}
  \inheritanchor[from=rectangle]{north east}
  \inheritanchor[from=rectangle]{center}
  \inheritanchor[from=rectangle]{west}
  \inheritanchor[from=rectangle]{east}
  \inheritanchor[from=rectangle]{mid}
  \inheritanchor[from=rectangle]{mid west}
  \inheritanchor[from=rectangle]{mid east}
  \inheritanchor[from=rectangle]{base}
  \inheritanchor[from=rectangle]{base west}
  \inheritanchor[from=rectangle]{base east}
  \inheritanchor[from=rectangle]{south}
  \inheritanchor[from=rectangle]{south west}
  \inheritanchor[from=rectangle]{south east}

  \savedmacro\cornerradius{%
    \edef\cornerradius{\pgfkeysvalueof{/pgf/rectangle corner radius}}%
  }

  \backgroundpath{%
    \northeast\advance\pgf@y-\cornerradius\relax
    \pgfpathmoveto{}%
    \pgfpatharc{0}{90}{\cornerradius}%
    \northeast\pgf@ya=\pgf@y\southwest\advance\pgf@x\cornerradius\relax\pgf@y=\pgf@ya
    \pgfpathlineto{}%
    \pgfpatharc{90}{180}{\cornerradius}%
    \southwest\advance\pgf@y\cornerradius\relax
    \pgfpathlineto{}%
    \pgfpatharc{180}{270}{\cornerradius}%
    \northeast\pgf@xa=\pgf@x\advance\pgf@xa-\cornerradius\southwest\pgf@x=\pgf@xa
    \pgfpathlineto{}%
    \pgfpatharc{270}{360}{\cornerradius}%
    \northeast\advance\pgf@y-\cornerradius\relax
    \pgfpathlineto{}%
  }

  \anchor{south west}{%
    \southwest
    \pgfmathparse{\cornerradius*(1-1/sqrt(2))}
    \advance\pgf@x\pgfmathresult pt
    \advance\pgf@y\pgfmathresult pt}

  \anchor{north east}{%
    \northeast
    \pgfmathparse{\cornerradius*(1-1/sqrt(2))}
    \advance\pgf@x-\pgfmathresult pt
    \advance\pgf@y-\pgfmathresult pt}

  \anchor{north west}{%
    \southwest\pgf@xa=\pgf@x
    \northeast\pgf@x=\pgf@xa
    \pgfmathparse{\cornerradius*(1-1/sqrt(2))}
    \advance\pgf@x\pgfmathresult pt
    \advance\pgf@y-\pgfmathresult pt}

  \anchor{south east}{%
    \southwest\pgf@ya=\pgf@y
    \northeast\pgf@y=\pgf@ya
    \pgfmathparse{\cornerradius*(1-1/sqrt(2))}
    \advance\pgf@x-\pgfmathresult pt
    \advance\pgf@y\pgfmathresult pt}

  \anchor{before north east}{\northeast\advance\pgf@y-\cornerradius}
  \anchor{after north east}{\northeast\advance\pgf@x-\cornerradius}
  \anchor{before north west}{\southwest\pgf@xa=\pgf@x\advance\pgf@xa\cornerradius
    \northeast\pgf@x=\pgf@xa}
  \anchor{after north west}{\northeast\pgf@ya=\pgf@y\advance\pgf@ya-\cornerradius
    \southwest\pgf@y=\pgf@ya}
  \anchor{before south west}{\southwest\advance\pgf@y\cornerradius}
  \anchor{after south west}{\southwest\advance\pgf@x\cornerradius}
  \anchor{before south east}{\northeast\pgf@xa=\pgf@x\advance\pgf@xa-\cornerradius
    \southwest\pgf@x=\pgf@xa}
  \anchor{after south east}{\southwest\pgf@ya=\pgf@y\advance\pgf@ya\cornerradius
    \northeast\pgf@y=\pgf@ya}

  \anchorborder{%
    \pgf@xb=\pgf@x
    \pgf@yb=\pgf@y%
    \southwest%
    \pgf@xa=\pgf@x
    \pgf@ya=\pgf@y%
    \northeast%
    \advance\pgf@x by-\pgf@xa%
    \advance\pgf@y by-\pgf@ya%
    \pgf@xc=.5\pgf@x
    \pgf@yc=.5\pgf@y%
    \advance\pgf@xa by\pgf@xc
    \advance\pgf@ya by\pgf@yc%
    \edef\pgf@marshal{%
      \noexpand\pgfpointborderrectangle
      {\noexpand\pgfqpoint{\the\pgf@xb}{\the\pgf@yb}}
      {\noexpand\pgfqpoint{\the\pgf@xc}{\the\pgf@yc}}%
    }%
    \pgf@process{\pgf@marshal}%
    \advance\pgf@x by\pgf@xa%
    \advance\pgf@y by\pgf@ya%
    \pgfextract@process\borderpoint{}%
    \pgf@rectanglewrc@donecorner@false
    \southwest\pgf@xc=\pgf@x\pgf@yc=\pgf@y
    \advance\pgf@xc\cornerradius\relax\advance\pgf@yc\cornerradius\relax 
    \borderpoint
    \ifdim\pgf@x<\pgf@xc\relax\ifdim\pgf@y<\pgf@yc\relax
      \pgf@rectanglewithroundedcorners@docorner{-\cornerradius}{0pt}{180}{270}%
    \fi\fi
    \ifpgf@rectanglewrc@donecorner@\else
      \southwest\pgf@yc=\pgf@y\relax\northeast\pgf@xc=\pgf@x\relax
      \advance\pgf@xc-\cornerradius\relax\advance\pgf@yc\cornerradius\relax
      \borderpoint
      \ifdim\pgf@x>\pgf@xc\relax\ifdim\pgf@y<\pgf@yc\relax
       \pgf@rectanglewithroundedcorners@docorner{0pt}{-\cornerradius}{270}{360}%
      \fi\fi
    \fi
    \ifpgf@rectanglewrc@donecorner@\else
      \northeast\pgf@xc=\pgf@x\relax\pgf@yc=\pgf@y\relax
      \advance\pgf@xc-\cornerradius\relax\advance\pgf@yc-\cornerradius\relax
      \borderpoint
      \ifdim\pgf@x>\pgf@xc\relax\ifdim\pgf@y>\pgf@yc\relax
       \pgf@rectanglewithroundedcorners@docorner{\cornerradius}{0pt}{0}{90}%
      \fi\fi
    \fi
    \ifpgf@rectanglewrc@donecorner@\else
      \northeast\pgf@yc=\pgf@y\relax\southwest\pgf@xc=\pgf@x\relax
      \advance\pgf@xc\cornerradius\relax\advance\pgf@yc-\cornerradius\relax
      \borderpoint
      \ifdim\pgf@x<\pgf@xc\relax\ifdim\pgf@y>\pgf@yc\relax
       \pgf@rectanglewithroundedcorners@docorner{0pt}{\cornerradius}{90}{180}%
      \fi\fi
    \fi
  }
}
\makeatother

\usepackage{pgfplots}
\usepackage{pgfplotstable}

\pgfplotsset{compat=newest}

\usetikzlibrary{pgfplots.groupplots}

\usepgflibrary{decorations.pathmorphing}
\usepgfplotslibrary{fillbetween,colormaps,external,colorbrewer}

\pgfplotsset{aspect ratio/.code args={#1:#2}{%
  \def\axisdefaultwidth{#1 cm}%
  \def\axisdefaultheight{#2 cm}},
}

\pgfplotsset{cycle list/Dark2-8}
\pgfplotsset{colormap/viridis}
\pgfkeys{/pgf/number format/.cd,1000 sep={\,}}

\AtEndPreamble{\ifthenelseproperty{compilation}{externalize}{
		\tikzexternalize[prefix=\getproperty{compilation}{externalfolder}]
		\typeout{----- USING TIKZ EXTERNALIZE -----}
	}{
		\typeout{----- NOT USING TIKZ EXTERNALIZE -----}
	}
}

\usepackage{listings}

\definecolor{javagray}{rgb}{0.55, 0.52, 0.54} 
\definecolor{javared}{rgb}{0.6,0,0} 
\definecolor{javagreen}{rgb}{0.25,0.5,0.35} 
\definecolor{javapurple}{rgb}{0.5,0,0.35} 
\definecolor{javadocblue}{rgb}{0.25,0.35,0.75} 
\definecolor{javaLila}{RGB}{127,0,85}

\let\origthelstnumber\thelstnumber

\makeatletter
\newcommand*\Suppressnumber{%
  \lst@AddToHook{OnNewLine}{%
    \let\thelstnumber\relax%
  }%
}
\newcommand\Reactivatenumber[1]{%
  \global\c@lstnumber#1%
  \global\advance\c@lstnumber\m@ne\relax%
  \lst@AddToHook{OnNewLine}{%
  \let\thelstnumber\origthelstnumber%
  }%
}
\makeatother

\makeatletter
\lst@InputCatcodes
\def\lst@DefEC{%
 \lst@CCECUse \lst@ProcessLetter
^^80^^81^^82^^83^^84^^85^^86^^87^^88^^89^^8a^^8b^^8c^^8d^^8e^^8f%
  ^^90^^91^^92^^93^^94^^95^^96^^97^^98^^99^^9a^^9b^^9c^^9d^^9e^^9f%
  ^^a0^^a1^^a2^^a3^^a4^^a5^^a6^^a7^^a8^^a9^^aa^^ab^^ac^^ad^^ae^^af%
  ^^b0^^b1^^b2^^b3^^b4^^b5^^b6^^b7^^b8^^b9^^ba^^bb^^bc^^bd^^be^^bf%
  ^^c0^^c1^^c2^^c3^^c4^^c5^^c6^^c7^^c8^^c9^^ca^^cb^^cc^^cd^^ce^^cf%
  ^^d0^^d1^^d2^^d3^^d4^^d5^^d6^^d7^^d8^^d9^^da^^db^^dc^^dd^^de^^df%
  ^^e0^^e1^^e2^^e3^^e4^^e5^^e6^^e7^^e8^^e9^^ea^^eb^^ec^^ed^^ee^^ef%
  ^^f0^^f1^^f2^^f3^^f4^^f5^^f6^^f7^^f8^^f9^^fa^^fb^^fc^^fd^^fe^^ff%
	^^^^03b8^^^^03c8^^^^03b7^^^^03bc^^^^03c3^^^^03b1^^^^03a9^^^^03b6%
	^^^^03c9^^^^03b4^^^^03c0%
^^00}
\lst@RestoreCatcodes
\makeatother

\RequirePackageWithOption{hyperref}

\hypersetup
		{
			pdfencoding=unicode,
			pdfpagemode=UseOutlines,
			bookmarksopenlevel=0,
			pdfpagelayout=TwoPageRight,
			hidelinks,
			pdfduplex=DuplexFlipLongEdge,
		}

\usepackage{epigraph}

\usepackage
		[
			math,
			toc,
		]
		{blindtext}

\usepackage
		[
			noabbrev
		]
		{cleveref}

\crefname{line}{line}{lines}
\creflabelformat{line}{#2#1#3}

\crefname{reaction}{reaction}{reactions}
\creflabelformat{reaction}{#2(#1)#3}

\crefname{listing}{code}{codes}

\usepackage
		[
			strict=true, 
			autostyle=true, 
			german=guillemets, 
		]
		{csquotes}

\setquotestyle{german} 

\usepackage{lettrine} 

\setpropertyifundefined{biblatex}{style}{numeric-comp}

\usepackage
		[
			hyperref=true,
			backend=biber, 
			url=true, 
			abbreviate=false,
			isbn=false, 
			doi=true, 
			backref=false, 
			urldate=long, 
			style=\getproperty{biblatex}{style},
			maxbibnames=3, 
			giveninits=true,
			block=none,
            sorting=none,
			defernumbers=true,
		]
		{biblatex}

\makeatletter
\@ifpackageloaded{seqsplit}{
	\DeclareFieldFormat{eid}{Art.\addnbspace\seqsplit{#1}}
}{}
\makeatother

\DeclareSourcemap{
  \maps[datatype=bibtex]{
    \map{
      \step[fieldsource=doi,final]
      \step[fieldset=url,null]
    }
  }
}

\DeclareFieldFormat{journaltitle}{\mkbibemph{#1},} 
\DeclareFieldFormat[inbook,thesis]{title}{\mkbibemph{#1}\addperiod} 
\DeclareFieldFormat[article,misc]{title}{\enquote{#1}} 
\DeclareFieldFormat[article,inproceedings,incollection]{volume}{Vol.~#1} 
\DeclareFieldFormat{edition}{\ifinteger{#1}{\mkbibordedition{#1}\addnbthinspace{}ed.}{#1\isdot}}

\DeclareFieldFormat{eprint:urn}{\textsc{URN}\addcolon\space\ifhyperref{\href{http://www.nbn-resolving.org/#1}{\nolinkurl{#1}}}{\nolinkurl{#1}}}

\DeclareSortingTemplate{ndymdt}{
 	\sort{
		\field{presort}
	}
	\sort[final]{
		\field{sortkey}
	}
	\sort[direction=descending]{
		\field{sortyear}
		\field{year}
		\literal{9999}
	}
	\sort[direction=descending]{
		\field[padside=left,padwidth=2,padchar=0]{month}
		\literal{99}
	}
	\sort[direction=descending]{
		\field[padside=left,padwidth=2,padchar=0]{day}
		\literal{99}
	}
}

\ifluatex
	\def\getAuthorBibCv{\FirstChar{\getproperty{author}{firstname}}.~\getproperty{author}{familyname}}
\else
	\def\getAuthorBibCv{\getproperty{author}{firstname}~\getproperty{author}{familyname}}
\fi

\newboolean{isCoauthorList}
\setboolean{isCoauthorList}{false}

\DefineBibliographyStrings{english}{
    andothers={%
        \ifthenelse{%
            \boolean{isCoauthorList}
        }{%
            et\addabbrvspace al\adddot\addcomma\addspace amongst them \textbf{\textsc{\getAuthorBibCv}}
        }{%
            et\addabbrvspace al\adddot}
    }
}

\DeclareBibliographyCategory{dontbib}
\let\printbibliographyold\printbibliography%
\renewcommand{\printbibliography}[1][]{%
    \typeout{----- printbibliography ------}%
    \printbibliographyold[notcategory=dontbib,#1]
}%
\makeatletter
\newcommand{\tempmaxup}[1]{\def\blx@maxcitenames{99}#1}
\DeclareCiteCommand{\fullcitecontribution}[\tempmaxup]{
    \usebibmacro{prenote}
    \addtocategory{dontbib}{\thefield{entrykey}}
}{
    \textbf{\usebibmacro{maintitle+title}}
    \newline\nopunct\newblock
    \usebibmacro{author}
    \newline\nopunct\newblock
    \usebibmacro{journal+issuetitle}, \usebibmacro{doi+eprint+url} \usebibmacro{addendum+pubstate}
}{
    \multicitedelim
}{
    \usebibmacro{postnote}
}
\makeatother

\usepackage
		[
			automark,
			headsepline, 
		]
		{scrlayer-scrpage}

\AtBeginDocument{%
    \ifKOMA
        \addtokomafont{pagehead}{\color{\getproperty{color}{main}}}
        \ifthenelseproperty{compilation}{clsdefineschapter}{%
            \addtokomafont{chapter}{\color{\getproperty{color}{main}}}
        }{}
        \addtokomafont{section}{\color{\getproperty{color}{main}}}
        \addtokomafont{subsection}{\color{\getproperty{color}{main}}}
        \addtokomafont{subsubsection}{\color{\getproperty{color}{main}}}
        \addtokomafont{paragraph}{\color{\getproperty{color}{main}}}

        \addtokomafont{footnoterule}{\color{\getproperty{color}{secondary}}}
        \addtokomafont{headsepline}{\color{\getproperty{color}{secondary}}}

    \fi
}

\usepackage{bookmark}

\bookmarksetup
			{
				open,
				addtohook={
					\ifnum\bookmarkget{level}<1
						\bookmarksetup{bold}
					\fi
				},
			}

\PassOptionsToPackage{automake=immediate}{glossaries-extra}

\usepackage%
		[
			nonumberlist, 
			acronym, 
			toc, 
		]
		{glossaries-extra}

\usepackage{glossary-mcols}

\setabbreviationstyle[acronym]{long-short}

\newglossary[slg]{symbols}{syi}{syg}{List of Symbols} 

\makeglossaries

\newglossarystyle{customGlossaryStyleForListOfSymbols}{%
    \renewenvironment{theglossary}{%
        \begin{longtable}{p{0.12\textwidth}p{\glsdescwidth}p{\glspagelistwidth}}
    }{%
        \end{longtable}
    }
    
    \renewcommand*{\glsgroupheading}[1]{}

}

\newglossarystyle{customListOfSymbols}{
    \setglossarystyle{customGlossaryStyleForListOfSymbols}
    
    \renewenvironment{theglossary}{%
        \begin{longtable}{p{0.12\textwidth}p{\glsdescwidth}p{\glspagelistwidth}}
    }{%
        \end{longtable}
    }
}

\newglossarystyle{customListOfAbbreviations}{%
    \renewenvironment{theglossary}{%
        \begin{longtable}{lp{\glsdescwidth}}
    }{%
        \end{longtable}
    }
    
    \renewcommand*{\glsgroupheading}[1]{}

}

\makeatletter
\newglossarystyle{customIndex}{
    \renewenvironment{theglossary}{%
        \setlength{\parindent}{0pt}
        \setlength{\parskip}{0pt plus 0.3pt}
        \let\item\@idxitem
    }{%
    }
    
    \renewcommand*{\glsgroupheading}[1]{}
    \renewcommand*{\glossaryentryfield}[5]{%
    \item\glstarget{##1}{##2}
        \ifx\relax##4\relax
        \else
            \space(##4)
        \fi
        \dotfill ##3\glspostdescription \space ##5
    }
    \renewcommand*{\glossarysubentryfield}[6]{%
        \ifcase##1\relax
        \item
        \or
            \subitem
        \else
            \subsubitem
        \fi
        \glstarget{##2}{##3}
        \ifx\relax##5\relax
        \else
            \space(##5)
        \fi
        \dotfill ##4\glspostdescription\space ##6
    }
    
    \renewcommand*{\glsgroupheading}[1]{%
        \item\textbf{\glsgetgrouptitle{##1}}\indexspace
    }
}
\makeatother

\newglossarystyle{documentationLabelStyle}{
    \renewenvironment{theglossary}{%
        \begin{longtable}{p{.2\textwidth}p{.3\textwidth}p{.5\textwidth}}
    }{%
        \end{longtable}
    }%

}

\newglossarystyle{documentationSymbolsStyle}{
    \renewenvironment{theglossary}{%
        \begin{longtable}{p{0.12\textwidth}p{.2\textwidth}p{.48\textwidth}p{.2\textwidth}}
    }{%
        \end{longtable}
    }

}

\newglossarystyle{publicationLabelStyle}{
	\renewenvironment{theglossary}{%
		\begin{longtable}{p{.2\textwidth}p{.8\textwidth}}
		}{%
		\end{longtable}
	}%

}

\newglossarystyle{publicationSymbolsStyle}{
}

\usepackage{scrtime}

\usepackage{ragged2e}

\usepackage{marginnote}

\DeclareDocumentCommand{\myMarginnote} 
					{
				 		O{0cm} O{c} m 
				 	}{
						\marginnote
								{
									\ifthispageodd
											{
												\RaggedRight 
											}{
												\RaggedLeft 
											}
									\raisebox{#1}[#1]
											{
												\begin{minipage}[#2]{\marginparwidth}
													\RaggedRight 
													\color{\getMainColor}
													\lineskiplimit=-\maxdimen
													\normalfont\sffamily
													#3\end{minipage}}
											}
					}

\ifluatex
	\usepackage[luatex]{pdflscape}
\else
\fi

\usepackage
		[
			final, 
		]
		{pdfpages}

\pretocmd{\includepdf}{%
    \ifthenelseproperty{compilation}{externalize}{%
        \tikzset{external/optimize=false}%
    }{}
}{}{}

\makeatletter
\@ifpackageloaded{cleveref}{%
    \newcounter{articlenumber}
    \crefname{articlenumber}{article}{articles}
    \Crefname{articlenumber}{Article}{Articles}

}
\makeatother

\usepackage{enumitem}

\setlist[description]{leftmargin=*,style=sameline}

\ifluatex

	\usepackage{luacode}

\usepackage
		[
			draft, 
		]
		{impnattypo}

\usepackage{selnolig}

    \usepackage{pdftexcmds}
    \makeatletter
    \let\pdfstrcmp\pdf@strcmp
    \makeatother
\else
\fi

\def\clap#1{\hbox to 0pt{\hss#1\hss}}

\newlength{\heightOfX}
\AtBeginDocument{\settoheight{\heightOfX}{X}} 

\newlength{\fsize}

\newcommand{\nobreakbefore}
	{%
		\relax
		\ifvmode
		\else
			\ifhmode
				\ifdim\lastskip > 0pt\relax
					\unskip\nobreakspace
				\fi
    		\fi
  		\fi
	}
\let\oldcite\cite
\renewcommand{\cite}{\nobreakbefore\oldcite}
\let\oldref\ref
\renewcommand{\ref}{\nobreakbefore\oldref}

\newcommand{\disabledprotrusion}[1]
		{
			\ifKOMA
                \ifthenelseproperty{compilation}{fontspec}{%
					\begingroup
						\addfontfeatures{Numbers={Lining,Monospaced}}
                }{}
			\fi
            \ifthenelseproperty{compilation}{microtype}{%
				\microtypesetup{protrusion=false}
            }{}
			#1
            \ifthenelseproperty{compilation}{microtype}{%
				\microtypesetup{protrusion=true}
            }{}
			\ifKOMA
                \ifthenelseproperty{compilation}{fontspec}{%
					\endgroup
                }{}
			\fi
		}

\makeatletter
\newcommand{\removeifnextchar}[3]
		{
			\begingroup
			\ltx@LocToksA{\endgroup#2}
			\ltx@LocToksB{\endgroup#3}
			\ltx@ifnextchar{#1}
				{
					\def\next{\the\ltx@LocToksA}
					\afterassignment\next
					\let\scratch= %
				}{
					\the\ltx@LocToksB
				}
		}
\makeatother

\newcommand{\Signature}[2]
		{%
			\vspace{2cm}%
			\noindent%
            \begin{tabular*}{\textwidth}{@{\extracolsep{0pt}} l @{\extracolsep{\fill}} r @{\extracolsep{0pt}}}%
#1, \getproperty{document}{date}	& \rule{0.33\textwidth}{1pt} 	\\
& \raggedleft{\textsc{#2}}
			\end{tabular*}
			\vspace{1cm}
		}

\AtBeginDocument{%
	\newif\ifKOMAandFancyChapterHeadings%
	\KOMAandFancyChapterHeadingsfalse%
    \ifKOMA%
        \ifthenelseproperty{compilation}{fancychapterheadings}{%
            \RedeclareSectionCommand[
                    beforeskip=\dimexpr3.3\baselineskip+1\parskip\relax,
                    innerskip=0pt,
                    afterskip=1.725\baselineskip plus .115\baselineskip minus .192\baselineskip,
                    afterindent=false
                ]%
                {chapter}%
            \KOMAandFancyChapterHeadingstrue%
        }{}%
    \fi%
    \ifKOMAandFancyChapterHeadings%
    \else%
    \fi%
}

\ifKOMA
    \ifthenelseproperty{compilation}{fontspec}{%
		\DeclareTOCStyleEntry[%
							entryformat={%
                                \addfontfeatures{Numbers={Lining,Monospaced}}\bfseries\sffamily\color{\getproperty{color}{main}}
										},
							pagenumberformat={%
											\addfontfeatures{Numbers={Lining,Monospaced}}\bfseries\sffamily
										},
						]
						{default}
						{chapter}
		\DeclareTOCStyleEntry[%
							entryformat={%
											\addfontfeatures{Numbers={Lining,Monospaced}}
										},
							pagenumberformat={%
											\addfontfeatures{Numbers={Lining,Monospaced}}
										},
						]
						{default}
						{section}
		\DeclareTOCStyleEntry[%
							entryformat={%
											\addfontfeatures{Numbers={Lining,Monospaced}}
										},
							pagenumberformat={%
											\addfontfeatures{Numbers={Lining,Monospaced}}
										},
						]
						{default}
						{subsection}
		\DeclareTOCStyleEntry[%
							entryformat={%
											\addfontfeatures{Numbers={Lining,Monospaced}}
										},
							pagenumberformat={%
											\addfontfeatures{Numbers={Lining,Monospaced}}
										},
						]
						{default}
						{subsubsection}
		\DeclareTOCStyleEntry[%
							entryformat={%
											\addfontfeatures{Numbers={Lining,Monospaced}}
										},
							pagenumberformat={%
											\addfontfeatures{Numbers={Lining,Monospaced}}
										},
						]
						{tocline}
						{figure}
		\DeclareTOCStyleEntry[%
							entryformat={%
											\addfontfeatures{Numbers={Lining,Monospaced}}
										},
							pagenumberformat={%
											\addfontfeatures{Numbers={Lining,Monospaced}}
										},
						]
						{tocline}
						{table}
    }{}
\fi

\makeatletter
\newcommand{\headlesschapter}[1]{%
  \begingroup
  \let\@makechapterhead\@gobble 
  \chapter{#1}
  \endgroup
}
\makeatother

\makeatletter
\newcommand\appendgraphicspath[1]{%
  \g@addto@macro\Ginput@path{#1}%
}
\makeatother

\RequirePackage{xkeyval}
\RequirePackage{xstring}
\RequirePackage{xfp}
\RequirePackage{ifthen}
\RequirePackage{tikzscale} 
\makeatletter
\define@key{igraph}{width}{\def\igraph@width{#1}}
\define@key{igraph}{svgwidth}{\def\igraph@svgwidth{#1}}
\define@key{igraph}{angle}{\def\igraph@angle{#1}}
\define@key{igraph}{draft}{\def\igraph@draft{#1}}
\define@key{igraph}{scale}{\def\igraph@scale{#1}}
\define@key{igraph}{height}{\def\igraph@height{#1}}
\define@boolkey{igraph}{scaletext}{\def\igraph@scaletext{#1}}
\newcommand{\igraph}[2][]{%
    \filename@parse{#2}
    \setkeys{igraph}{width=\linewidth} 
    \setkeys{igraph}{svgwidth=\linewidth} 
    \setkeys{igraph}{height=\empty} 
    \setkeys{igraph}{scaletext=false} 
    \setkeys{igraph}{#1}
    \ifthenelse{%
        \equal{\filename@ext}{pgf}%
    }{%
        \typeout{----- INCLUDE pgf @ #2 ------}%
        \let\pgfimageWithoutPath\pgfimage%
        \renewcommand{\pgfimage}[2][]{\typeout{----- INCLUDE pgfimgage @ ##2 ------}\pgfimageWithoutPath[##1]{\filename@area/##2}}%
        \resizebox{\igraph@width}{!}{\input{#2}}%
    }{%
        \ifnum\pdfstrcmp{\filename@ext}{pdf_tex}=0
            \typeout{----- INCLUDE pdf_tex @ #2 ------}%
            \IfSubStr{#1}{height}{%
                \PackageError{igraph}{pdf_tex does not allow the height attribute}{}%
                }{}%
            \ifthenelse{%
                \boolean{\igraph@scaletext}%
            }{%
                \def\svgwidth{\igraph@svgwidth}%
                \resizebox{\igraph@width}{!}{%
                    \import{\filename@area}{\filename@base.\filename@ext}%
                }%
            }{%
                \def\svgwidth{\igraph@width}%
                \import{\filename@area}{\filename@base.\filename@ext}%
            }%
        \else
            \ifthenelse{%
                \equal{\filename@ext}{svg}%
            }{%
                \typeout{----- INCLUDE svg @ #2 ------}%
                \PackageError{igraph}{svg not implemented!}{}%
            }{%
                \ifthenelse{%
                    \equal{\filename@ext}{tikz}%
                }{%
                    \typeout{----- INCLUDE tikz @ #2 ------}%
	                \tikzsetnextfilename{\filename@base}%
                    \begin{minipage}{\igraph@width}%
                        \input{#2}%
                    \end{minipage}%
                }{%
                    \typeout{----- INCLUDE with includegraphics @ #2 ------}%
                    \IfSubStr{#1}{height}{%
                        \includegraphics[#1]{#2}%
                    }{%
                        \includegraphics[width=\igraph@width, #1]{#2}%
                    }%
                }%
            }%
        \fi
    }%
}%
\makeatother

\newlength{\imageh}
\newlength{\imaged}
\newlength{\imagew}
\newcommand{\setimageh}[1]{
 \settoheight{\imageh}{\usebox{#1}}
}
\newcommand{\setimagew}[1]{
 \settowidth{\imagew}{\usebox{#1}}
}
\newcommand{\setimaged}[1]{
 \settodepth{\imaged}{\usebox{#1}}
}
\newsavebox{\imagesavebox}
\newcommand{\setimagedimensions}[1]{
    \savebox{\imagesavebox}{\igraph{#1}}
    \setimageh{\imagesavebox}
    \setimagew{\imagesavebox}
    \setimaged{\imagesavebox}
}

\makeatletter
\newcommand*{\storedata}[2]{%
  \count@=0 %
  \@tfor\@tmp:=#2\do{%
    \advance\count@\@ne
    \expandafter\let\csname data:\the\count@:#1\endcsname\@tmp
  }%
  \expandafter\edef\csname data:0:#1\endcsname{\the\count@}%
}
\newcommand*{\getdata}[2]{%
  \@ifundefined{data:0:#2}{%
    \@latex@error{Undefined data `#2'}\@ehc
  }{%
    \expandafter\@getdata\expandafter{%
      \the\numexpr
        \ifnum\numexpr(#1)<\z@
          \@nameuse{data:0:#2}+1+%
        \fi
        (#1)%
      \relax
    }{#2}{#1}%
  }%
}
\newcommand*{\@getdata}[3]{%
  \ifnum#1<\z@
    \@getdata@error{\the\numexpr(#3)\relax}{#2}%
  \else
    \ifnum#1>\@nameuse{data:0:#2} %
      \@getdata@error{#1}{#2}%
    \else
      \@nameuse{data:#1:#2}%
    \fi
  \fi
}
\newcommand*{\@getdata@error}[2]{%
  \@latex@error{%
    Wrong data selector #1 for `#2',\MessageBreak
    which only contains \@nameuse{data:0:#2} item(s)%
  }\@ehc
}
\makeatother

\storedata{mydata}{{one}{two}{three}}
\storedata{mylongdata}{
  {one} {two} {three} {four} {five} {six} {seven} {eight} {nine} {ten}
  {eleven} {twelve} {thirteen} {fourteen} {fifteen} {sixteen}
  {seventeen} {eighteen} {nineteen} {twenty} {twenty-one} {twenty-two}
  {twenty-three} {twenty-four}
}

\makeatletter
\RequirePackage{forarray}
\RequirePackage{calc}
\RequirePackage{listings}
\newlength\figratiosum
\newlength\figratio
\newlength\figheight
\newlength\figwidth
\newcounter{npaths}
\newcommand{\multigraph}{\begingroup
  \catcode`_=12 \domultigraph}
\newcommand{\catcodeigraph}{\begingroup
  \catcode`_=11 \igraph}
\define@key{multigraph}{width}{\def\multigraph@width{#1}}
\define@key{multigraph}{labels}{\def\multigraph@labels{#1}}
\newcommand{\domultigraph}[3][]{
    \setkeys{multigraph}{width=\linewidth - 1em} 
    \setkeys{multigraph}{labels=\empty} 
    \setkeys{multigraph}{#1}

    \setcounter{npaths}{0}
    \ForEachX{;}{%
        \stepcounter{npaths}
    }{#2}

    \IfSubStr{#1}{labels}{%
        \def\subfigurelabels{}
        \ForEachX{;}{%
            \edef\subfigurelabels{\subfigurelabels"\thislevelitem"}
            \ifnum\thislevelcount=\value{npaths}
            \else
                \edef\subfigurelabels{\subfigurelabels,}
            \fi
        }{\multigraph@labels}
    }{}

    \def\paths{}
    \setlength\figratiosum{0pt}
    \def\figratios{}
    \ForEachX{;}{%
        \edef\paths{\paths"\thislevelitem"}
        \ifnum\thislevelcount=\value{npaths}
        \else
            \edef\paths{\paths,}
        \fi

        \setimagedimensions{\thislevelitem}
        \setlength\figratio{1pt*\ratio{\imagew}{\imageh}}
        \addtolength{\figratiosum}{\figratio}
        \edef\figratios{\figratios"\the\figratio"}
        \ifnum\thislevelcount=\value{npaths}
        \else
            \edef\figratios{\figratios,}
        \fi
    }{#2}

    \storedata{subcaptions}{#3}

    \setlength{\figheight}{1pt*\ratio{\multigraph@width}{\figratiosum}}

    \addtocounter{npaths}{-1}  
    \foreach \i in {0,...,\value{npaths}}{%
        \pgfmathparse{{\figratios}[\i]}
        \setlength\figwidth{\figheight}
        \pgfmathparse{\figwidth * \pgfmathresult}
        \setlength{\figwidth}{\pgfmathresult pt}
        \begin{subfigure}{\figwidth}
            \centering
            \pgfmathparse{{\paths}[\i]}
            \typeout{----- Width for multigraph figure \pgfmathresult : \the\figwidth ------}
            \catcodeigraph{\pgfmathresult}\endgroup
            \subcaption{\getdata{\i+1}{subcaptions}}
            \IfSubStr{#1}{labels}{%
                \pgfmathparse{{\subfigurelabels}[\i]}
                \label{\pgfmathresult}
            }{}
        \end{subfigure}
    }
    \endgroup
}
\makeatother

\newcommand{\makeup}[1]{%
    \ensuremath{
        \ifluatex
            \symup{#1}
        \else
            \mathrm{#1}
        \fi
    }
}
\newcommand{\makebf}[1]{%
    \ensuremath{
        \ifluatex
            \symbf{#1}
        \else
            \mathbf{#1}
        \fi
    }
}

\newcommand{\eg}{e.\,g.\xspace}

\newcommand{\ie}{i.\,e.\xspace}

\newcommand{\real}{\ensuremath{\mathbb{R}}}

\newcommand{\naturals}{\ensuremath{\mathbb{N}}}  

\newcommand{\vect}[1]{\ensuremath{\vec{#1}}} 
\newcommand{\matr}[1]{\ensuremath{\makebf{#1}}}

\makeatletter
\newcommand{\Grad}[2][\@nil]{%
    \def\tmp{#1}%
    \ifx\tmp\@nnil
    	\ensuremath{\vect{\nabla} #2}
    \else
    	\ensuremath{\vect{\nabla}_{\!\!#1} #2}
    \fi}
\makeatother

\newcommand{\dx}[3][\empty]
		{
			\if{#1}\equal{\empty}
				\frac{\mathrm{d}#2}{\mathrm{d}#3}
			\else
				\frac{\mathrm{d}^{#1}#2}{\mathrm{d}#3^{#1}}
		}
\newcommand{\pdx}[3][\empty]
		{
			\if{#1}\equal{\empty}
				\frac{\partial#3}{\partial#2}
			\else
				\frac{\partial^{#1}#3}{\partial#2^{#1}}
		}
\makeatletter
\providecommand*{\diff}{\@ifnextchar^{\DIfF}{\DIfF^{}}}
\def\DIfF^#1{\mathop{\mathrm{\mathstrut d}}\nolimits^{#1}\gobblespace}
\def\gobblespace{\futurelet\diffarg\opspace}
\def\opspace
		{
			\let\DiffSpace\!
			\ifx\diffarg(
				\let\DiffSpace\relax
			\else
				\ifx\diffarg[%
					\let\DiffSpace\relax
				\else
					\ifx\diffarg\{%
						\let\DiffSpace\relax
					\fi
				\fi
			\fi
			\DiffSpace
		}
\makeatother

\newcommand{\orderof}[1]{\ensuremath{\mathcal{O}(#1)}}

\newcommand{\tightoverset}[2]{\mathop{#2}\limits^{\vbox to -.5ex{\kern-0.75ex\hbox{$\! #1$}\vss}}} 

\DeclareSIUnit[number-unit-product = \,]{\permille}{\textperthousand}

\newacronym{statistics}{Statistics}{\nopostdesc}
\newacronym[parent=statistics,plural=MCs]{MC}{MC}{Monte Carlo}
\newacronym[parent=statistics,plural=MCMCs]{MCMC}{MCMC}{Markov chain Monte Carlo}
\newacronym[parent=statistics]{MAP}{MAP}{maximum a posteriori}
\newacronym[parent=statistics,plural=LAs,longplural={Laplace approximations}]{LA}{LA}{Laplace approximation}
\newacronym[parent=statistics,first=kernel density estimate (KDE) \cite{Parzen1962,Silverman1986},plural=KDEs,firstplural=kernel density estimates (KDEs) \cite{Parzen1962,Silverman1986},longplural={kernel density estimates}]{KDE}{KDE}{kernel density estimate}
\newacronym[parent=statistics,plural=PDFs,longplural={probability density functions}]{PDF}{PDF}{probability density function}
\newacronym[parent=statistics,plural=GPs,longplural={Gaussian processes}]{GP}{GP}{Gaussian process}
\newacronym[parent=statistics,plural=LGIs,longplural={linear Gaussian inversions}]{LGI}{LGI}{linear Gaussian inversion}
\newacronym[parent=statistics,plural=LSs,longplural={least-squares}]{LS}{LS}{least-square}
\newacronym[parent=statistics]{MaLi}{Max-$\mathcal{L}$}{maximum likelihood}
\newacronym[parent=statistics,plural=iid,longplural={independent and identically distributed}]{iid}{iid}{independent and identically distributed}
\newacronym[parent=statistics]{FWHM}{FWHM}{full-width half-maximum}
\newacronym[parent=statistics]{HDI}{HDI}{highest density interval}
\newacronym[parent=statistics]{MHA}{MHA}{Metropolis-Hastings algorithm}
\newacronym[parent=statistics]{KS}{KS}{Kolmogorov–Smirnov}
\newglossaryentry{PearsonCorrelationSample}{%
	parent=statistics,
	name=\ensuremath{r_{\text{xy}}},
	type=symbols,
	sort=correlation,
	description={Pearson sample correlation coefficient},
	symbol={},
}
\newacronym[parent=statistics,plural=HPs,longplural={hyper-parameters}]{HP}{HP}{hyper-parameter}

\newacronym{dynamicalSystems}{Dynamical systems}{\nopostdesc}
\newacronym[parent=dynamicalSystems]{KAM}{KAM}{Kolmogorov–Arnold–Moser}

\newacronym{methods}{Methods}{\nopostdesc}
\newacronym[parent=methods]{FP}{FP}{Function parametrization}
\newacronym[parent=methods]{MISO}{MISO}{multiple-input single-output}
\newacronym[parent=methods]{MIMO}{MIMO}{multiple-input multiple-output}

\newacronym{functionalAnalysis}{Functional analysis}{\nopostdesc}
\newacronym[parent=functionalAnalysis,plural=FCs,longplural={Fourier coefficients}]{FC}{FC}{Fourier coefficient}

\newacronym{differentialEquations}{Differential Equations}{\nopostdesc}
\newacronym[parent=differentialEquations,plural=PDEs,longplural={Partial Differential Equations}]{PDE}{PDE}{Partial Differential Equation}

\newacronym{GeneralPhysics}{General Physics}{\nopostdesc}
\newacronym[parent=GeneralPhysics]{IR}{IR}{Infrared}
\newacronym[parent=GeneralPhysics]{EM}{EM}{electro-magnetic}

\newacronym{Organisational}{Organisational}{\nopostdesc}
\newacronym[parent=Organisational]{OP}{OP}{operation phase}

\newacronym[parent=Organisational]{KKS}{KKS}{\emph{Kraftwerk-Kennzeichensystem} Reference Designation System for Power Plants}
\newacronym[parent=Organisational]{PLM}{PLM}{Project Lifecycle Management}

\newacronym{plasmaTheory}{Plasma Theory}{\nopostdesc}
\newacronym[parent=plasmaTheory,plural=tokamaks]{tokamak}{tokamak}{тороидальная камера в магнитных катушках}
\newacronym[parent=plasmaTheory]{MHD}{MHD}{magnetohydrodynamic}
\newacronym[parent=plasmaTheory]{LCFS}{LCFS}{last closed flux surface}
\newacronym[parent=plasmaTheory]{LCMS}{LCMS}{last closed magnetic surface}
\newacronym[parent=plasmaTheory]{ISS95}{ISS95}{international stellarator scaling 1995}
\newacronym[parent=plasmaTheory]{ISS04}{ISS04}{international stellarator scaling 2004}
\newacronym[parent=plasmaTheory]{NTM}{NTM}{neoclassical tearing modes}
\newacronym[parent=plasmaTheory]{HPP}{HPP}{heat pulse propagation}

\newacronym{plasmaPhysics}{Plasma Physics}{\nopostdesc}

\newacronym[parent=plasmaPhysics]{Omode}{O~mode}{ordinary mode}
\newacronym[parent=plasmaPhysics]{Xmode}{X~mode}{extraordinary mode}
\newacronym[parent=plasmaPhysics]{EBW}{EBW}{electron Bernstein wave}
\newacronym[parent=plasmaPhysics]{TM}{TM}{transverse magnetic}
\newacronym[parent=plasmaPhysics]{TEM}{TEM}{transverse electromagnetic}
\newacronym[parent=plasmaPhysics,plural=EEDFs,firstplural=electron energy distribution functions (EEDFs)]{EEDF}{EEDF}{electron energy distribution function}
\newacronym[parent=plasmaPhysics]{ECE}{ECE}{electron cyclotron emission}

\newacronym[parent=plasmaPhysics]{LFS}{LFS}{low field side}
\newacronym[parent=plasmaPhysics]{HFS}{HFS}{high field side}
\newacronym[parent=plasmaPhysics]{ELM}{ELM}{edge localized mode}
\newacronym[parent=plasmaPhysics]{MARFE}{MARFE}{multifaceted asymmetric radiation from the edge}
\newacronym[parent=plasmaPhysics]{Serpens}{Serpens}{self-regulated plasma edge beneath the last-closed-flux-surface}
\newacronym[parent=plasmaPhysics,plural=SOLs,firstplural=scrape-off layers]{SOL}{SOL}{scrape-off layer}
\newacronym[parent=plasmaPhysics]{ECRH}{ECRH}{electron cyclotron resonance heating}
\newacronym[parent=plasmaPhysics]{ECCD}{ECCD}{electron cyclotron current drive}
\newacronym[parent=plasmaPhysics]{NBCD}{NBCD}{neutral beam current drive}
\newacronym[parent=plasmaPhysics]{ICRH}{ICRH}{ion cyclotron radiation heating}
\newacronym[parent=plasmaPhysics]{FL}{FL}{field line}
\newacronym[parent=plasmaPhysics]{PFR}{PFR}{private flux region}

\newacronym[parent=plasmaPhysics]{see}{\textsc{SEE}}{secondary electron emission}
\newacronym[parent=plasmaPhysics]{SE}{\textsc{SE}}{sheath edge}
\newacronym[parent=plasmaPhysics]{SL}{SL}{strike line}
\newacronym[parent=plasmaPhysics]{PWI}{PWI}{plasma wall interaction}

\newacronym{fusionReactorComponent}{Nuclear Fusion Machine Components}{\nopostdesc}
\newacronym[parent=fusionReactorComponent,plural=PFCs,firstplural=plasma facing components (PFCs)]{PFC}{PFC}{plasma facing component}
\newacronym[parent=fusionReactorComponent,]{VV}{VV}{vaccum vessel}
\newacronym[parent=fusionReactorComponent,plural=IVCs,firstplural=in-vessel components (IVCs)]{IVC}{IVC}{in-vessel component}
\newacronym[parent=fusionReactorComponent,plural=TDUs,firstplural=test divertor units (TDUs)]{TDU}{TDU}{test divertor unit}
\newacronym[parent=fusionReactorComponent,plural=HHFDs, firstplural=high heatflux divertors (HHFDs)]{HHFD}{HHFD}{high heatflux divertor}
\newacronym[parent=fusionReactorComponent,]{TDUSE}{TDUSE}{test divertor unit scraper element}
\newacronym[parent=fusionReactorComponent,]{PG}{PG}{Pumping Gap}
\newacronym[parent=fusionReactorComponent,]{TE}{TE}{Target Element}

\newacronym[parent=fusionReactorComponent]{NBI}{NBI}{neutral beam injection}
\newacronym[parent=fusionReactorComponent,plural=PIs,firstplural=pellet injections (PIs)]{PI}{PI}{pellet injection}

\newacronym{plasmaTheoryCode}{Plasma Theory Codes}{\nopostdesc}
\newacronym[parent=plasmaTheoryCode]{FLD}{FLD}{field line diffusion}
\newacronym[parent=plasmaTheoryCode]{FLT}{FLT}{field line tracing}
\newacronym[parent=plasmaTheoryCode]{LOWENDEL}{LOWENDEL}{``loads on Wendelstein''}
\newacronym[parent=plasmaTheoryCode]{VMEC}{VMEC}{variational moments equilibrium code}
\newacronym[parent=plasmaTheoryCode]{EXTENDER}{EXTENDER}{virtual casing principle code}
\newacronym[parent=plasmaTheoryCode]{PIES}{PIES}{Princeton iterative equilibrium solver}
\newacronym[parent=plasmaTheoryCode]{SPEC}{SPEC}{stepped-pressure equilibrium code}
\newacronym[parent=plasmaTheoryCode]{EMC3E}{EMC3-Eirene}{Edge Monte Carlo 3D code coupled to the neutral transport code EIRENE}
\newacronym[parent=plasmaTheoryCode]{TRAVIS}{TRAVIS}{tracing visualized}
\newacronym[parent=plasmaTheoryCode]{SPECE}{SPECE}{solver for plasma electron cyclotron emission} 
\newacronym[parent=plasmaTheoryCode]{EFIT}{EFIT}{equilibrium fitting}
\newacronym[parent=plasmaTheoryCode]{Theo}{THEODOR}{Thermal Energy onto Divertor}
\newacronym[parent=plasmaTheoryCode]{STRAHL}{STRAHL}{Radial impurity transport and emission code}
\newacronym[parent=plasmaTheoryCode]{ROSE}{ROSE}{ROSE Optimises Stellarator Equilibria}
\newacronym[parent=plasmaTheoryCode]{V3FIT}{V3FIT}{three-dimensional equilibrium reconstruction}
\newacronym[parent=plasmaTheoryCode]{SIMPLE}{SIMPLE}{symplectic integration methods for particle loss estimation}

\newacronym[parent=plasmaTheoryCode]{IDA}{IDA}{integrated data analysis}
\newacronym[parent=plasmaTheoryCode]{BSM}{BSM}{Bayesian scientific modeling}
\newacronym[parent=plasmaTheoryCode]{SBI}{SBI}{simulation-based inference}

\newacronym{database}{Databases}{\nopostdesc}
\newacronym[parent=database]{ADAS}{ADAS}{Atomic Data and Analysis Structure}

\newacronym{plasmaDiagnostic}{Plasma Diagnostics}{\nopostdesc}
\newacronym[parent=plasmaDiagnostic]{MPM}{MPM}{Multi-purpose manipulator}
\newacronym[parent=plasmaDiagnostic]{TS}{TS}{Thomson scattering}
\newacronym[parent=plasmaDiagnostic]{CTS}{CTS}{collective Thomson scattering}
\newacronym[parent=plasmaDiagnostic]{XICS}{XICS}{X-ray imaging crystal spectroscopy}
\newacronym[parent=plasmaDiagnostic]{CXRS}{CXRS}{charge exchange recombination spectroscopy}
\newacronym[parent=plasmaDiagnostic]{LiBES}{LiBES}{lithium beam emission spectroscopy}
\newacronym[parent=plasmaDiagnostic]{DCI}{DCI}{deuterium cyanide laser interferometry}
\newacronym[parent=plasmaDiagnostic,plural=RFMs,firstplural=reflectometers (RFMs)]{RFM}{RFM}{reflectometer}
\newacronym[parent=plasmaDiagnostic]{LP}{LP}{Langmuir probe}
\newacronym[parent=plasmaDiagnostic,sort=Ha-Camera]{HAC}{$H_\alpha$-C}{$H_\alpha$-Camera}
\newacronym[parent=plasmaDiagnostic,sort=Ha-Spectrometer]{HAS}{$H_\alpha$-S}{$H_\alpha$-Spectrometer}
\newacronym[parent=plasmaDiagnostic]{IRC}{IRC}{Infrared-Camera}
\newacronym[parent=plasmaDiagnostic]{HB}{HeBS}{Helium beam spectroscopy}
\newacronym[parent=plasmaDiagnostic]{Hexos}{HEXOS}{High efficiency extreme ultraviolet overview spectrometer}

\newacronym{gasDiagnostic}{Gas Diagnostics}{\nopostdesc}
\newacronym[parent=gasDiagnostic]{DRGA}{DRGA}{Diagnostic residual gas analyser}
\newacronym[parent=gasDiagnostic]{LIF}{LIF}{Laser induced fluorescence}

\newacronym{fusionExperiment}{Nuclear Fusion Experiments}{\nopostdesc}

\newacronym[parent=fusionExperiment,sort=W7]{W7}{\protect\mbox{W7}}{Wendelstein~\protect\mbox{7}}

\newacronym[parent=fusionExperiment,sort=W7-A]{W7A}{\protect\mbox{W7-A}}{Wendelstein~\protect\mbox{7-A}}

\newacronym[parent=fusionExperiment,sort=W7-AS]{W7AS}{\protect\mbox{W7-AS}}{Wendelstein~\protect\mbox{7-AS}}

\newacronym[parent=fusionExperiment,sort=W7-X]{W7X}{\protect\mbox{W7-X}}{Wendelstein~\protect\mbox{7-X}}

\newacronym[parent=fusionExperiment]{NCSX}{NCSX}{National Compact Stellarator Experiment}

\newacronym[parent=fusionExperiment]{CNT}{CNT}{Columbia Non-neutral Torus}

\newacronym[parent=fusionExperiment,sort=LHD]{LHD}{\protect\mbox{LHD}}{large helical device}

\newacronym[parent=fusionExperiment]{WEGA}{WEGA}{Wendelstein Experiment in Greifswald für die Ausbildung} 

\newacronym[parent=fusionExperiment]{HSX}{HSX}{Helically Symmetric Experiment}

\newacronym[parent=fusionExperiment]{JET}{JET}{Joint European Torus}

\newacronym[parent=fusionExperiment]{MAST}{MAST}{Mega-Ampere Spherical Tokamak}

\newacronym[parent=fusionExperiment]{AUG}{AUG}{ASDEX Upgrade}

\newacronym[parent=fusionExperiment]{ASDEX}{ASDEX}{axialsymmetrisches Divertor-Experiment}

\newacronym[parent=fusionExperiment]{EAST}{EAST}{experimental advanced superconducting tokamak}

\newacronym[parent=fusionExperiment]{ITER}{ITER}{lat.~\textit{the way} (formerly International Thermonuclear Experimental Reactor)}

\newacronym[parent=fusionExperiment]{AlcCM}{Alcator C-Mod}{Alto Campo Toro C-Mod} 

\newacronym[parent=fusionExperiment,sort=D]{D3D}{\mbox{DIII-D}}{Doublet III-D}

\newacronym[parent=fusionExperiment,sort=JT60]{JT60}{\mbox{JT-60}}{Japan Torus-60}
\newacronym[parent=fusionExperiment,sort=JT60SA]{JT60SA}{\mbox{JT-60SA}}{Japan Torus-60 Super Upgrade}

\newacronym[parent=fusionExperiment,sort=WEST]{WEST}{\mbox{WEST}}{Tungsten Environment in Steady-state Tokamak (formerly Tore Supra)}

\newacronym[parent=fusionExperiment,sort=TCV]{TCV}{\mbox{TCV}}{Tokamak à configuration variable}

\newacronym[parent=fusionExperiment,sort=TEXT]{TEXT}{\mbox{TEXT}}{Texas Tokamak}
\newacronym[parent=fusionExperiment,sort=TEXTU]{TEXTU}{\mbox{TEXT-U}}{Texas Tokamak - Upgrade}

\newacronym[parent=fusionExperiment,sort=Textor]{Textor}{\mbox{Textor}}{Tokamak Experiment for Technology Oriented Research}

\newacronym[parent=fusionExperiment,sort=Helias]{Helias}{\mbox{Helias}}{helical advanced stellarator}

\newacronym{stellaratorSymmetry}{Stellarator Symmetries}{\nopostdesc}

\newacronym[parent=stellaratorSymmetry,sort=QI]{QI}{QI}{quasi-isodynamic}
\newacronym[parent=stellaratorSymmetry,sort=QO]{QO}{QO}{quasi-omnigeneous}
\newacronym[parent=stellaratorSymmetry,sort=QA]{QA}{QA}{quasi-axisymmetric}
\newacronym[parent=stellaratorSymmetry,sort=QH]{QH}{QH}{quasi-helical}

\newacronym{plasmaLab}{Plasma Physics Laboratories}{\nopostdesc}

\newacronym[parent=plasmaLab]{PPPL}{PPPL}{Princeton Plasma Physics Laporatory}

\newacronym[parent=plasmaLab]{IPP}{IPP}{Max-Planck Institute for Plasma Physics}

\newacronym[sort=W7-X]{magneticConfig}{\mbox{W7-X} Magnetic Configurations}{\nopostdesc}
\newacronym[parent=magneticConfig]{EIM}{\textsc{EIM}}{'Standard' magnetic configuration in the centre of accessible space}
\newacronym[parent=magneticConfig]{DBM}{\textsc{DBM}}{Low iota magnetic configuration}

\newacronym{nuclearReaction}{Nuclear Reactions}{\nopostdesc}
\newacronym[parent=nuclearReaction]{DT}{DT}{Deuterium-Tritium}

\newglossaryentry{physics}{%
    name={General physics quantitites},
    type=symbols,
    description={\nopostdesc},
    symbol={},
}
\newglossaryentry{n}{%
    parent=physics,
    name=\ensuremath{n},
	type=symbols,
	sort=density,
	description={Particle density, $n = N / V$},
    symbol={\si{\per\cubic\meter}}
}
\newglossaryentry{T}{%
    parent=physics,
    name=\ensuremath{T},
	type=symbols,
	sort=temperature,
	description={Temperature},
    symbol={\si{\eV}}
}
\newglossaryentry{energy}{%
    parent=physics,
    name=\ensuremath{W},
	type=symbols,
	sort=energy,
	description={Plasma kinetic energy},
    symbol={\si{\kg\square\meter\per\square\second}}
}
\newglossaryentry{cs}{%
    parent=physics,
    name=\ensuremath{\sigma},
	type=symbols,
	sort=cross-section,
	description={Reaction cross-section},
    symbol={\si{\meter\square}}
}
\newglossaryentry{v}{%
    parent=physics,
    name=\ensuremath{\vect{v}},
	type=symbols,
	sort=velocity,
	description={Particle velocity},
    symbol={\si{\meter\per\second}}
}
\newglossaryentry{me}{%
    parent=physics,
	name=\ensuremath{m_{\text{e}}},
	type=symbols,
	sort= mass ,
	description={Electron mass},
    symbol={\si{\kilo\gram}},
}
\newglossaryentry{mp}{%
    parent=physics,
	name=\ensuremath{m_{\text{p}}},
	type=symbols,
	sort= mass ,
	description={Proton mass},
    symbol={\si{\kilo\gram}},
}

\newglossaryentry{B}{%
    parent=physics,
    name=\ensuremath{\vect{B}},
	type=symbols,
	sort=electromagnetism,
	description={Magnetic field},
    symbol={\si{\tesla} = \si{\kg\per\ampere\per\square\second}}
}
\newglossaryentry{E}{%
    parent=physics,
    name=\ensuremath{\vect{E}},
	type=symbols,
	sort=electromagnetism,
	description={Electric field},
    symbol={\si{\kg\meter\per\ampere\per\cubic\second}}
}
\newglossaryentry{chargedensity}{%
    parent=physics,
    name=\ensuremath{\rho},
	type=symbols,
	sort=electromagnetism,
	description={Electric charge density},
    symbol={\si{\ampere\second\per\cubic\meter}}
}
\newglossaryentry{currentdensity}{%
    parent=physics,
    name=\ensuremath{\vect{j}},
	type=symbols,
	sort=electromagnetism,
	description={Electric current density},
    symbol={\si{\ampere\per\square\meter}}
}
\newglossaryentry{resistance}{%
    parent=physics,
    name=\ensuremath{R},
	type=symbols,
	sort=impedance,
	description={Electric resistance},
    symbol={\si{\ohm}}
}
\newglossaryentry{inductance}{%
    parent=physics,
    name=\ensuremath{L},
	type=symbols,
	sort=inductance,
	description={Inductance},
    symbol={\si{\henry} = \si{\kg\square\meter\per\square\ampere\per\square\second}}
}
\newglossaryentry{larmor}{%
    parent=physics,
	name=\ensuremath{\rho},
	type=symbols,
	sort={Larmor radius},
	description={Larmor radius},
	symbol={\si{\meter}}
}

\newglossaryentry{plasmaphysics}{%
    name={Plasma physics quantitites},
    type=symbols,
    description={\nopostdesc},
    symbol={},
}

\newglossaryentry{vperp}{%
    parent=plasmaphysics,
	name=\ensuremath{v_{\perp}},
	type=symbols,
	sort=velocity,
	description={Field-perpendicular velocity},
	symbol={\si{\meter\per\second}}
}
\newglossaryentry{flowvelocity}{%
    parent=plasmaphysics,
    name=\ensuremath{\vect{v_{\text{f}}}},
	type=symbols,
	sort=velocity,
	description={Plasma flow velocity},
    symbol={\si{\meter\per\second}}
}
\newglossaryentry{p}{%
    parent=plasmaphysics,
    name=\ensuremath{p},
	type=symbols,
	sort=pressure,
	description={Plasma pressure},
    symbol={\si{\kg\per\meter\per\square\second}}
}
\newglossaryentry{V}{%
    parent=plasmaphysics,
    name=\ensuremath{V},
	type=symbols,
	sort=volume,
	description={Plasma volume},
    symbol={\si{\cubic\meter}}
}
\newglossaryentry{viscosity}{%
    parent=plasmaphysics,
    name=\ensuremath{\matr{\pi}},
	type=symbols,
	sort=viscosity,
	description={Plasma viscosity tensor},
    symbol={\si{\kg\per\meter\per\second}}
}
\newglossaryentry{Te}{%
    parent=plasmaphysics,
    name=\ensuremath{T_{\mathrm{e}}},
	type=symbols,
	sort=temperature,
	description={Electron temperature},
    symbol={\si{\eV}}
}
\newglossaryentry{Ti}{%
    parent=plasmaphysics,
	name=\ensuremath{T_{\mathrm{i}}},
	type=symbols,
	sort=temperature,
	description={Ion temperature},
    symbol={\si{\eV}}
}
\newglossaryentry{Tn}{%
    parent=plasmaphysics,
	name=\ensuremath{T_{\mathrm{n}}},
	type=symbols,
	sort=temperature,
	description={Neutral temperature},
	symbol={\si{\eV}}
}
\newglossaryentry{ni}{%
    parent=plasmaphysics,
    name=\ensuremath{n_{\mathrm{i}}},
	type=symbols,
	sort=density,
	description={Ion density},
    symbol={\si{\per\cubic\meter}}
}
\newglossaryentry{ne}{%
    parent=plasmaphysics,
    name=\ensuremath{n_{\mathrm{e}}},
	type=symbols,
	sort=density,
	description={Electron density},
    symbol={\si{\per\cubic\meter}}
}
\newglossaryentry{nn}{%
    parent=plasmaphysics,
	name=\ensuremath{n_{\mathrm{n}}},
	type=symbols,
	sort=density,
	description={Neutral density},
	symbol={\si{\per\cubic\meter}}
}

\newglossaryentry{nimp}{%
    parent=plasmaphysics,
	name=\ensuremath{n_{\mathrm{imp}}},
	type=symbols,
	sort=density,
	description={Impurity density},
	symbol={\si{\per\cubic\meter}}
}

\newglossaryentry{te}{%
    parent=plasmaphysics,
    name=\ensuremath{\tau_{\mathrm{E}}},
	type=symbols,
	sort=time,
	description={Energy confinement time},
    symbol={\si{\second}}
}
\newglossaryentry{radialShift}{%
    parent=plasmaphysics,
    name=\ensuremath{\makeup{\Delta} R},
	type=symbols,
	sort=confinement,
	description={Radial shift of the magnetic axis},
    symbol={\si{\meter}}
}
\newglossaryentry{plasmaBeta}{%
    parent=plasmaphysics,
    name=\ensuremath{\beta},
	type=symbols,
	sort=confinement,
	description={Plasma beta},
    symbol={}
}
\newglossaryentry{Itor}{%
    parent=plasmaphysics,
    name=\ensuremath{I_{\mathrm{tor}}},
	type=symbols,
	sort=confinement,
	description={Toroidal plasma current},
    symbol={\si{\ampere}}
}
\newglossaryentry{Ibs}{%
    parent=plasmaphysics,
    name=\ensuremath{I_{\mathrm{bs}}},
	type=symbols,
	sort=confinement,
	description={Bootstrap current},
    symbol={\si{\ampere}}
}
\newglossaryentry{Ips}{%
    parent=plasmaphysics,
    name=\ensuremath{I_{\mathrm{ps}}},
	type=symbols,
	sort=confinement,
	description={Pfirsch-Schlueter current},
    symbol={\si{\ampere}}
}
\newcommand{\quer}[1]{\mathrel{\hbox{-}\mkern-6.55mu #1}}
\newglossaryentry{ibar}{%
    parent=plasmaphysics,
    name=\ensuremath{\quer{\iota}},
	type=symbols,
	sort=confinement,
	description={Rotational transform},
    symbol={}
}

\newglossaryentry{shear}{%
	parent=plasmaphysics,
	name=\ensuremath{s},
	type=symbols,
	sort=confinement,
	description={Shear, radial derivative of \ensuremath{\quer{\iota}}},
	symbol={}
}

\newglossaryentry{ibarCF}{%
    parent=plasmaphysics,
    name=\ensuremath{\quer{\iota}_{\mathrm{CF}}},
	type=symbols,
	sort=confinement,
	description={Current free rotational transform},
    symbol={}
}

\newglossaryentry{Vp}{%
    parent=plasmaphysics,
    name=\ensuremath{V_\mathrm{p}},
	type=symbols,
	sort=potential,
	description={Plasma potential},
    symbol={\si{\volt}}
}
\newglossaryentry{Vf}{%
    parent=plasmaphysics,
    name=\ensuremath{V_\mathrm{f}},
	type=symbols,
	sort=potential,
	description={Floating potential},
    symbol={\si{\volt}}
}
\newglossaryentry{Isat}{%
    parent=plasmaphysics,
    name=\ensuremath{I_\text{sat}},
	type=symbols,
	sort=current,
	description={Ion saturation current},
    symbol={\si{\ampere}}
}	
\newglossaryentry{esat}{%
	parent=plasmaphysics,
	name=\ensuremath{I_{e,\text{sat}}},
	type=symbols,
	sort=current,
	description={Electron saturation current},
	symbol={\si{\ampere}}
}	
\newglossaryentry{jsat}{%
    parent=plasmaphysics,
    name=\ensuremath{j_\text{sat}},
	type=symbols,
	sort=current,
	description={Ion saturation current density},
    symbol={\si{\ampere\per\square\meter}}
}
\newglossaryentry{csound}{%
    parent=plasmaphysics,
    name=\ensuremath{c_\mathrm{s}},
	type=symbols,
	sort=velocity,
	description={Ion sound speed},
    symbol={\si{\meter\per\second}}
}
\newglossaryentry{Vbias}{%
    parent=plasmaphysics,
    name=\ensuremath{V_\mathrm{bias}},
	type=symbols,
	sort=potential,
	description={Bias voltage},
    symbol={\si{\volt}}
}
\newglossaryentry{r}{%
    parent=plasmaphysics,
    name=\ensuremath{r},
	type=symbols,
	sort=radius,
	description={Minor radius},
    symbol={\si{\meter}}
}
\newglossaryentry{ra}{%
    parent=plasmaphysics,
    name=\ensuremath{r_a},
	type=symbols,
	sort=radius,
	description={Minor radius of the last closed flux surface},
    symbol={\si{\meter}}
}
\newglossaryentry{reff}{%
    parent=plasmaphysics,
    name=\ensuremath{r_{\mathrm{eff}}},
	type=symbols,
	sort=radius,
	description={Effective minor radius},
    symbol={\si{\meter}}
}
\newglossaryentry{R}{%
    parent=plasmaphysics,
    name=\ensuremath{R},
	type=symbols,
	sort=radius,
	description={Major radius},
    symbol={\si{\meter}}
}
\newglossaryentry{normminrad}{%
	parent=plasmaphysics,
	name=\ensuremath{\varrho},
	type=symbols,
	sort=radius,
	description={Normalised minor radius},
	symbol={}
}
\newglossaryentry{aspectRatio}{%
    parent=plasmaphysics,
    name=\ensuremath{\epsilon},
	type=symbols,
	sort=radius,
	description={Aspect ratio},
    symbol={\si{}}
}
\newglossaryentry{Nfp}{%
    parent=plasmaphysics,
    name=\ensuremath{N_{\text{fp}}},
	type=symbols,
	sort=number,
	description={Number of field periods},
    symbol={\si{}},
}
\newglossaryentry{phiEdge}{%
    parent=plasmaphysics,
    name=\ensuremath{\phi_{\mathrm{edge}}},
	type=symbols,
	sort=magnetic flux,
	description={Total enclosed magnetic toroidal flux},
    symbol={\si{\volt\second}}
}

\newglossaryentry{Pconv}{%
    parent=plasmaphysics,
    name=\ensuremath{P_{\mathrm{conv}}},
	type=symbols,
	sort=power,
	description={Convective power},
    symbol={\si{\watt}}
}
\newglossaryentry{Prad}{%
    parent=plasmaphysics,
    name=\ensuremath{P_{\mathrm{rad}}},
	type=symbols,
	sort=power,
	description={Radiated power},
    symbol={\si{\watt}}
}
\newglossaryentry{Pheat}{%
    parent=plasmaphysics,
    name=\ensuremath{P_{\mathrm{heat}}},
	type=symbols,
	sort=power,
	description={Total heating power},
    symbol={\si{\watt}}
}
\newglossaryentry{Lc}{%
    parent=plasmaphysics,
    name=\ensuremath{L_{\mathrm{c}}},
	type=symbols,
	sort=length,
	description={Connection length},
    symbol={\si{\meter}}
}
\newglossaryentry{heatflux}{%
    parent=plasmaphysics,
    name=\ensuremath{q},
	type=symbols,
	sort=power flux,
	description={Heat flux or heat load},
    symbol={\si{\MW\per\square\meter}}
}
\newglossaryentry{Zav}{%
    parent=plasmaphysics,
    name=\ensuremath{Z_{\mathrm{av}}},
	type=symbols,
	sort=ion charge,
	description={Average ion charge},
    symbol={\si{}},
}
\newglossaryentry{Zeff}{%
    parent=plasmaphysics,
    name=\ensuremath{Z_{\mathrm{eff}}},
	type=symbols,
	sort=ion charge,
	description={Effective ion charge},
    symbol={\si{}},
}
\newglossaryentry{iongyro}{%
    parent=plasmaphysics,
    name=\ensuremath{\rho_{\text{i}}},
	type=symbols,
	sort=radius,
	description={Ion gyro (Lamor) radius},
    symbol={\si{\meter}},
}
\newglossaryentry{mi}{%
    parent=plasmaphysics,
    name=\ensuremath{m_{\text{i}}},
	type=symbols,
	sort=mass,
	description={Ion mass},
    symbol={\si{\kilo\gram}},
}
\newglossaryentry{ndl}{%
	parent=plasmaphysics,
	name=\ensuremath{n\text{d}\ell},
	type=symbols,
	sort=density,
	description={Line integrated density},
	symbol={\si{\per\square\meter}},
}
\newglossaryentry{Wdia}{%
	parent=plasmaphysics,
	name=\ensuremath{W_\text{dia}},
	type=symbols,
	sort=energy,
	description={Diamagnetic energy},
	symbol={\si{\joule}},
}

\newglossaryentry{epseff}{%
	parent=plasmaphysics,
	name=\ensuremath{\epsilon_\text{eff}},
	type=symbols,
	sort=effective ripple,
	description={Effective ripple},
	symbol={\si{}},
}

\newglossaryentry{Halpha}{%
    parent=plasmaphysics,
	name=\ensuremath{\text{H}_{α}\xspace},
	type=symbols,
	sort= Wavelengths ,
	description={Hydrogen $\alpha$ transition line, Balmer series transition $n=3 \rightarrow n=2$, \SI{656.5}{\nano\meter}},
	symbol=\si{\nano\meter},
}
\newglossaryentry{sxb}{%
    parent=plasmaphysics,
	name=\ensuremath{\text{S/XB}},
	type=symbols,
	sort= Coefficients,
	description={Ratio of ionisation, excitation and branching ratio. Inverse photons per neutral.},
	symbol=,
}
\newglossaryentry{particleFlux}{%
    parent=plasmaphysics,
	name=\ensuremath{\Gamma},
	type=symbols,
	sort= Particle flux,
	description={Particle flux},
	symbol=\si{\per\second\per\square\meter},
}
\newglossaryentry{stf}{%
    parent=plasmaphysics,
	name=\ensuremath{\gamma_{\text{s}}},
	type=symbols,
	sort=Sheath transmission coefficient,
	description={Sheath transmission coefficient},
    symbol={\si{}},
}
\newglossaryentry{frad}{%
    parent=plasmaphysics,
	name=\ensuremath{f_{\text{rad}}},
	type=symbols,
	sort=Fraction,
	description={Radiated power fraction},
    symbol={\si{}},
}
\newglossaryentry{mfpi}{%
	parent=plasmaphysics,
	name=\ensuremath{\lambda_\text{mfp,i}},
	type=symbols,
	sort=Length,
	description={Mean free path length of ionisation},
	symbol={\si{\meter}},
}
\newglossaryentry{frec}{%
	parent=plasmaphysics,
	name=\ensuremath{f_\text{rec}},
	type=symbols,
	sort=Fraction,
	description={Fraction of ions recycled as neutrals at PFCs},
	symbol={\si{}},
}
\newglossaryentry{Pecrh}{%
	parent=plasmaphysics,
	name=\ensuremath{P_\text{ECRH}},
	type=symbols,
	sort=Power,
	description={ECRH Power},
	symbol={\si{\watt}},
}

\newcommand{\plasmaBeta}{\gls{plasmaBeta}\xspace}
\newcommand{\Itor}{\gls{Itor}\xspace}
\newcommand{\ibar}{\gls{ibar}\xspace}
\newcommand{\epseff}{\gls{epseff}\xspace}

\newacronym{engineering}{Engineering}{\nopostdesc}
\newacronym[parent=engineering]{MPCDF}{MPCDF}{Max Planck computing and data facility}

\newacronym[parent=engineering]{CAD}{CAD}{computer-aided design}
\newacronym[parent=engineering]{SOA}{SOA}{service-oriented architecture}
\newacronym[parent=engineering]{ESB}{ESB}{enterprise service bus}
\newacronym[parent=engineering]{HPC}{HPC}{high-performance computing}
\newacronym[parent=engineering,plural=GPUs,longplural={graphical processing units}]{GPU}{GPU}{graphical processing unit}

\newacronym[parent=engineering]{DAQ}{DAQ}{data acquisition}
\newacronym[parent=engineering]{ADC}{ADC}{analog-to-digital converter}
\newacronym[parent=engineering]{LO}{LO}{local oscillator}
\newacronym[parent=engineering]{PLL}{PLL}{phase locked loop}
\newacronym[parent=engineering]{CPU}{CPU}{central processing unit}
\newacronym[parent=engineering]{OAMP}{OAmp}{operational amplifier}
\newacronym[parent=engineering]{DAMP}{DAmp}{differential amplifier}
\newacronym[parent=engineering]{PCB}{PCB}{printed circuit board}
\newacronym[parent=engineering]{CMRR}{CMRR}{common mode rejection ratio}
\newacronym[parent=engineering]{IC}{IC}{integrated circuit}
\newacronym[parent=engineering]{CCD}{CCD}{charge-coupled device}
\newacronym[parent=engineering]{PMT}{PMT}{photo-multiplyer tube}
\newacronym[parent=engineering]{ROI}{ROI}{region of interest}
\newacronym[parent=engineering,plural=LOSs,longplural={lines of sight}]{LOS}{LOS}{line of sight}
\newacronym[parent=engineering]{EDICAM}{EDICAM}{Event Detection Intelligent Camera}

\newacronym[parent=engineering]{CFC}{CFC}{carbon fibre reinforced carbon}

\newacronym{machineLearning}{Machine Learning}{\nopostdesc}
\newacronym[parent=machineLearning]{ML}{ML}{machine learning}
\newacronym[parent=machineLearning]{RL}{RL}{reinforcement learning}
\newacronym[parent=machineLearning,plural=NNs,firstplural=artificial neural networks (NNs)]{NN}{NN}{artificial neural network}
\newacronym[parent=machineLearning]{EI}{EI}{expected improvement}

\newacronym[parent=machineLearning,plural=FF-FCs,firstplural=feed-forward fully-connected neural networks (FF-FC)]{FFFC}{FF-FC}{feed-forward fully-connected neural network}
\newacronym[parent=machineLearning,plural=MLPs,firstplural=multilayer perceptrons (MLP)]{MLP}{MLP}{multilayer perceptron}
\newacronym[parent=machineLearning,plural=CNNs,firstplural=convolutional neural networks (CNNs)]{CNN}{CNN}{convolutional neural network}
\newacronym[parent=machineLearning,plural=DCNNs,firstplural=deep convolutional neural networks (DCNNs)]{DCNN}{DCNN}{deep convolutional neural network}
\newacronym[parent=machineLearning,plural=IRNNs,firstplural=Inception-ResNets (IRNNs)]{IRNN}{IRNN}{Inception-ResNet}
\newacronym[parent=machineLearning,plural=DINNs,firstplural=deep inception neural networks (DINNs)]{DINN}{DINN}{deep inception neural network}
\newacronym[parent=machineLearning,plural=GANs,firstplural=generative adversarial neural networks (GANs)]{GAN}{GAN}{generative adversarial neural network}
\newacronym[parent=machineLearning,plural=DQNs,firstplural=deep Q-networks]{DQN}{DQN}{deep Q-network}
\newacronym[parent=machineLearning,plural=AEs,firstplural=autoencoders]{AE}{AE}{autoencoder}
\newacronym[parent=machineLearning,plural=VAEs,firstplural=variational autoencoders]{VAE}{VAE}{variational autoencoder}
\newacronym[parent=machineLearning,plural=PiNNs,firstplural=physics informed neural networks]{PiNN}{PiNN}{physics informed neural network}
\newacronym[parent=machineLearning,plural=DeepONets,firstplural=deep operator networks]{DeepONet}{DeepONet}{deep operator network}

\newacronym[parent=machineLearning]{MNIST}{MNIST}{MNIST} 

\newacronym[parent=machineLearning]{ReLU}{ReLU\xspace}{rectified linear unit}
\newacronym[parent=machineLearning]{SeLU}{SeLU\xspace}{scaled exponential linear unit}
\newacronym[parent=machineLearning]{SiLU}{SiLU\xspace}{sigmoid linear unit}

\newacronym[parent=machineLearning]{rmse}{rmse\xspace}{root-mean-square error}
\newacronym[parent=machineLearning]{mse}{mse\xspace}{mean-squared error}
\newacronym[parent=machineLearning]{msd}{msd\xspace}{mean-squared deviation}
\newacronym[parent=machineLearning]{nmse}{nmse\xspace}{normalized mean-squared error}
\newacronym[parent=machineLearning]{rmsd}{rmsd\xspace}{root-mean-squared deviation}
\newacronym[parent=machineLearning]{nrmsd}{nrmsd\xspace}{normalized root-mean-squared deviation}
\newacronym[parent=machineLearning]{nrmse}{nrmse\xspace}{normalized root-mean-squared error}
\newacronym[parent=machineLearning]{mae}{mae\xspace}{mean absolute error}
\newacronym[parent=machineLearning]{mape}{mape\xspace}{mean absolute percentage error}

\newacronym[parent=machineLearning]{TPE}{TPE}{tree-structured Parzen estimator}
\newacronym[parent=machineLearning]{NAS}{NAS}{neural architecture search}
\newacronym[parent=machineLearning]{NES}{NES}{neural ensemble search}
\newacronym[parent=machineLearning]{SMBO}{SMBO}{sequential model-based optimization}
\newacronym[parent=machineLearning]{BFGS}{BFGS}{Broyden–Fletcher–Goldfarb–Shanno}
\newacronym[parent=machineLearning]{LBFGS}{L-BFGS}{limited-memory Broyden–Fletcher–Goldfarb–Shanno}
\newacronym[parent=machineLearning]{AD}{AD}{automatic differentiation}

\newglossaryentry{ml}{%
    name={Machine learning quantitites},
    type=symbols,
    description={\nopostdesc},
    symbol={},
}
\newglossaryentry{lossfunction}{%
    name=\ensuremath{\mathcal{L}},
	type=symbols,
	parent=ml,
	sort=loss,
	description={Loss function},
    symbol={\si{}}
}

\newcommand{\averagePlasmaBeta}{\ensuremath{\langle \gls{plasmaBeta} \rangle}\xspace}

\newcommand{\lCore}{\ensuremath{l_{\text{core}}}\xspace}
\newcommand{\lEdge}{\ensuremath{l_{\text{edge}}}\xspace}
\newcommand{\tLocation}{\ensuremath{s_0}\xspace}
\newcommand{\tWidth}{\ensuremath{s_w}\xspace}

\newcommand{\Bmax}{\ensuremath{B_{\text{max}}}\xspace}
\newcommand{\Bmin}{\ensuremath{B_{\text{min}}}\xspace}

\newcommand{\psiedge}{\ensuremath{\psi_{\text{edge}}}\xspace}
\newcommand{\phiedge}{\ensuremath{\Phi_{\text{edge}}}\xspace}
\newcommand{\epseffproxy}{\ensuremath{\hat{\epsilon}_\text{eff}}\xspace}

\newcommand\twoNorm[1]{\ensuremath{\lVert#1\rVert_2}}
\newcommand{\quotes}[1]{``#1''}
\newcommand{\average}[1]{\overline{#1}}

\definecolor{lightNullColor}{HTML}{d1e5f0}
\definecolor{nullColor}{HTML}{67a9cf}
\definecolor{darkNullColor}{HTML}{2166ac}
\definecolor{lightFiniteColor}{HTML}{fddbc7}
\definecolor{finiteColor}{HTML}{ef8a62}
\definecolor{darkFiniteColor}{HTML}{b2182b}

\DeclareSIUnit{\nothing}{\relax}

\DeclareSIUnit{\arbitraryunit}{a.u.}

\setproperty{document}{title}{Physics-regularized neural network of the ideal-MHD solution operator in Wendelstein 7-X configurations}
\setproperty{author}{firstname}{Andrea}
\setproperty{author}{familyname}{Merlo}
\setproperty{coauthor1}{firstname}{Daniel}
\setproperty{coauthor1}{familyname}{Böckenhoff}
\setproperty{coauthor1}{affiliationindices}{1}
\setproperty{coauthor2}{firstname}{Jonathan}
\setproperty{coauthor2}{familyname}{Schilling}
\setproperty{coauthor2}{affiliationindices}{1}
\setproperty{coauthor8}{firstname}{Samuel Aaron}
\setproperty{coauthor8}{familyname}{Lazerson}
\setproperty{coauthor8}{affiliationindices}{1}
\setproperty{coauthor9}{firstname}{Thomas Sunn}
\setproperty{coauthor9}{familyname}{Pedersen}
\setproperty{coauthor9}{affiliationindices}{1}
\setproperty{coauthors}{W7Xteaminclude}{True}
\setproperty{affiliations}{1}{Max-Planck-Institute for Plasma Physics, 17491 Greifswald, Germany}

\definechangesauthor[name={Andrea Merlo}, color=red]{AM}
\definechangesauthor[name={Daniel Böckenhoff}, color=blue]{DB}
\definechangesauthor[name={Jonathan Schilling}, color=orange]{JS}
\definechangesauthor[name={Thomas Sunn Pedersen}, color=magenta]{TP}
\definechangesauthor[name={Samuel Aaron Lazerson}, color=green]{SL}

\makeatletter
\let\blx@rerun@biber\relax
\makeatother

\DeclareSIUnit{\nothing}{\relax}

\DeclareSIUnit{\arbitraryunit}{a.u.}

\let\pgfimage=\includegraphics
\graphicspath{{./}}

\begin{document}

    \typeout{----- BEGIN DOCUMENT -----}
    \typeout{}
    \typeout{--------------------------------}
    \typeout{----- Document properties: -----}
    \typeout{--------------------------------}
    \typeout{Font size: \the\fsize}
    \typeout{Text width: \the\textwidth}
    \typeout{--------------------------------}
    \typeout{--------------------------------}

    \ifthenelseproperty{compilation}{frontmatter}{%
	\frontmatter
}{}

\ifthenelseproperty{compilation}{titlepage}{%
    \ifKOMA
    \pagestyle{empty}%
    \title{\getproperty{document}{title}%
    \ifpropertydefined{document}{status}{%
    \thanks{
        This is the Accepted Manuscript version of an article accepted for publication in \getproperty{document}{journal}. \getproperty{document}{editor} is not responsible for any errors or omissions in this version of the manuscript or any version derived from it. This Accepted Manuscript is published under a \getproperty{document}{license} licence. The Version of Record is available online at \url{\getproperty{document}{doi}}
    }%
    }}%
    \author[\getproperty{author}{affiliationindices}]{\getproperty{author}{firstname} \getproperty{author}{familyname}}
    \affil[1]{\getproperty{affiliations}{1}}
    
    \ifpropertydefined{coauthor1}{familyname}{
    \author[\getproperty{coauthor1}{affiliationindices}]{\getproperty{coauthor1}{firstname} \getproperty{coauthor1}{familyname}}}{}

    \ifpropertydefined{coauthor2}{familyname}{
    \author[\getproperty{coauthor2}{affiliationindices}]{\getproperty{coauthor2}{firstname} \getproperty{coauthor2}{familyname}}}{}

    \ifpropertydefined{coauthor3}{familyname}{
    \author[\getproperty{coauthor3}{affiliationindices}]{\getproperty{coauthor3}{firstname} \getproperty{coauthor3}{familyname}}}{}

    \ifpropertydefined{coauthor4}{familyname}{
    \author[\getproperty{coauthor4}{affiliationindices}]{\getproperty{coauthor4}{firstname} \getproperty{coauthor4}{familyname}}}{}

    \ifpropertydefined{coauthor5}{familyname}{
    \author[\getproperty{coauthor5}{affiliationindices}]{\getproperty{coauthor5}{firstname} \getproperty{coauthor5}{familyname}}}{}

    \ifpropertydefined{coauthor6}{familyname}{
    \author[\getproperty{coauthor6}{affiliationindices}]{\getproperty{coauthor6}{firstname} \getproperty{coauthor6}{familyname}}}{}

    \ifpropertydefined{coauthor7}{familyname}{
    \author[\getproperty{coauthor7}{affiliationindices}]{\getproperty{coauthor7}{firstname} \getproperty{coauthor7}{familyname}}}{}

    \ifpropertydefined{coauthor8}{familyname}{
    \author[\getproperty{coauthor8}{affiliationindices}]{\getproperty{coauthor8}{firstname} \getproperty{coauthor8}{familyname}}}{}

    \ifpropertydefined{coauthor9}{familyname}{
    \author[\getproperty{coauthor9}{affiliationindices}]{\getproperty{coauthor9}{firstname} \getproperty{coauthor9}{familyname}}}{}

    \ifpropertydefined{coauthor10}{familyname}{
    \author[\getproperty{coauthor10}{affiliationindices}]{\getproperty{coauthor10}{firstname} \getproperty{coauthor10}{familyname}}}{}

    \ifpropertydefined{coauthors}{W7Xteaminclude}{
    \author[ ]{the W7-X team}}{}
    
    \ifpropertydefined{affiliations}{2}{
        \affil[2]{\getproperty{affiliations}{2}}
    }{}
    \ifpropertydefined{affiliations}{3}{
        \affil[3]{\getproperty{affiliations}{3}}
    }{}
    \ifpropertydefined{affiliations}{4}{
        \affil[4]{\getproperty{affiliations}{4}}
    }{}
    \ifpropertydefined{affiliations}{5}{
        \affil[5]{\getproperty{affiliations}{5}}
    }{}
    \ifpropertydefined{affiliations}{*}{
        \affil[*]{\getproperty{affiliations}{*}}
    }{}
        
    \date{\today}%
    \maketitle%
    \pagestyle{scrheadings}
\else
	\pagestyle{empty}
	\title{\getproperty{document}{title}}
	\author{\getproperty{author}{firstname} \getproperty{author}{familyname}}
	\date{\getproperty{document}{date}}
	\maketitle
\fi

}{}

\ifthenelseproperty{compilation}{abstract}{%
	\section*{Abstract}

\sisetup{scientific-notation = false}

\deleted[id=AM]{The stellarator is a promising concept to produce energy from nuclear fusion by magnetically confining a high-pressure plasma.}
\replaced[id=AM]{
    The computational cost of constructing \num{3}D \gls{MHD} equilibria is one of the limiting factors in stellarator research and design.
}{%
    In a stellarator,
    the confining field is three-dimensional,
    and the computational cost of solving the \num{3}D \gls{MHD} equations currently limits stellarator research and design.
}
Although data-driven approaches have been proposed to provide fast \num{3}D \gls{MHD} equilibria,
the accuracy with which equilibrium properties are reconstructed is unknown.
In this work,
we describe an \gls{NN} that quickly approximates the ideal-\gls{MHD} solution operator in \gls{W7X} configurations.
This model fulfils equilibrium symmetries by construction.
The \gls{MHD} force residual regularizes the solution of the \gls{NN} to satisfy the ideal-\gls{MHD} equations.
The model predicts the equilibrium solution with high accuracy,
and it faithfully reconstructs global equilibrium quantities and proxy functions used in stellarator optimization.
\deleted[id=AM]{
    The regularization term enforces that the \gls{NN} reduces the ideal-\gls{MHD} force residual,
    and solutions that are better than ground truth equilibria can be obtained at inference time.
}
We also optimize \gls{W7X} magnetic configurations,
where desiderable configurations can be found in terms of fast particle confinement.
This work demonstrates with which accuracy \gls{NN} models can approximate the \num{3}D ideal-\gls{MHD} solution operator and reconstruct equilibrium properties of interest,
and it suggests how they might be used to optimize stellarator magnetic configurations.
\glsresetall%

}{}

\ifthenelseproperty{compilation}{glossaries}{
	\glsresetall
}{}

\ifthenelseproperty{compilation}{toc}{%
    \disabledprotrusion{\tableofcontents}
}{}

\makeatletter
\@ifundefined{mainmatter}{}{\mainmatter}
\makeatother



\section{Introduction}\label{sec:introduction}

\deleted[id=AM]{
\glsdisp{MHD}{Magnetohydrodynamics (MHD)} describes how plasma pressure, current density and magnetic field interact.
In magnetic confinement fusion,
an externally applied magnetic field confines the hot plasma.
The plasma reacts its own magnetic field.
A stellarator fusion reactor will operate at high plasma pressure,
namely,
\mbox{$\averagePlasmaBeta\sim$\SI{5}{\percent}}~\mbox{\cite{Warmer2016}},
where \averagePlasmaBeta is the volume averaged plasma beta (\mbox{$\plasmaBeta = \frac{2 \mu_0 p}{B^2}$}),
which is the ratio between the kinetic plasma pressure and the magnetic field pressure.
}

The computational cost of \replaced[id=AM]{constructing}{solving} \num{3}D \gls{MHD} equilibria \replaced[id=AM]{is one major limiting factor in}{currently limits} stellarator theory, experimentation and design.
Depending on the desired resolution and accuracy of the solution,
such computations require up to \orderof{\si{10}{}} CPUh~\cite{Seal2017,Panici2022}.

Fast and accurate equilibrium reconstructions are crucial to interpret experimental results in magnetically confined devices.
At \gls{W7X},
the Bayesian \deleted[id=AM]{\gls{SBI} }framework MINERVA~\cite{Svensson2007,Svensson2010} reconstructs the plasma state (\eg, ion and electron temperature).
For each forward evaluation of the simulation model,
a free-boundary ideal-\gls{MHD} equilibrium has to be \replaced[id=AM]{constructed}{solved}.
For each reconstructed state (\ie, a single timestamp),
between \orderof{\num{e3}} and \orderof{\num{e6}} forward evaluations are required~\cite{Schilling2018,Hoefel2019}.
Due to the computational cost of \replaced[id=AM]{constructing}{solving} free-boundary ideal-\gls{MHD} equilibria,
a self-consistent Bayesian inversion of a single plasma state has been attempted only once~\cite{Schilling2018} and is not part of the standard data evaluation procedure.

\gls{W7X} is currently the largest stellarator experiment in operation.
Its scientific mission is to assess the stellarator line on the path towards a fusion reactor~\cite{Grieger1993}.
The configuration space of \gls{W7X} provides a large experimental test bed,
however,
only nine reference configurations are mainly investigated~\cite{Andreeva2002}.
Experimental time is limited,
and priorities must be established to fit financial and human resources.
Flight simulators provide a parallel and cheaper path to explore and exploit current and future experiments~\cite{Fable2021,Morris2022}.
In a stellarator,
\num{3}D \gls{MHD} equilibria bound the accuracy and cost of simulation.

\replaced[id=AM]{Differentiable and less expensive \gls{MHD} equilibrium solutions}{Faster and differentiable \gls{MHD} solvers} would allow a more extensive exploration of the stellarator optimization space
\replaced[id=AM]{, in which t}{.
Stellarator optimization involves the search of an optimal magnetic field to confine the plasma.
T}he number of degrees of freedom is usually large \mbox{$\sim\orderof{\num{e2}}$}~\cite{Xanthopoulos2014,Landreman2022a}.
The computational cost of the objective function,
which is largely dominated by \replaced[id=AM]{deriving}{solving the} \added[id=AM]{ideal-}\gls{MHD} \replaced[id=AM]{equilibria}{equations},
limits the extent to which we can explore the optimization space.
\deleted[id=AM]{
Moreover,
when using a finite-differences method to approximate the gradient of the objective function,
the number of function evaluations scales linearly with the degrees of freedom.
}
According to recent works,
each optimization run requires up to \orderof{\num{e3}} function evaluations~\cite{Xanthopoulos2014,Paul2018,Paul2021}.

Data-driven approaches (\eg, \glspl{NN}) can provide fast 3D ideal-\gls{MHD} equilibria~\cite{VanMilligen1995,Callaghan1999,Callaghan2000,Sengupta2004,Sengupta2007,Merlo2021}.
When benchmarked against equilibria from non-linear \replaced[id=AM]{codes}{solvers} (\eg, the \gls{VMEC}),
the geometry of the flux surfaces is in excellent agreement.
However,
it is unclear how equilibrium properties (\eg, plasma stability) can be faithfully reproduced by the \gls{NN} model.
Furthermore,
it is unclear how well the \gls{NN} equilibria satisfy the \gls{MHD} equations.
To make use of such \gls{NN} models in downstream tasks
(\eg, \replaced[id=AM]{Bayesian inference}{\gls{SBI}}, optimization),
it is necessary to investigate how well \glspl{NN} capture equilibrium properties.

The ideal-\gls{MHD} equilibrium problem is a system of \glspl{PDE}.
In general,
initial and boundary conditions,
as well as input parameters define the \gls{PDE} problem.
The solution operator maps these variable terms to the corresponding \gls{PDE} solution.

In this paper,
we extend~\cite{Merlo2021} by learning the solution operator of the ideal-\gls{MHD} equilibrium problem in the subspace of \gls{W7X} magnetic configurations with a \gls{DeepONet}\cite{Lu2020a}.
A \gls{DeepONet} is a \gls{NN} macro-architecture with two pathways:
a branch network to encode the \gls{PDE} problem inputs,
and a trunk network to represent the domain of solution function.
Here,
the \gls{NN} model maps an externally applied field (defined by the currents in the coil system) and a plasma state (defined by the pressure and toroidal current profiles) to \added[id=AM]{an approximation of }the ideal-\gls{MHD} equilibrium solution.
Within the boundary of the training data (\ie, \gls{W7X} magnetic configurations),
the \gls{NN} model approximates the solution of unseen equilibrium problems at a fraction of the computational cost currently required by a traditional \replaced[id=AM]{code}{solver}.
Summarizing the paper's contributions,
we:
\begin{itemize}
\item{
  propose a \gls{DeepONet} like architecture to represent the solution operator of the ideal-\gls{MHD} equilibrium problem (see~\Cref{sec:mhdnet});
}
\item{
  train the model to \replaced[id=AM]{approximate}{predict} the solution of ideal-\gls{MHD} equilibrium problem obtained from a non-linear \replaced[id=AM]{code}{solver} (\eg, \gls{VMEC}),
  while the ideal-\gls{MHD} force residual regularizes the model's solution (see~\Cref{sec:data});
}
\item{
  investigate how well the model's solution satisfies the ideal-\gls{MHD} equations,
  and propose a strategy to improve the solution at inference time by \replaced[id=AM]{approximating}{solving} an equivalent fixed-boundary equilibrium \replaced[id=AM]{solution}{problem} (see~\Cref{sec:mhd-loss});
}
\item{
  investigate to which degree equilibrium properties of interest are faithfully reproduced,
  in the context of ideal-\gls{MHD} stability (\eg, magnetic well),
  neoclassical transport (\eg, the effective ripple),
  and fast particle confinement (\eg, extrema of the magnetic field strength along a field line) (see~\Cref{sec:mhd-stability,sec:neo-transport,sec:fast-ions});
}
\item{
  show the use of the model in the a posteriori optimization of \gls{W7X} magnetic configurations (see~\Cref{sec:optimization});
}
\end{itemize}

\section{Ideal-\gls{MHD} equilibria}

\subsection{The ideal-\gls{MHD} force balance}\label{sec:mhd}

The equilibrium magnetic field under the ideal-\gls{MHD} model is characterized by the force balance equation,
Ampere's and Gauss's law:

\begin{gather}
  \vec{F} = - \vec{J} \times \vec{B} + \vec{\nabla} p = 0,\label{eq:f-mhd}\\
  \vec{\nabla} \times \vec{B} = \mu_0 \vec{J}, \\
  \vec{\nabla} \cdot \vec{B} = 0,
\end{gather}

where $\vec{B}$ is the magnetic field,
$\vec{J}$ is the current density,
$p$ is the plasma pressure,
and $\mu_0$ is the permeability of free space.

In this work,
we assume that a set of nested toroidal flux surfaces exists,
and that the pressure is a flux function.
Let \mbox{$\vec{x}^* = (R, \phi, Z)$} be a cylindrical coordinate system where $R$ is the major radius,
$\phi$ is the cylindrical toroidal angle,
and $Z$ is the height above mid-plane.
\mbox{$\vec{\alpha} = (s, \theta, \varphi)$} is a flux coordinate system where \mbox{$s = \frac{\psi}{\psiedge}$} is a radial-like coordinate,
$\theta$ is a poloidal-like angle,
and $\varphi$ is a toroidal-like angle
(in this work, $\varphi = \phi$).
$2 \pi \psi$ is the toroidal flux enclosed by a flux surface,
and $\phiedge = 2 \pi \psiedge$ is the toroidal flux enclosed by the plasma boundary
(\ie, $s=1$).
In the inverse equilibrium formulation~\cite{Bauer1978},
$\vec{\alpha}$ describes the independent variables of the computational grid,
and the flux surfaces locations are described by the coordinate transformation \mbox{$f : \vec{\alpha} \rightarrow \vec{x}^*$}.

Because magnetic fields are divergence free
(\mbox{$\vec{\nabla} \cdot \vec{B} = 0$})
and flux surfaces are assumed to be nested
(\mbox{$\vec{B} \cdot \vec{\nabla} s = B^s = 0$}),
the magnetic field can be written in contravariant form following~\cite{Kruskal1958}:

\begin{gather}
  \vec{B} = \vec{\nabla} \varphi \times \vec{\nabla} \chi + \vec{\nabla} \psi \times \vec{\nabla} \theta^* = B^{\theta} \vec{e}_{\theta} + B^{\varphi} \vec{e}_{\varphi},\label{eq:b-up}
\end{gather}

where $2 \pi \chi$ is the poloidal flux,
\mbox{$\theta^* = \theta + \lambda(s, \theta, \varphi)$} is the poloidal angle for which the magnetic field lines are straight in $(s, \theta^*, \varphi)$,
and $\lambda$ is a periodic function of $\theta$ and $\varphi$ with zero average.
\mbox{$\vec{e}_i = \frac{\partial \vec{x}^*}{\partial \alpha_i}$} are the covariant basis vectors,
and $\vec{e}^i = \vec{\nabla} \alpha_i$ are the contravariant basis vectors.

The contravariant components of the magnetic field
(\mbox{$B^i = \vec{B} \cdot \vec{e}^i$})
are then:

\begin{gather}
  B^{\theta} = \frac{1}{\sqrt{g}} \Phi^{\prime} (\gls{ibar} - \frac{\partial \lambda}{\partial \varphi}),\label{eq:b-theta-up}\\
  B^{\varphi} = \frac{1}{\sqrt{g}} \Phi^{\prime}(1 + \frac{\partial \lambda}{\partial \theta}),\label{eq:b-zeta-up}
\end{gather}

where \gls{ibar} is the rotational transform,
the prime denotes $\partial / \partial s$,
and \mbox{$\sqrt{g} = (\vec{e}_s \cdot \vec{e}_{\theta} \times \vec{e}_{\varphi}) = (\vec{\nabla} s \cdot \vec{\nabla} \theta \times \vec{\nabla} \varphi)^{-1}$} is the Jacobian of the coordinate transformation $f$.

The covariant representation of $\vec{B}$
(\mbox{$B_i = \vec{B} \cdot \vec{e}_{i}$})
can be obtained from~\cref{eq:b-theta-up,eq:b-zeta-up} and the metric tensor \mbox{$g_{ij} = \vec{e}_i \cdot \vec{e}_j = \frac{\partial \vec{x}^*}{\partial \alpha_i} \cdot \frac{\partial \vec{x}^*}{\partial \alpha_j}$} as follows:

\begin{gather}
  B_s = B^{\theta} g_{\theta s} + B^{\varphi} g_{\varphi s},\label{eq:b-s-down}\\
  B_{\theta} = B^{\theta} g_{\theta \theta} + B^{\varphi} g_{\varphi \theta},\label{eq:b-theta-down}\\
  B_{\varphi} = B^{\theta} g_{\theta \varphi} + B^{\varphi} g_{\varphi \varphi}.\label{eq:b-zeta-down}
\end{gather}

The contravariant components of the current density
(\mbox{$J^i = \vec{J} \cdot \vec{e}^i = \frac{1}{\mu_0} (\vec{\nabla} \times \vec{B}) \cdot \vec{e}^i$})
are:

\begin{gather}
  J^s = \frac{1}{\mu_0 \sqrt{g}} (\frac{\partial B_{\varphi}}{\partial \theta} - \frac{\partial B_{\theta}}{\partial \varphi}),\label{eq:j-s-up}\\
  J^{\theta} = \frac{1}{\mu_0 \sqrt{g}} (\frac{\partial B_s}{\partial \varphi} - \frac{\partial B_{\varphi}}{\partial s}),\label{eq:j-theta-up}\\
  J^{\varphi} = \frac{1}{\mu_0 \sqrt{g}} (\frac{\partial B_{\theta}}{\partial s} - \frac{\partial B_s}{\partial \theta}).\label{eq:j-zeta-up}
\end{gather}

Finally,
inserting~\cref{eq:b-theta-up,eq:b-zeta-up,eq:j-s-up,eq:j-theta-up,eq:j-zeta-up} in~\cref{eq:f-mhd},
the covariant form of the \gls{MHD} force residual is given by:

\begin{gather}
  \vec{F} = F_s \vec{\nabla} s + F_{\beta} \replaced[id=AM]{\vec{\beta}}{\vec{\nabla} \beta},\\
  F_s = \frac{1}{\mu_0} (B^{\theta} \frac{\partial B_{\theta}}{\partial s} - B^{\theta} \frac{\partial B_s}{\partial \theta} - B^{\varphi} \frac{\partial B_s}{\partial \varphi} + B^{\varphi} \frac{\partial B_{\varphi}}{\partial s}) + p^{\prime},\\
  F_{\beta} = J^s = \frac{1}{\mu_0 \sqrt{g}} (\frac{\partial B_{\varphi}}{\partial \theta} - \frac{\partial B_{\theta}}{\partial \varphi}),
\end{gather}

where \mbox{$\vec{\beta} = \sqrt{g} (B^{\varphi} \vec{\nabla} \theta - B^{\theta} \vec{\nabla} \varphi)$}.

If we define $\vec{x}=(R, \lambda, Z)$,
the \gls{MHD} force residual can be then computed from the mapping $\vec{x}(s, \theta, \varphi)$,
and the $p(s)$ and $\gls{ibar}(s)$ flux functions\added[id=AM]{ (see~\cref{sec:mhd-residual})}.

\subsection{Fixed- and free-boundary equilibria}\label{sec:free-boundary}

The ideal-\gls{MHD} problem admits two kinds of boundary conditions: fixed- and free-boundary conditions.
In the fixed-boundary case,
the shape ($R$ and $Z$) of the outermost flux surface (\ie, $s=1$) is fixed.
In the free-boundary case,
the shape of the outermost flux surface is determined by the continuity of the total pressure $B^2 / 2 \mu_0 + p$ and by the vanishing of the normal component of the vacuum field at the plasma-vacuum interface
(thus enforcing the $s=1$ surface to be a flux surface).

Multiple quantities parametrize the equilibrium problem:
$\phiedge \in \real$  defines the total toroidal flux enclosed by the plasma,
$p: [0, 1] \rightarrow \real$ defines the pressure profile,
and $\gls{ibar}: [0, 1] \rightarrow \real $ defines the rotational transform profile.
The toroidal current profile $\gls{Itor}: [0, 1] \rightarrow \real$ can be also specified in place of the iota profile:
since \mbox{$\gls{Itor}(s)=\int_0^s \int_0^{2 \pi} J^{\varphi} \sqrt{g} ds^{\prime} d \theta$},
and $J^{\varphi}$ linearly depends on \gls{ibar},
the rotational transform profile consistent with a given toroidal current profile can be computed by solving a linear algebraic equation.

In case of free-boundary equilibria,
a vacuum field,
which is usually generated by a set of external coils,
can be provided.
A reference field generated by each coil is commonly precomputed on the cylindrical grid,
and the total vacuum field is given as the superposition of each coil field \mbox{$\vec{B}_v = \sum_{i=1}^{N_c} i_i \vec{B}_i^0$},
where $\vec{B}_i^0$ is the reference field generated by each coil given a reference current $I_i^0$,
$N_c$ is the number of independent coils,
and $i_i = I_i / I_i^0$ is the current ratio between the actual and reference coil current.
The set of current ratios $\vec{i} \in \real^{N_c}$ additionally parametrizes the equilibrium problem.

In this work,
since in equilibrium reconstruction routines the plasma boundary is not known a priori,
we are only interested in free-boundary equilibria.
In addition,
no diagnostics to probe the iota profile are installed at \gls{W7X},
whereas continuous Rogowski coils measure the net toroidal current.
Therefore,
the toroidal current profile is specified instead of the iota profile.

\subsection{The ideal-\gls{MHD} solution operator}\label{sec:mhd-solution-operator}

In this work,
a set of scalar parameters ($i_i$ and \phiedge) and plasma profiles ($p$ and \gls{Itor}),
represented by one-dimensional functions of the radial coordinate $s$,
define a instance of the ideal-\gls{MHD} equilibrium problem.
The equilibrium solution is represented by the mapping \mbox{$(s, \theta, \varphi) \rightarrow \vec{x}(s, \theta, \varphi)$} (where $\vec{x}=(R, \lambda, Z)$) and by the one-dimensional function $s \rightarrow \gls{ibar}(s)$.

Let us define the solution operator as the operator that maps the set of scalar parameters and plasma profiles to the \gls{MHD} solution:

\begin{gather}\label{eq:solution-operator}
  \mathcal{O}(\vec{i}, \phiedge, p(s), \gls{Itor}(s))(s, \theta, \varphi) = [ \vec{x}(s, \theta, \varphi), \gls{ibar}(s) ]^T,
\end{gather}

where $\vec{i} \in \real^{N_c}$,
$\phiedge \in \real$,
\mbox{$\{p, \gls{Itor}, \gls{ibar} \} : [0, 1] \rightarrow \real$} are one-dimensional functions defined on the radial plasma domain,
and \mbox{$\vec{x}: [0, 1] \times [0, 2 \pi]\deleted[id=AM]{,} \times [0, 2 \pi] \rightarrow \real^3$} is the mapping between flux coordinates and $(R, \lambda, Z)$.
\section{\glspl{NN} to approximate non-linear operators}\label{sec:mhdnet}

\glspl{NN} can approximate non-linear continuous functions~\cite{Cybenko1989,Hornik1991,Pinkus1999,Eldan2016,Lu2017}\added[id=AM]{, discontinuous functions~\cite{Llanas2008,Jagtap2020},} and non-linear continuous operators~\cite{Chen1995,Chen1995a,Lu2021}.
In this work,
we propose a \gls{NN} model to learn the non-linear ideal-\gls{MHD} operator.
A set of free parameters $\Theta \in \real^N$ (\ie, trainable weights) parametrizes the \gls{NN} model:

\begin{gather}\label{eq:nn-operator}
  \text{NN}(\vec{i}, \phiedge, p(s), \gls{Itor}(s); \Theta)(s, \theta, \varphi) = [ \vec{\tilde{x}}(s, \theta, \varphi), \tilde{\gls{ibar}}(s) ]^T,
\end{gather}

where \mbox{$\vec{\tilde{x}}(s, \theta, \varphi)$} and $\tilde{\gls{ibar}}(s)$ compose the equilibrium solution provided by the \gls{NN} model.

Three sources of error characterize the accuracy of a \gls{NN}:
approximation,
optimization and generalization errors~\cite{Poggio2019}.
The approximation error describes how well a \gls{NN} represents the target function or operator,
the optimization error describes how well the \gls{NN} training procedure minimizes the training loss (\ie, empirical risk),
and the generalization error describes how well the \gls{NN} generalizes to unseen data.

The approximation theorem in~\cite{Chen1995} shows how any \gls{NN} with a specific structure and enough capacity (\ie, number of free parameters) guarantees a small approximation error.
However,
no guarantees are given on the optimization and generalization errors.
Also, from a practical standpoint, estimating the required number of free parameters to meet "enough capacity" is difficult.
Fulfilling known symmetries of the solution by construction should ease the training process
(\ie, it should reduce the optimization error),
and the addition of a physics constraint loss should improve generalization
(\ie, it should reduce the generalization error).
The next section will introduce them separately.

\subsection{Imposing symmetries in the \gls{NN} model}\label{sec:symmetries}

In a learning algorithm,
the available computational budget is divided into three baskets:
the capacity (\ie, size) of the model sets the approximation error,
the length of the training affects the optimization error,
and the size of the data set impacts the generalization error.

The space of possible functions that the model can represent is large.
The training process is essentially a \quotes{search} in this space.
An inductive bias \quotes{allows a learning algorithm to prioritize one solution over another, independent of the observed data}~\cite{Battaglia2018}.
An inductive bias encodes the solution's prior knowledge
(\eg, a solution's symmetry);
it allows budget to be relocated from the capacity basket to the other two baskets,
thus,
reducing the optimization and generalization errors.

The equilibrium solution should satisfy multiple symmetries.
Let us define \mbox{$(X^1, X^2, X^3) = (R, \lambda, Z)$}.
Each quantity $X^j$ (with $j \in \{ 1, 2, 3 \}$) should be a periodic function of the poloidal and toroidal angles:

\begin{gather}
  X^j(s, \theta, \varphi) = X^j(s, \theta + 2 \pi, \varphi),\label{eq:poloidal-symmetry}\\
  X^j(s, \theta, \varphi) = X^j(s, \theta, \varphi + 2 \pi / \gls{Nfp}),\label{eq:toroidal-symmetry}
\end{gather}

where \gls{Nfp} is the number of toroidal field period.

Moreover,
if stellarator symmetry is assumed
(which is the case for most experimental devices, including \gls{W7X}),
the solution should satisfy the following relationships~\cite{Hirshman1983}:

\begin{gather}
  R(s, \theta, \varphi) = R(s, - \theta, -\varphi),\label{eq:r-stellarator-symmetry}\\
  \lambda(s, \theta, \varphi) = - \lambda(s, - \theta, -\varphi),\label{eq:lambda-stellarator-symmetry}\\
  Z(s, \theta, \varphi) = - Z(s, - \theta, -\varphi).\label{eq:z-stellarator-symmetry}
\end{gather}

For example,
the periodicity of $X^j$ and stellarator symmetry can be imposed if the solution is expanded in Fourier series along the poloidal and toroidal directions:

\begin{gather}
  R(s, \theta, \varphi) = \sum_{mn} R_{mn} \left( s \right) \cos \left( m \theta - \gls{Nfp} n \varphi \right),\label{eq:r}\\
  \lambda(s, \theta, \varphi) = \sum_{mn} \lambda_{mn} \left( s \right) \sin \left( m \theta - \gls{Nfp} n \varphi \right),\label{eq:lambda}\\
  Z(s, \theta, \varphi) = \sum_{mn} Z_{mn} \left( s \right) \sin \left( m \theta - \gls{Nfp} n \varphi \right),\label{eq:z}
\end{gather}

where $X_{mn}^j$ are the associated Fourier coefficients,
$m$ is the poloidal mode number,
and $n$ is the toroidal mode number.
Since $\lambda$ is a periodic function with zero average,
$\lambda_{00} = 0$.

At the magnetic axis ($s=0$),
the flux surface geometry must be regular (\ie, infinitely differentiable).

If it is expanded in Fourier series,
the Fourier coefficients $X_{mn}^j$ must have the form~\cite{Lewis1990}:

\begin{gather}
  X_{mn}^j = \sqrt{s}^m \sum_{k\in\naturals\cup\{0\}}X_{kmn}^j s^k.
\end{gather}

\Cref{sec:architecture} introduces how the \gls{NN} model inherently satisfies this property.

\subsection{\gls{DeepONet}-like architecture}\label{sec:architecture}

\begin{figure}[h!]
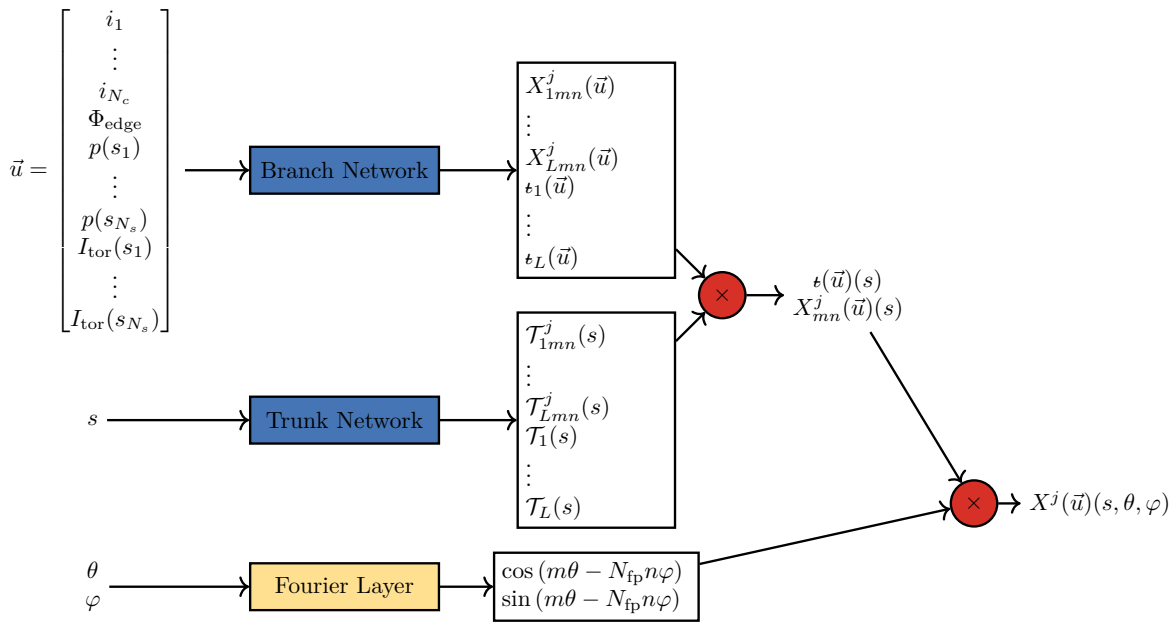

  \centering
  \igraph[]{content_figures_mhdnet_paper.pgf}%
  \caption{%
    The macro architecture of the \gls{NN} model to approximate the ideal-\gls{MHD} solution operator.
    In case of free-boundary equilibria,
    a set of parameters
    (the normalized coil currents $[i_1, \hdots, i_{N_c}]^T$ and the toroidal flux enclosed by the plasma \phiedge)
    and two functions (the pressure $p(s)$ and toroidal current $\Itor(s)$ plasma profiles)
    define the ideal-\gls{MHD} equilibrium problem.
    The variable $u$ combines the set of parameters and the plasma profiles values at selected flux surfaces,
    $[s_1, \hdots, s_{N_s}]^T$.
    The solution is represented by the set of flux surface coordinates $(R, Z)$,
    the renormalization parameter $\lambda$,
    and the iota profile \gls{ibar}.
    The solution domain spans the radial $s$,
    poloidal $\theta$,
    and toroidal $\varphi$ directions.
    The quantities $X^j \in \{R, \lambda, Z\}$ are expanded in Fourier series along the poloidal and toroidal directions.
    A trunk network constructs a set of non-linear \quotes{basis} functions of $s$,
    which radially expand the $X^j$ Fourier coefficients and iota.
    A branch network parametrizes the set of trunk basis coefficients as a function of $u$.
  }%
  \label{fig:mhdnet}
\end{figure}

\Cref{fig:mhdnet} shows the \gls{NN} macro architecture.
The plasma profiles are observed at a set of equally spaced radial locations,
$s_i$ for $i \in \{ 1, \hdots, N_s \}$,
also called sensors.
In addition to the independent variables ($s$, $\theta$, and $\varphi$),
the model has two sets of inputs:
the set of scalar equilibrium parameters ($[i_1, \hdots, i_{N_c}]^T$ and \phiedge) and the plasma profiles ($[p(s_0), \hdots, p(s_{N_s})]^T$ and $[\gls{Itor}(s_0), \hdots, \gls{Itor}(s_{N_s})]^T$).
Those inputs are concatenated into $\vec{u} = [i_1, \hdots, i_{N_c}, \phiedge, p(s_1), \hdots, p(s_{N_s}), \gls{Itor}(s_1), \hdots, \gls{Itor}(s_{N_s})]^T$.
The model represents the \gls{MHD} equilibrium solution as:

\begin{gather}
  X^j(\vec{u})(s, \theta, \varphi) = \sum_{mn} X_{mn}^j(\vec{u})(s) \replaced[id=AM]{\mathcal{F}_{mn}}{\mathcal{F}}(\theta, \varphi) ,\\
  X_{mn}^j(\vec{u})(s) = \sqrt{s}^m \sum_{l=1}^L X_{lmn}^j (\vec{u}) \mathcal{T}_{lmn}^j(s) ,\\
  \gls{ibar}(\vec{u})(s) = \sum_{l=1}^L \gls{ibar}_l(\vec{u}) \mathcal{T}_l(s) ,
\end{gather}

where $X^j \in \{ R, \lambda, Z \}$,
\mbox{$\replaced[id=AM]{\mathcal{F}_{mn}}{\mathcal{F}}(\theta, \varphi) = \cos (m \theta - \gls{Nfp} n \varphi)$} in case of $R$ and \mbox{$\replaced[id=AM]{\mathcal{F}_{mn}}{\mathcal{F}}(\theta, \varphi) = \sin (m \theta - \gls{Nfp} n \varphi)$} in case of $\lambda$ or $Z$.
The Fourier terms $\replaced[id=AM]{\mathcal{F}_{mn}}{\mathcal{F}}(\theta, \varphi)$ expand each $X^j$,
where $X_{mn}^j$ represent the Fourier coefficients.

Up to $m=8$ poloidal and $|n|=11$ toroidal modes represent each $X^j$.
For \gls{W7X},
previous works showed that up to \num{6} modes are sufficient to represent the geometry of the flux surfaces~\cite{Sengupta2004,Sengupta2007,Merlo2021}.
However,
using \gls{VMEC} equilibria with up to $m=11$ and $|n|=12$ as a \quotes{high-fidelity} reference,
a Fourier resolution scan revealed that up to $m=8$ and $|n|=11$ modes are required to represent the equilibrium field with a field error below \SI{1}{\percent} (see~\cref{sec:fourier-scaling}).

For each Fourier coefficients and for iota,
a \quotes{trunk} network parametrizes a set of non-linear functions of $s$,
$[\mathcal{T}_{1mn}^j(s), \hdots, \mathcal{T}_{Lmn}^j(s)]^T \in \real^L$ and $[\mathcal{T}_1(s), \hdots, \mathcal{T}_L(s)]^T \in \real^L$,
respectively.
A \quotes{branch} network outputs a set of scalar coefficients as a function of $u$,
$[X_{1mn}^j, \hdots, X_{Lmn}^j]^T \in \real^L$ and $[\gls{ibar}_1, \hdots, \gls{ibar}_L]^T \in \real^L$.
Finally,
the iota and the Fourier coefficients are obtained as a linear combination of trunk network functions,
which are weighted by the branch network coefficients.

Iota highly depends on the toroidal current~\cite{Strand2001}.
To let the model exploit this dependency,
the toroidal current profile substitutes the last trunk network function in the expansion of the iota profile:
$\mathcal{T}_L(s) = \gls{Itor}(s)$.

The trunk network expands the radial coordinate $s$ into a set of \quotes{basis} functions,
independent for each Fourier coefficient $X_{mn}^j$ and iota $\gls{ibar}$.
We inappropriately refer to them as a set of basis functions on the interval $[0, 1]$
since
these functions are not guaranteed to form a basis (a set of orthogonal functions).
To obey the required form needed to represent $X^j$ (see~\cref{sec:symmetries}),
each function \footnote{only the functions that expand the Fourier coefficients $X_{mm}^j$} is multiplied by a $\sqrt{s}^m$ factor,
and the two first trunk network functions are the constant and identity functions, respectively:
$\mathcal{T}_{1mn}^j(s) = \mathcal{T}_1(s) = 1$ and $\mathcal{T}_{2mn}^j(s) = \mathcal{T}_2(s) = s$.
Imposing the identity function as one of the trunk network functions effectively bypasses the trunk network layers,
and it preserves the model ability to build simple functions (\eg, the identity function).
This construction is known as a \quotes{skip connection} in machine learning~\cite{He2016}.
A \gls{MLP} parametrizes the trunk network.
Please refer to~\cref{sec:appendix-hp} for a detailed description of the trunk network architecture.

The branch network maps the scalar parameters and the plasma profiles of the equilibrium problem to the set of coefficients associated to the trunk functions.
The one-dimensional plasma profiles are observed at a set of sensors $[s_1, \hdots, s_{N_s}]^T$.
A set of one-dimensional convolutional layers extract high-level features from the plasma profiles into a compressed latent representation,
which is then concatenated to the set of scalar parameters.
Finally,
a \gls{MLP} outputs the coefficients associated to the trunk functions.
Please refer to~\cref{sec:appendix-hp} for a detailed description of the branch network architecture.

The model can represent a global ideal-\gls{MHD} solution
(\ie, a solution that is defined everywhere in the plasma volume)
and fulfils the required symmetries and regularity condition at the magnetic axis by construction.
Because of the Fourier expansion along the poloidal and toroidal directions,
the periodic boundary conditions and stellarator symmetry are satisfied.
The required radial form of each Fourier coefficients is satisfied due to the $\sqrt{s}^m$ factor and the trunk network skip connection.
\section{Training}\label{sec:data}

The model is trained in a supervised-learning fashion:
\gls{VMEC} provides the ground-truth ideal-\gls{MHD} equilibria.
To restrict the space in which to approximate the ideal-\gls{MHD} solution operator,
only \gls{W7X} configurations are considered.

\subsection{Data set}\label{sec:training-data}

The boundaries of the training data limit the application of any data-driven method.
In this work,
the training data are sampled from a multivariate distribution that approximates the experimental distribution (\cref{tab:distribution}).
The coil current ratios define the vacuum magnetic configuration,
the plasma volume determines the total toroidal flux at the edge.
One-dimensional \glspl{GP} on the $[0,1]$ interval represent the normalized pressure and toroidal current profiles.
The volume averaged plasma beta informs the scale of the pressure profile,
and the total integrated toroidal current defines the scale of the current profile.
See~\cite[Section~2.2]{Merlo2021} for a more in depth description of the methodology.

\begin{table}[h!]
  \caption{%
    Distributions from which the training data have been sampled.
    $U(a, b)$ represents a uniform distribution on the $[a, b]$ interval,
    and $N(\mu, \sigma)$ represents a normal distribution with $\mu$ mean and $\sigma$ standard variation.
    One-dimensional \glspl{GP} with a non-stationary covariance function model the plasma profiles.
    The \gls{GP} covariance function scale factor $\sigma_f$,
    core length scale \lCore,
    edge length scale \lEdge,
    core-edge transition location \tLocation,
    and core-edge transition width \tWidth parametrize each \gls{GP}~\cite[Section~2.2]{Merlo2021}.
    The references from which the distribution parameters have been informed are listed.
  }%
  \label{tab:distribution}
  \centering
  \begin{tabular}{lccc}
    \toprule
    Quantity                                    & Distribution                & Unit              & References                                              \\
    \multicolumn{4}{l}{Magnetic Configuration}                                                                                                              \\
    \midrule
    Non-planar coil currents $i_{[2 \ldots 5]}$ & $U(\num{0.6}, \num{1.2})$   & -                 & \cite{Andreeva2002,Sengupta2004,Sengupta2007,Merlo2021} \\
    Planar coil currents $i_{[A,B]}$            & $U(\num{-0.8}, \num{0.8})$  & -                 & \cite{Andreeva2002,Sengupta2004,Sengupta2007,Merlo2021} \\
    \midrule
    \multicolumn{4}{l}{Plasma Parameters}                                                                                                                   \\
    \midrule
    Plasma volume $V_p$                         & $N(\num{30}, \num{3.0})$    & $\si{\meter^3}$   & \cite{Geiger2014}                                       \\
    Plasma beta $\beta$                         & $U(\num{0}, \num{5})$       & \si{\percent}     & \cite{Erckmann1997}                                     \\
    Total toroidal current $I_{tor}$            & $U(\num{-20}, \num{20})$    & \si{\kilo\ampere} & \cite{Geiger2014}                                       \\
    \midrule
    \multicolumn{4}{l}{Pressure Profile \gls{GP}}                                                                                                           \\
    \midrule
    Covariance function scale factor $\sigma_f$ & $U(\num{0.1}, \num{5.0})$   & -                 & \cite{Merlo2021}                                        \\
    Core length scale \lCore                    & $U(\num{0.1}, \num{1.0})$   & -                 & \cite{Merlo2021}                                        \\
    Edge length scale \lEdge                    & $U(\num{0.1}, \num{2.0})$   & -                 & \cite{Merlo2021}                                        \\
    Core-edge transition location \tLocation    & $U(\num{0.1}, \num{1.0})$   & -                 & \cite{Merlo2021}                                        \\
    Core-edge transition width \tWidth          & $U(\num{0.1}, \num{0.3})$   & -                 & \cite{Merlo2021}                                        \\
    \midrule
    \multicolumn{4}{l}{Toroidal Current Profile \gls{GP}}                                                                                                   \\
    \midrule
    Covariance function scale factor $\sigma_f$ & $U(\num{1.0}, \num{5.0})$   & -                 & \cite{Merlo2021}                                        \\
    Core length scale \lCore                    & $U(\num{0.1}, \num{1.0})$   & -                 & \cite{Merlo2021}                                        \\
    Edge length scale \lEdge                    & $U(\num{0.5}, \num{1.0})$   & -                 & \cite{Merlo2021}                                        \\
    Core-edge transition location \tLocation    & $U(\num{0.001}, \num{0.8})$ & -                 & \cite{Merlo2021}                                        \\
    Core-edge transition width \tWidth          & $U(\num{0.1}, \num{0.5})$   & -                 & \cite{Merlo2021}                                        \\
    \bottomrule
  \end{tabular}
\end{table}

\begin{description}
  \item[Vacuum configuration]{%
        The current ratios between the \gls{W7X} coils define the vacuum magnetic configuration.
        Two symmetric sets of coils compose each half toroidal field period:
        five non-planar coils,
        two planar (tilted) coils,
        and one control coil constitute each set.
        Defining $I_1$ as the current flowing in the first non-planar coil,
        $i_i = \frac{I_i}{I_1}$ for $i \in \{1, \hdots, N_c \}$ are the coil current ratios.
        In the data set,
        the current of the first non-planar coil is held fixed at $I_1=\SI{13068}{\ampere}$
        (current value to obtain $B_{\text{axis}}(\varphi=0) \simeq \SI{2.52}{\tesla}$ in case of the standard configuration),
        and the current in the control coil is null.
        The coil current ratios are uniformly sampled to cover a broad configuration space around the nine reference \gls{W7X} configurations~\cite{Andreeva2002}.
        }
  \item[Plasma volume]{%
        For a given magnetic field strength,
        \phiedge sets the plasma volume:\newline
        \mbox{$\phiedge = \int_0^1 \int_0^{2 \pi} \vec{B} \cdot \vec{\nabla} \varphi \sqrt{g} ds d\theta \propto B a \pi^2 \simeq B V_p / R$},
        where $a$ is the effective minor radius,
        $R$ the major radius,
        and $V_p$ the plasma volume.
        The plasma volume is sampled from a normal distribution with mean $\SI{30}{\meter^3}$ and standard variation $\SI{3}{\meter^3}$~\cite{Geiger2014}.
        Once a plasma volume has been sampled,
        a linear fit estimates \phiedge.
        }
  \item[Plasma beta]{%
        For a given magnetic field strength and pressure profile shape,
        the pressure on axis sets the volume averaged plasma beta:
        \mbox{$\averagePlasmaBeta = 2 \mu_0 \langle p / B^2 \rangle_{\text{Vol}} = 2 \mu_0 p_0 \langle \average{p} / B^2 \rangle_{\text{Vol}}$},
        where $p_0$ is the pressure on axis and $\average{p} = p / p_0$ is the normalized pressure profile.
        In previous operational campaigns,
        \gls{W7X} has reached a maximum \averagePlasmaBeta of roughly \SI{1}{\percent}.
        However,
        \gls{W7X} has been optimized for higher plasma beta (up to $\averagePlasmaBeta \simeq \SI{5}{\percent}$~\cite{Grieger1993}).
        Therefore,
        the volume averaged plasma beta is sampled from a uniform distribution between \SI{0}{\percent} and \SI{5}{\percent}.
        Once a plasma beta has been sampled,
        a linear fit estimates $p_0$.
        }
  \item[Net toroidal current]{%
        The plasma bootstrap current and the externally induced current (\eg, via \gls{ECCD}) mainly define the net toroidal current.
        The net toroidal current should be minimized for the robust operation of an island divertor.
        In \gls{W7X} operations,
        two current-free scenarios are envisioned~\cite{Geiger2014}:
        a scenario in which the bootstrap current is minimized via the magnetic configuration,
        and a scenario in which the bootstrap current is balanced by a strong \gls{ECCD} current.
        During the first operational campaign,
        the measured bootstrap current ranged between \SI{-7}{\kilo\ampere} and \SI{17}{\kilo\ampere}~\cite{Neuner2021}.
        Therefore,
        the total toroidal current is sampled from a uniform distribution between \SI{-20}{\kilo\ampere} and \SI{20}{\kilo\ampere}.
        }
  \item[Pressure profile shape]{%
        For electrons and ions,
        the gradient of the density and temperature (\mbox{$p \propto nT$}) profiles can be substantially different in the core and edge regions.
        To account for such difference,
        a \gls{GP} with a non-stationary covariance function models the pressure profile.
        The \gls{GP} is parametrized by a set of \glspl{HP}~\cite[Section~2.2]{Merlo2021}:
        the covariance function scale factor $\sigma_f$,
        the core length scale \lCore,
        the edge length scale \lEdge,
        the transition location \tLocation,
        and the transition width \tWidth.
        In \gls{W7X},
        the pressure profile is generally flat,
        however,
        high-performance operations are correlated with a peaked density profile~\cite{Bozhenkov2020}.
        To account for both,
        flat and peaked profiles are included in the data set.
        Profiles with a non-physical positive pressure gradient
        (\ie, $\nabla p > 0$)
        are discarded.
        }
  \item[Toroidal current profile shape]{%
        In W7-X,
        the toroidal current has two main contributions:
        the \gls{ECCD} induced current and the bootstrap current.
        The bootstrap current density is generally flat across the plasma volume,
        however,
        the \gls{ECCD} current density is peaked at the deposition location.
        To capture both shapes,
        a \gls{GP} with a non-stationary covariance function is used to model the toroidal current profile.
        }
\end{description}


\num{14281} equilibria have been sampled in the data set,
but equilibria for which \gls{VMEC} was not able to minimize the \gls{MHD} variational forces up to a tolerance of \SI{1e-7}{\newton\henry\per\meter} were discarded~\footnote{$\text{ftol}=\num{1e-14}$ in VMEC}.
The data set thus contains \num{9857} equilibria.
\Cref{fig:beta-hist,fig:pressure-peaking-factor-hist,fig:peak-jtor} show the histograms of a subset equilibrium properties for all converged equilibria:
the volume averaged beta \averagePlasmaBeta,
the pressure peaking factor \mbox{$p_0 / \langle p \rangle$},
and the peak toroidal current density \mbox{$\max_s | j_{\text{tor}}(s) |$}.
The \averagePlasmaBeta distribution is not uniform due to the not converged equilibria at high beta (\cref{fig:beta-hist}).
Nevertheless,
the data set holds both flat (\ie, small peaking factor) and peaked (\ie, large peaking factor) pressure profiles (\cref{fig:pressure-peaking-factor-hist}).

The data set has been split into a training set (\SI{80}{\percent}),
a validation set (\SI{10}{\percent}),
and a test set (\SI{10}{\percent}).
The training set has been used to train the model,
the validation set has been used to assess the best model \glspl{HP},
and the test set has been used to assess the model performance on held-out data.

To improve the training procedure,
the model inputs have been normalized with the mean and standard deviation computed on \SI{20}{\percent} of the training data.

\begin{figure}[h!]
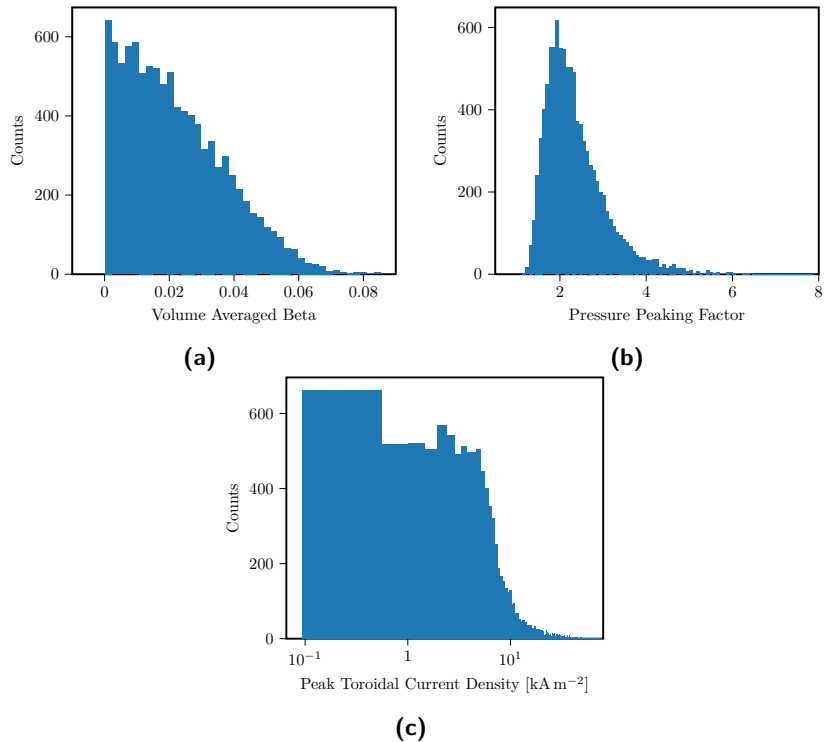

  \centering
  \multigraph[labels={fig:beta-hist}{fig:pressure-peaking-factor-hist}{fig:peak-jtor}]{content_figures_hist_beta.pgf;content_figures_hist_pressure_peaking_factor.pgf;content_figures_hist_peak_jtor.pgf}%
  {%
    {}{}{}%
  }%
  \caption{%
    Histograms of the volume averaged beta~\subref{fig:beta-hist},
    the pressure peaking factor~\subref{fig:pressure-peaking-factor-hist},
    and the peak toroidal current density~\subref{fig:peak-jtor} for all converged equilibria in the data set.
  }
\end{figure}

\subsection{Loss function}\label{sec:training}

Three terms form the model loss function:
a data term,
a gradient term,
and a physics-based regularization term.
The data term biases the model to provide equilibria that are close to the ground truth
(computed by \gls{VMEC}),
the gradient term regularizes the model equilibria to exhibit the same radial derivatives as the ground truth,
and the physics-based term regularizes the model equilibria to satisfy the ideal-\gls{MHD} equations.

\subsubsection{Data loss}\label{sec:data-loss}

The model equilibrium solution is globally defined in the whole plasma volume.
However,
to numerically compute a residual measure against the ground truth,
an equally spaced grid is adopted:
$N_s=99$ flux surfaces,
$N_{\theta}=32$ poloidal locations,
and $N_{\varphi}=36$ toroidal locations compose the loss grid.
The iota profile,
being a flux surface quantity,
is evaluated only at the radial grid locations.
On each equilibrium,
the data loss term is:

\begin{gather}
  L_d = \sum_{j=1}^3 \alpha_{X^j} \sum_{i=1}^{N_s} \sum_{t=1}^{N_{\theta}} \sum_{z=1}^{N_{\varphi}} [\tilde{X}^j(s_i, \theta_t, \varphi_z) - X^j(s_i, \theta_t, \varphi_z)]^2 + \alpha_{\gls{ibar}} \sum_{i=1}^{N_s} [\tilde{\gls{ibar}}(s_i) - \gls{ibar}(s_i)]^2,
\end{gather}

where $\tilde{X}^j$ and $\tilde{\gls{ibar}}$ are the model outputs,
and $\alpha_{X^j}$ and $\alpha_{\gls{ibar}}$ are the set of coefficients that weight the data loss terms.

\subsubsection{Gradient loss}\label{sec:gradient-loss}

An equilibrium solution is fully defined by the mapping $X^j(s, \theta, \varphi)$ for $j \in \{1, 2, 3\}$ and by $\gls{ibar}(s)$.
However,
many equilibrium properties depend also on equilibrium first and second radial, poloidal and toroidal derivatives.
For example,
the flux surface Jacobian $\sqrt{g}$,
which is the local linear approximation of the coordinate transformation \mbox{$(s, \theta, \varphi) \rightarrow (R, \phi, Z)$},
is:

\begin{gather}
  \sqrt{g} = R (\frac{\partial R}{\partial \theta} \frac{\partial Z}{\partial s} - \frac{\partial R}{\partial s} \frac{\partial Z}{\partial \theta}).
\end{gather}

To faithfully reconstruct the equilibrium properties,
the gradients of the equilibrium solution need to match the gradients of the ground truth.
Due to the Fourier expansion of the solution in the poloidal and toroidal directions,
the poloidal and toroidal derivatives are already constrained by the Fourier coefficients:
for example,
\mbox{$\frac{\partial R}{\partial \theta} = \frac{\partial }{\partial \theta} [ \sum_{mn} R_{mn} \cos (m \theta - \gls{Nfp} n \varphi) ]= - \sum_{mn} m R_{mn} \sin (m \theta - \gls{Nfp} n \varphi)$}.
A similar relationship holds for any poloidal and toroidal derivatives of $(R, \lambda, Z)$.

However,
the radial derivatives of the equilibrium solution are not guaranteed to match those of the ground truth.
The radial derivatives depend on the learnt trunk network functions,
which are derived during training to minimize a $L^2$ residual with the ground truth on a finite-spacing radial grid.
Even if the $L^2$ residual vanishes,
numerical artefacts
(\eg, numerical oscillations)
are known to occur~\cite{Gresho1981}.
Therefore,
the radial derivatives of the model solution can substantially differ from the ground truth.

To ameliorate these phenomena,
a gradient term regularizes the radial derivatives of the model equilibrium:

\begin{gather}
  L_g = \sum_{j=1}^3 \alpha_{X^j_s} \sum_{i=1}^{N_s} \sum_{t=1}^{N_{\theta}} \sum_{z=1}^{N_{\varphi}} [ \frac{\partial \tilde{X}^j}{\partial s}(s_i, \theta_t, \varphi_z) - \frac{\partial X^j}{\partial s}(s_i, \theta_t, \varphi_z)]^2 + \alpha_{\gls{ibar}^{\prime}} \sum_{i=1}^{N_s} [\frac{\partial \tilde{\gls{ibar}}}{\partial s}(s_i) - \frac{\partial \gls{ibar}}{\partial s}(s_i)]^2,
\end{gather}

where $\alpha_{X^j_s}$ and $\alpha_{\gls{ibar}^{\prime}}$ are a set of coefficients that weight the gradient loss terms.
The radial derivatives are approximated with a central finite difference scheme for both the model and ground truth solutions.

\subsubsection{Physics regularization}\label{sec:physics-loss}

The ideal-\gls{MHD} equilibrium problem exhibits a high condition number~\cite{Hirshman1991}.
In the framework of \gls{MHD} equilibria,
the condition number is defined as ratio between the largest and the smallest eigenvalue of the \gls{MHD} force residual linearized around an equilibrium solution.
Due to the presence of second-order radial derivatives in the residual,
the condition number scales with the square of the number of radial grid locations~\cite{Hirshman1991},
namely,
$\mathcal{P} \sim N_s^2$
($\orderof{\num{e4}}$ in this work).

A high condition number means that the \gls{MHD} force residual is highly sensitive to the equilibrium solution:
small deviations of the solution away from the ground truth may lead to a large \gls{MHD} residual.
Even if the model equilibria are close (in the $L^2$ sense) to the ground truth equilibria,
they are not guaranteed to well minimize the residual.

The \gls{MHD} force residual in the loss function regularizes the model to provide equilibria that better satisfy the ideal-\gls{MHD} equations.
For an equilibrium solution,
\mbox{$\twoNorm{\vec{F}} = 0$} everywhere in the domain.
The volume-averaged residual norm can be computed as:

\begin{gather}
  \langle \twoNorm{\vec{F}} \rangle_{\text{Vol}} = \int_0^1 \int_0^{2\pi} \int_0^{2\pi} \twoNorm{\vec{F}} \sqrt{g} ds d\theta d\varphi,
\end{gather}

where \mbox{$\langle A \rangle_{\text{Vol}} = \int_0^1 \int_0^{2\pi} \int_0^{2\pi} A \sqrt{g} ds d\theta d\varphi$} denotes a volume-averaged operator.

In this work,
a proxy for the \gls{MHD} force residual is used instead.
\gls{VMEC} (the code that provides the ground truth equilibria) does not directly minimize the force residual,
instead,
it minimizes the total energy of the system (\ie, it uses a variational approach).
It has been shown that equilibria computed by \gls{VMEC} only poorly minimize the ideal-\gls{MHD} force residual~\cite{Panici2022}.
Therefore,
adding the full \mbox{$\langle \twoNorm{\vec{F}} \rangle_{\text{Vol}}$} as regularization term may cause instabilities in the training procedure.
The volume averaged radial force balance assuming a vanishing helical component (\mbox{$F_{\beta} = 0$}) is used instead.
This quantity is also used in \gls{VMEC} to determine the goodness of a converged solution.
Let us firstly define

\begin{gather}\label{eq:f-star}
  f_* = \frac{\mu_0}{(2 \pi)^2} \langle F_s \vert_{F_{\beta}=0} \rangle_{\text{Fs}},
\end{gather}

where \mbox{$\langle A \rangle_{\text{Fs}} = \int_0^{2\pi} \int_0^{2\pi} A \sqrt{g} d\theta d\varphi$} is a flux surface averaged operator.
Then,
the model is regularized with the following term:

\begin{gather}
  L_{\text{MHD}} = \alpha_{\text{MHD}} \int_{s_{min}}^1 f_*^2 ds ,
\end{gather}

where $s_{min} = 0.02$ to avoid the sensitivity of the \gls{MHD} force residual at the magnetic axis.
Please refer to~\cref{sec:f-star} for the derivation of $f_*$.

\subsection{Training stages}\label{sec:training-stages}

Three subsequent training stages compose the training procedure.
In every training stage,
early stopping is employed during training~\cite{Morgan1989},
and the best model accordingly to the validation loss is used as a warm start for the next stage.

\begin{description}
  \item[Data stage]{%
        At the beginning of the training,
        only the data loss $L_d$ is used.
        A set of flux surfaces randomly sampled across all equilibria in the training set compose each batch
        (\ie, individual flux surfaces from different equilibria are randomly shuffled into the same batch).
        }
  \item[Gradient stage]{%
        In this stage,
        the gradient loss $L_g$ is included in the overall loss.
        To compute the radial derivatives of the solution,
        the flux surfaces of the same equilibrium are aggregated together in the same batch.
        However,
        multiple equilibria are still randomly shuffled into each batch.
        }
  \item[Physics regularization stage]{%
        Finally,
        the physics regularization term $L_{\text{MHD}}$ is included in the loss function.
        As in the gradient stage,
        and to compute $f_*$,
        the flux surfaces of the same equilibrium are aggregated together in the same batch.
        Still,
        multiple equilibria form each batch.
        }
\end{description}

During the physics regularization stage,
a curriculum learning approach is used~\cite{Bengio2009}.
The high condition number limits the convergence rate and imposes a superior bound on the learning rate
(\ie, the step size):
even if the model commits a finite,
but small error on the equilibrium solution,
the regularization term might lead to a diverging loss.

To mitigate this issue,
the model is gradually regularized,
starting with well-predicted equilibria.
At every batch,
the physics regularization term is included into the loss function only for the subset of equilibria where the \gls{mse} on $(R, Z)$
(the key quantities affecting the \gls{MHD} residual)
is below a given threshold.
When the regularization term has not decreased in the previous \num{50} epochs,
the threshold is progressively increased by a factor of \num{10}.
More formally,
the physics regularization term for the generic batch $\mathcal{B}$ at the $k$-th step is:

\begin{gather}
  L_{\text{MHD},\mathcal{B}}^k = \frac{1}{|\mathcal{B}|} \sum_{i=1}^{|\mathcal{B}|} w_i^k L_{\text{MHD},i}, \\
  w_i^k = H(L^k - L_{(R,Z),i}) ,
\end{gather}

where $L_{\text{MHD},i}$ is the physics regularization term for the $i$-th equilibrium in the batch,
$H(\cdot)$ is the Heaviside function,
$L^k$ is the loss threshold value at the $k$-th step,
$L_{(R,Z),i}$ is the \gls{mse} on $(R, Z)$ for the $i$-th equilibrium.
The loss threshold is progressively increased,
\ie, $L^k = \si{10}^k L^0$.
An initial threshold of $L^0=\SI{3e-7}{\meter\squared}$ is used.
At the fourth step,
the physics regularization term is computed on all the equilibria in the batch,
namely,
$L^4 = +\infty$.

\subsection{\glsdisp{HP}{Hyper-parameters}}\label{sec:hp}

Due to the computational cost of the whole training procedure
($\simeq\num{8}$ days on a single NVIDIA Tesla-V100 \gls{GPU}),
a simplified scenario is used to identify the best model \glspl{HP}.
Given a fixed budget of $\simeq\SI{6}{\mega\nothing}$ free parameters,
\gls{HP} grid search is performed for the depth and width of both trunk and branch networks.
In addition,
only half of the training data is used.
The optimum \glspl{HP} combination is chosen based on the loss on the validation set in the data stage.

The trunk and branch networks are then proportionally upscaled to reach $\simeq\SI{40}{\mega\nothing}$ free parameters.
\Cref{tab:trunk-hp,tab:branch-hp} summarize the \glspl{HP} of the final model.

The AdamW~\cite{Kingma2015,Loshchilov2017} optimizer is used to train the model.
In each stage,
due to the different loss function,
the learning rate,
the learning rate scheduler,
and the $L_2$ regularization term are derived with a limited grid search.
The loss on the validation set is used to inform the search.
In addition,
early stopping~\cite{Morgan1989} is employed during training.
\Cref{tab:data-stage-hp,tab:gradient-stage-hp,tab:physics-stage-hp} report the \glspl{HP} used in each training stage.
\section{Results}\label{sec:results}

In the following chapter, the model's accuracy is evaluated by how well it predicts the equilibrium problem solution and reconstructs equilibrium properties of interest (which are not explicitly included in the loss function) across many physics domains (\cref{tab:results}):
flux surface geometry (\cref{sec:geo});
iota profile, shear profile and magnetic field structure (\cref{sec:mag});
\gls{MHD} force residual (\cref{sec:mhd-loss});
ideal-\gls{MHD} stability (\cref{sec:mhd-stability});
neoclassical transport (\cref{sec:neo-transport});
fast particle confinement (\cref{sec:fast-ions}).

The \gls{rmse} and \gls{mape} summarize the error on the test set.
Given a quantity of interest $y \in \real^K$ (\eg, the magnetic well),
the \gls{rmse} evaluates the average residual between the ground truth and predicted values for that quantity.
If $\mathcal{Y} = \{y \in \real^K\}$ is the set of true values,
and $\hat{\mathcal{Y}} = \{\hat{y} \in \real^K\}$ is the set of related model predictions,
the \gls{rmse} on $y$ is:

\begin{gather}\label{eq:rmse}
  \text{rmse}_y = \sqrt{\frac{1}{K|\mathcal{Y}|} \sum_{i=1}^{|\mathcal{Y}|} \sum_{k=1}^{K} (\hat{y}_i^k - y_i^k)^2} .
\end{gather}

Given a quantity of interest $y \in \real^K$,
the \gls{mape} measures the average relative deviation of ground truth with respect to predicted values for that quantity.
Using the same notation as for the \gls{rmse},
the \gls{mape} on $y$ is:

\begin{gather}\label{eq:mape}
  \text{mape}_y = \frac{1}{K|\mathcal{Y}|} \sum_{i=1}^{|\mathcal{Y}|} \sum_{k=1}^{K} \vert \frac{ \hat{y}_i^k - y_i^k }{y_i^k} \vert .
\end{gather}

\begin{table}[h!]
  \caption{
    Model error metrics on the test set across multiple equilibrium properties (which are not explicitly included in the loss function).
    For the description of each quantity,
    please see~\cref{sec:geo,sec:mag,sec:mhd-loss,sec:mhd-stability,sec:neo-transport,sec:fast-ions}.
    Each row represents either an error metric
    (\eg, the flux surface error $\epsilon_{\text{Fs}}$)
    or an equilibrium property
    (\eg, the proxy for the radial force balance $f_*$).
    In both cases,
    the table shows the average over the whole test set.
    In case of an equilibrium property,
    the bracket indicates the ground truth value as computed by \gls{VMEC}.
  }%
  \label{tab:results}
  \centering
  \begin{tabular}{lc}
    \toprule
    \multicolumn{2}{c}{Flux Surfaces Geometry}                                                                                     \\%
    \midrule
    Flux Surface Error $\epsilon_{\text{Fs}}$ [\si{\milli\meter}]                                  & \num{6.16e-1}                 \\%
    $R_{\text{axis}}$ \gls{rmse} [\si{\milli\meter}]                                               & \num{5.84e-1}                 \\%
    $\lambda$ \gls{rmse} [\si{\milli\radian}]                                                      & \num{3.90e+0}                 \\%
    \midrule
    \multicolumn{2}{c}{Magnetics}                                                                                                  \\%
    \midrule
    Iota \gls{mape}                                                                                & \num{9.31e-4}                 \\%
    Shear \gls{mape}                                                                               & \num{8.44e-1}                 \\%
    Mean Shear \gls{mape}                                                                          & \num{8.16e-2}                 \\%
    Magnetic Field \gls{rmse} [\si{\milli\tesla}]                                                  & \num{2.36e+1}                 \\%
    Volume Averaged Plasma Beta \gls{mape}                                                         & \num{1.37e-3}                 \\%
    \midrule
    \multicolumn{2}{c}{Ideal-\gls{MHD} Force Balance}                                                                              \\%
    \midrule
    \gls{MHD} force residual proxy $\varepsilon_{f_*}$ (\gls{VMEC}) [\si{\newton\henry\per\meter}] & \num{1.97e-2} (\num{2.43e-5}) \\%
    Normalized \gls{MHD} force residual proxy $\eta_{f_*} $ (\gls{VMEC})                           & \num{5.01e-1} (\num{5.29e-4}) \\%
    \midrule
    \multicolumn{2}{c}{Ideal-\gls{MHD} Stability}                                                                                  \\%
    \midrule
    Vacuum Magnetic Well Depth $\average{W}_{\text{vacuum}}$ \gls{mape}                            & \num{3.43e-2}                 \\%
    Finite-\averagePlasmaBeta Magnetic Well Depth $\average{W}$ \gls{mape}                         & \num{2.61e-1}                 \\%
    Vacuum Mangetic Well $W_{\text{vacuum}}$                                                       & \num{5.12e-1}                 \\%
    Finite-\averagePlasmaBeta Magnetic Well $W$ \gls{mape}                                         & \num{1.93e+0}                 \\%
    Mercier Well $D_{W}$ \gls{mape}                                                                & \num{1.12e+0}                 \\%
    \midrule
    \multicolumn{2}{c}{Neoclassical Transport}                                                                                     \\%
    \midrule
    Effective Ripple \epseff ($s \leq 0.33$) \gls{mape}                                            & \num{5.46e-1}                 \\%
    Effective Ripple Proxy \epseffproxy \gls{mape}                                                 & \num{1.33e-1}                 \\%
    \midrule
    \multicolumn{2}{c}{Fast particle confinement}                                                                                  \\%
    \midrule
    Standard deviation of the maxima of $B$ $\sigma_{\Bmax}$ \gls{rmse} [\si{\milli\tesla}]        & \num{9.43e+0}                 \\%
    Standard deviation of the minima of $B$ $\sigma_{\Bmin}$ \gls{rmse} [\si{\milli\tesla}]        & \num{6.57e+0}                 \\%
    \bottomrule
  \end{tabular}
\end{table}

To further investigate the model accuracy,
and to validate its use in downstream applications,
\cref{sec:optimization} shows the a posteriori optimization of \gls{W7X} coil currents to search for ideal-\gls{MHD} unstable and improved fast particle confinement configurations.
In this task,
the model replaces a traditional non-linear ideal-\gls{MHD} \replaced[id=AM]{code}{solver} (\eg, \gls{VMEC}) in \replaced[id=AM]{constructing}{solving} free-boundary equilibria.

The results of the model on the test set are now presented.
Because of the adopted finite differences scheme,
numerical noise affects quantities that rely on the radial derivative of $(R, \lambda, Z)$ at the magnetic axis and on the nearest flux surface%
; those values are invalid.
A note is present in the caption of all affected figures.

\subsection{Flux Surface Geometry}\label{sec:geo}

The geometry of the flux surfaces $(R, \lambda, Z)$ and \ibar represent the solution of the ideal-\gls{MHD} equilibrium problem:
they fully describe the equilibrium magnetic field and its properties.
Therefore,
the accuracy with which the model predicts $(R, \lambda, Z)$ and \ibar bounds the accuracy with which it reconstructs equilibrium properties.

\begin{figure}[h!]
  \igraph[]{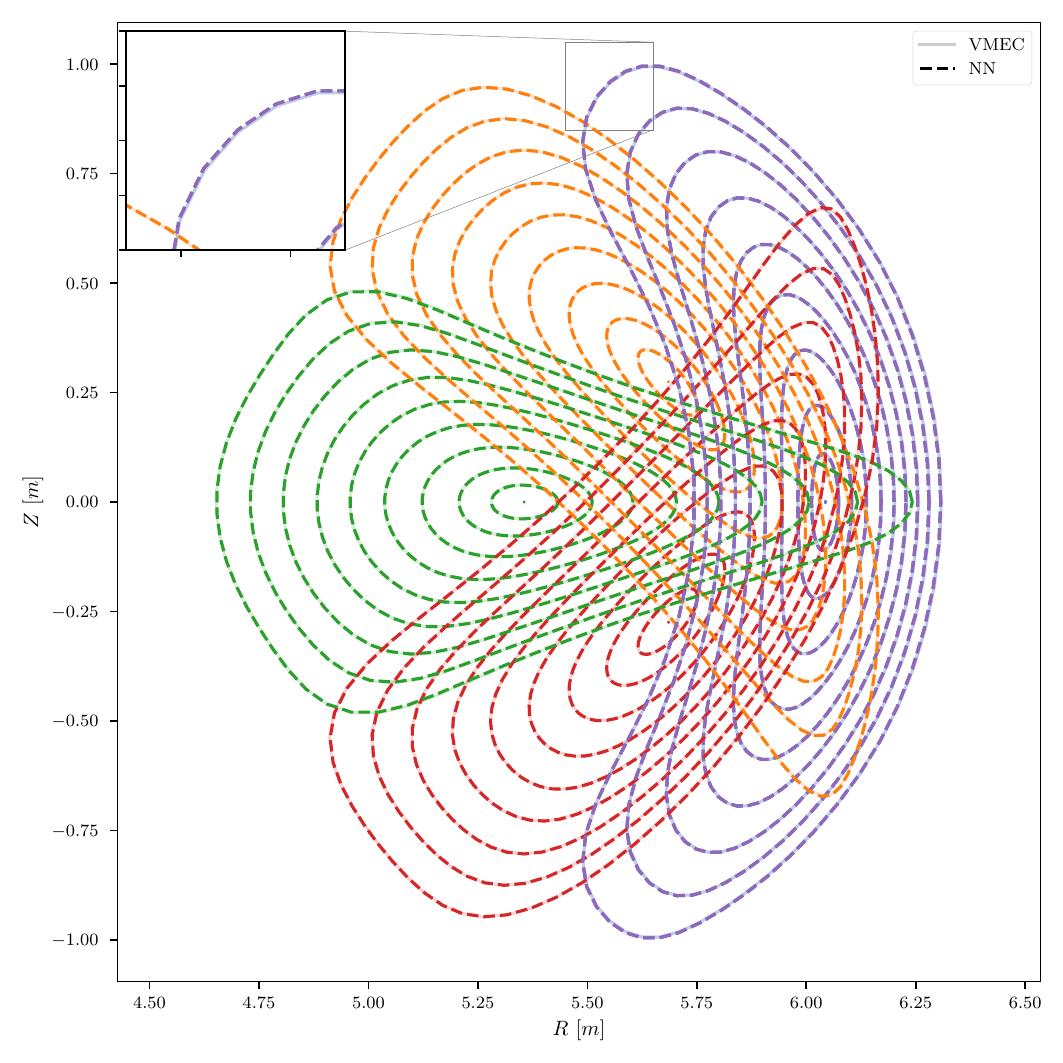}%
  \caption{
    The ground truth (solid) and predicted (dashed) flux surfaces geometry,
    $(R, Z)$,
    for the median predicted equilibrium.
    Four toroidal locations at
    $\varphi=\SI{0}{\degree}$ (pink),
    $\varphi=\SI{18}{\degree}$ (orange),
    $\varphi=\SI{36}{\degree}$ (green),
    and $\varphi=\SI{54}{\degree}$ (red) are shown.
    For each toroidal location,
    ten flux surfaces equally spaced in $\rho=\sqrt{s}$ are shown.
    The model approximates the geometry of the flux surfaces with high accuracy:
    the average distance from the ground truth solution is below $\SI{1}{\milli\meter}$.
    To highlight the model prediction,
    the region in which the flux surfaces have the largest curvature is zoomed in.
    Indeed,
    the two curves overlap each other.
  }%
  \label{fig:flux-surfaces-median}
\end{figure}

The flux surface geometry is predicted with an exceptional accuracy (\cref{fig:flux-surfaces-median}).
The flux surface error between the ground truth and the model prediction is evaluated as:

\begin{gather}\label{eq:flux-surface-error}
  \varepsilon_{\text{Fs}} = \langle \twoNorm{[\Delta R_i, \Delta Z_i]^T} \rangle_i ,
\end{gather}

where $\Delta R_i$ and $\Delta Z_i$ are the differences in flux surfaces relative to the ground truth,
and the brackets denote the mean across all equilibria and grid locations.
\added[id=AM]{
  In other words,
  $\varepsilon_{\text{Fs}}$ represents the average absolute distance $\sqrt{(\Delta R)^2 + (\Delta Z)^2}$ across all equilibria and grid locations.
}
The model achieves $\varepsilon_{\text{Fs}}=\SI{616}{\micro\meter}$.
Similarly,
the model successfully predicts the angle renormalization parameter $\lambda$,
achieving $\text{rmse}_{\lambda}=\SI{2.41}{\milli\radian}$.

\begin{figure}[h!]
  \centering
  \igraph[]{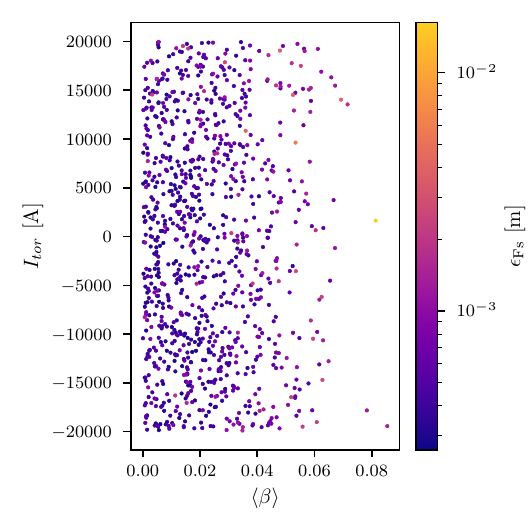}%
  \caption{
    Distribution of the flux surface error $\varepsilon_{\text{Fs}}$ as a function of the equilibrium volume averaged beta $\langle \beta \rangle$ and the net toroidal current enclosed by the plasma \Itor in the test set.
  }%
  \label{fig:metrics}
\end{figure}

\added[id=AM]{
  Low-beta equilibria are better regressed than high-beta equilibria (\cref{fig:metrics}).
  As expected from the distribution of the volume averaged beta in the training data set (see~\cref{fig:beta-hist}),
  the flux surface error increases towards the boundary of the training set
  (\ie, where the plasma beta increases).
  No clear patterns are visible in terms of the net toroidal current enclosed by the plasma.
  The flux surface error does not considerably change when the physics regularization is applied (see~\cref{tab:mhd-loss}).
}

As the plasma beta increases,
the plasma column moves outwards.
This effect is called the Shafranov shift,
and it is usually measured as $\Delta / a$,
where \mbox{$\Delta = R_{\text{axis}}(\plasmaBeta) - R_{\text{axis}}(0)$} and $a$ is the effective minor radius.
The Shafranov shift defines the equilibrium beta limit (\mbox{$\Delta / a \simeq \num{0.5}$})~\cite{Hayashi1990},
and it causes an outward movement of the \gls{LCFS}, which affects the edge magnetic field topology.
In \gls{W7X},
the minimization of the Shafranov shift was an optimization criterion~\cite{Beidler1990}.
Nevertheless,
a Shafranov shift of \mbox{$\Delta\simeq\SI{5}{\centi\meter}$} has been observed in high-\plasmaBeta operations~\cite{Bozhenkov2018}.

\begin{figure}[h!]
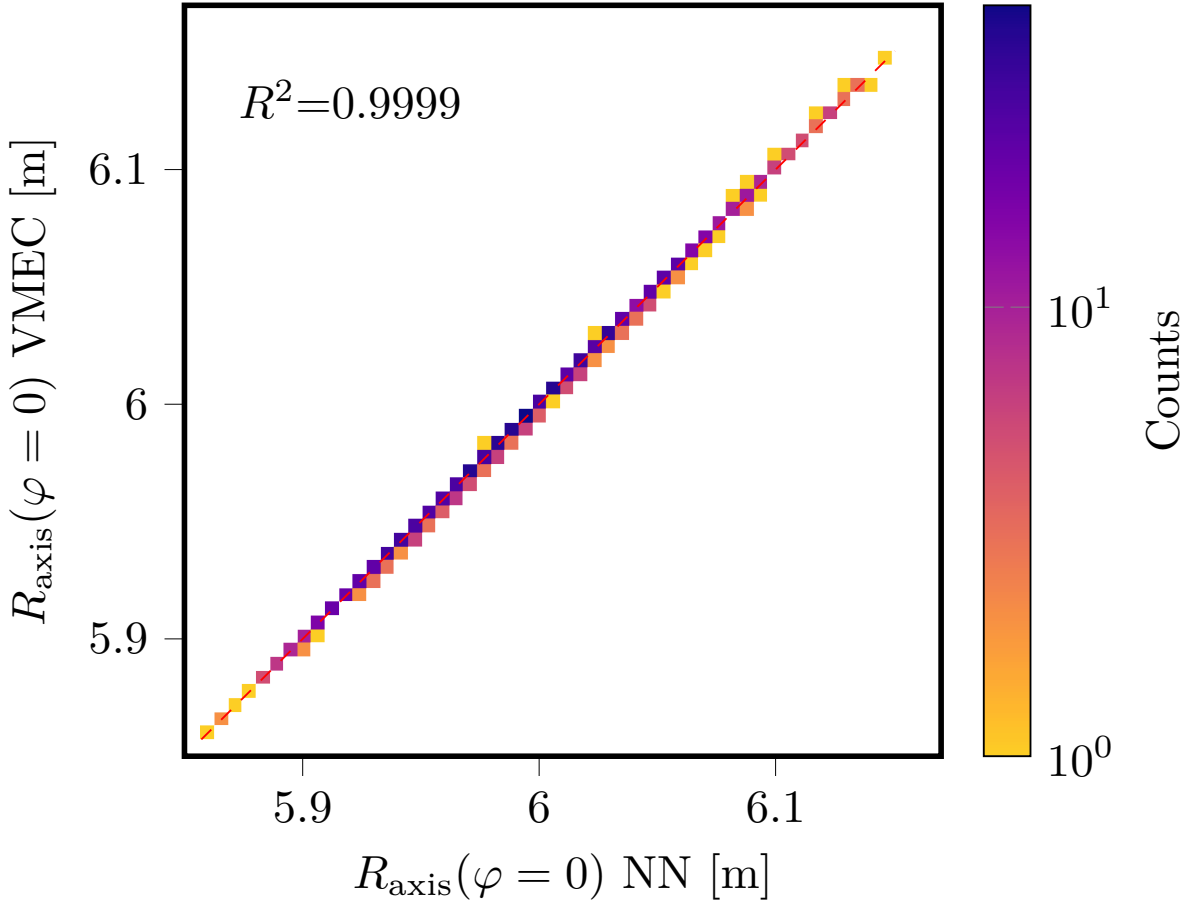

  \igraph[]{content_figures_scatter_r_axis.pgf}%
  \caption{
    Reconstruction of the location of the magnetic axis at $R_{\text{axis}}(\varphi=0)$ across all equilibria in the test set.
    The color-bar indicates the counts in each histogram bin.
    To guide the eye,
    a red dashed line indicates the case of perfect reconstruction.
  }%
  \label{fig:r-axis-scatter}
\end{figure}

The model faithfully predicts this finite-beta effect (\cref{fig:r-axis-scatter}):
the axis location at $\varphi=0$ is resolved with a relative error of $\text{mape}_{R_{\text{axis}}} < \SI{0.01}{\percent}$.
A two-dimensional histogram visualizes the distribution of ground truth and predicted values.
The color-bar indicates the counts in each two-dimensional histogram bins.
To guide the eye,
a red dashed line indicates the case of perfect regression:
the more the density of points lies on the line,
the more the model provides accurate equilibrium properties.
The coefficient of determination $R^2$ captures the reconstruction quality.

\subsection{Magnetics}\label{sec:mag}

\ibar sets the location of the resonant surfaces in the plasma volume.
It also defines the island chain and magnetic edge topology,
which affects how the \gls{W7X} island divertor operates.
Although equilibria with the assumption of nested flux surfaces cannot resolve islands,
\gls{VMEC} equilibria are an important requirement in the investigation of plasma phenomena involving resonant surfaces~\cite{Lazerson2016,Zanini2020,Zanini2021}.

The model effectively predicts the \ibar profile (\cref{fig:iota-scatter}).
A relative error of $\text{mape}_{\ibar}<\SI{0.1}{\percent}$ is achieved. 
In case of the median regressed equilibrium,
the predicted and ground truth \ibar profiles overlap (\cref{fig:iota-median}).

\begin{figure}[h!]
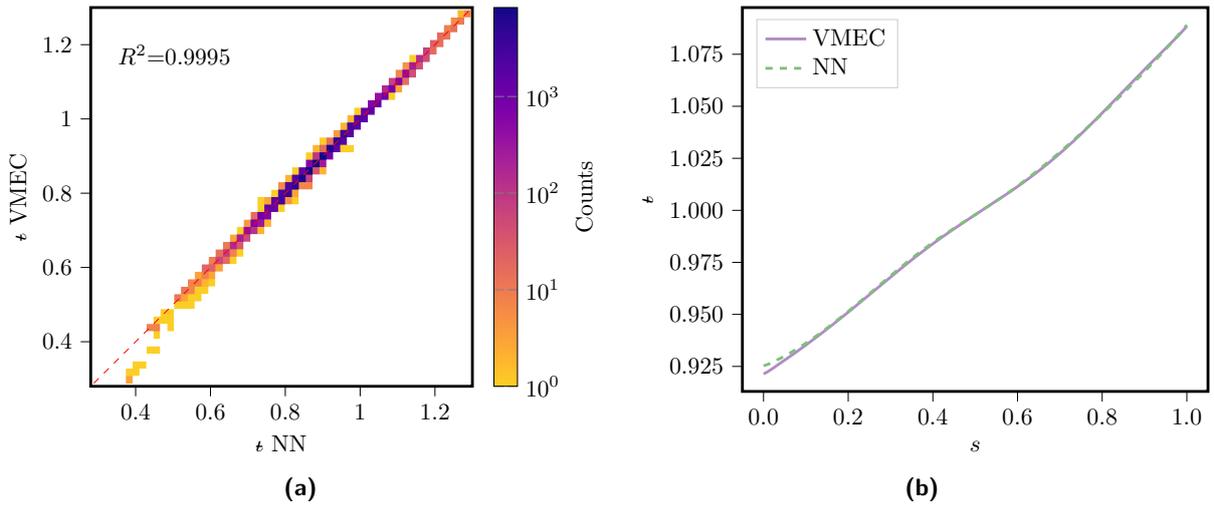

  \centering
  \multigraph[labels={fig:iota-scatter}{fig:iota-median}]{content_figures_scatter_iota.pgf;content_figures_iota.pgf}%
  {%
    {}{}%
  }%
  \caption{
    \subref{fig:iota-scatter} Reconstruction of the iota profile \gls{ibar} across all equilibria in the test set.
    The color-bar indicates the counts in each histogram bin.
    To guide the eye,
    a red dashed line indicates the case of perfect reconstruction.
    \subref{fig:iota-median} The true (solid pink) and predicted (dashed green) iota profile for the median regressed equilibrium in the test set.
  }
\end{figure}

The magnetic shear is the radial derivative of \ibar:

\begin{gather}\label{eq:shear}
  \ibar^{\prime} = \frac{d \ibar}{d \psi} = \frac{\partial \ibar}{\partial s} \frac{\partial s}{\partial \psi} = \frac{1}{\psiedge} \frac{\partial \ibar}{\partial s} ,
\end{gather}

where $2 \pi \psi(s) = \phiedge s$ is the toroidal flux.
It represents the change of direction of the magnetic field lines from one flux surface to another.
The magnetic shear influences multiple properties of the equilibrium:
it affects \gls{MHD} stability~\cite{Boozer2005,Aleynikova2022},
it is inversely proportional to the island width
(\ie, large islands are present at low shear)~\cite{Boozer2005},
and a low shear leads to long connection lengths allowing efficient divertor operation~\cite{Feng2006}.

The model qualitatively reconstructs the shear profile (\cref{fig:shear-scatter}).
\Cref{fig:shear-median} shows that the ground truth and predicted shear profile only partially overlap.
Across all equilibria in the test set,
the relative error on the magnetic shear is high $\text{mape}_{\ibar^{\prime}}=\SI{84.4}{\percent}$.

\begin{figure}[h!]
  \centering
  \multigraph[labels={fig:shear-scatter}{fig:shear-median}]{content_figures_scatter_shear_.pgf;content_figures_shear_.pgf}%
  {%
    {}{}%
  }%
  \caption{
    \subref{fig:shear-scatter} Reconstruction of the local magnetic shear $\gls{ibar}^{\prime}$ across all equilibria in the test set.
    The color-bar indicates the counts in each histogram bin.
    To guide the eye,
    a red dashed line indicates the case of perfect reconstruction.
    \subref{fig:shear-median} The true (solid pink) and predicted (dashed green) magnetic shear profile for the median regressed equilibrium in the set.
  }
\end{figure}

\begin{figure}[h!]
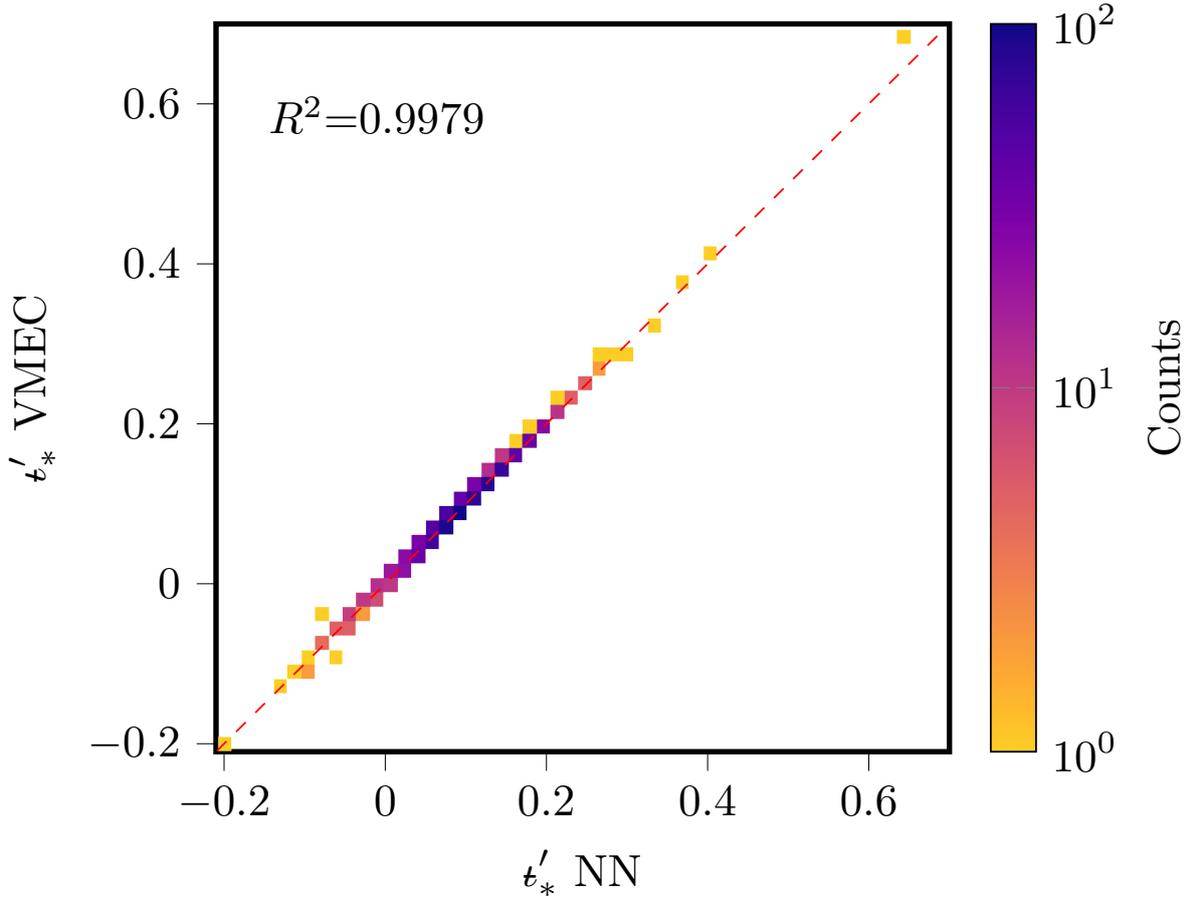

  \centering
  \igraph[]{content_figures_scatter_mean_shear.pgf}%
  \caption{
    Reconstruction of the average magnetic shear across all equilibria in the test set.
    The color-bar indicates the counts in each histogram bin.
    To guide the eye,
    a red dashed line indicates the case of perfect reconstruction.
    The average magnetic shear is defined as the shear of the least squares linear fit of the iota profile~\cite{Landreman2021b}.
  }%
  \label{fig:mean-shear-scatter}%
\end{figure}

However,
the model captures the average magnetic shear $\ibar^{\prime}_*$ (\cref{fig:mean-shear-scatter}):
$\text{mape}_{\ibar^{\prime}_*}=\SI{8.16}{\percent}$,
where the average magnetic shear is the shear of the least squares linear fit of the iota profile.

The structure of the magnetic field plays a crucial role in many physics aspects:
neoclassical and turbulence transport,
confinement of energetic particles,
plasma instabilities,
and deposition of \gls{ECRH} heating power.

\begin{figure}[h!]
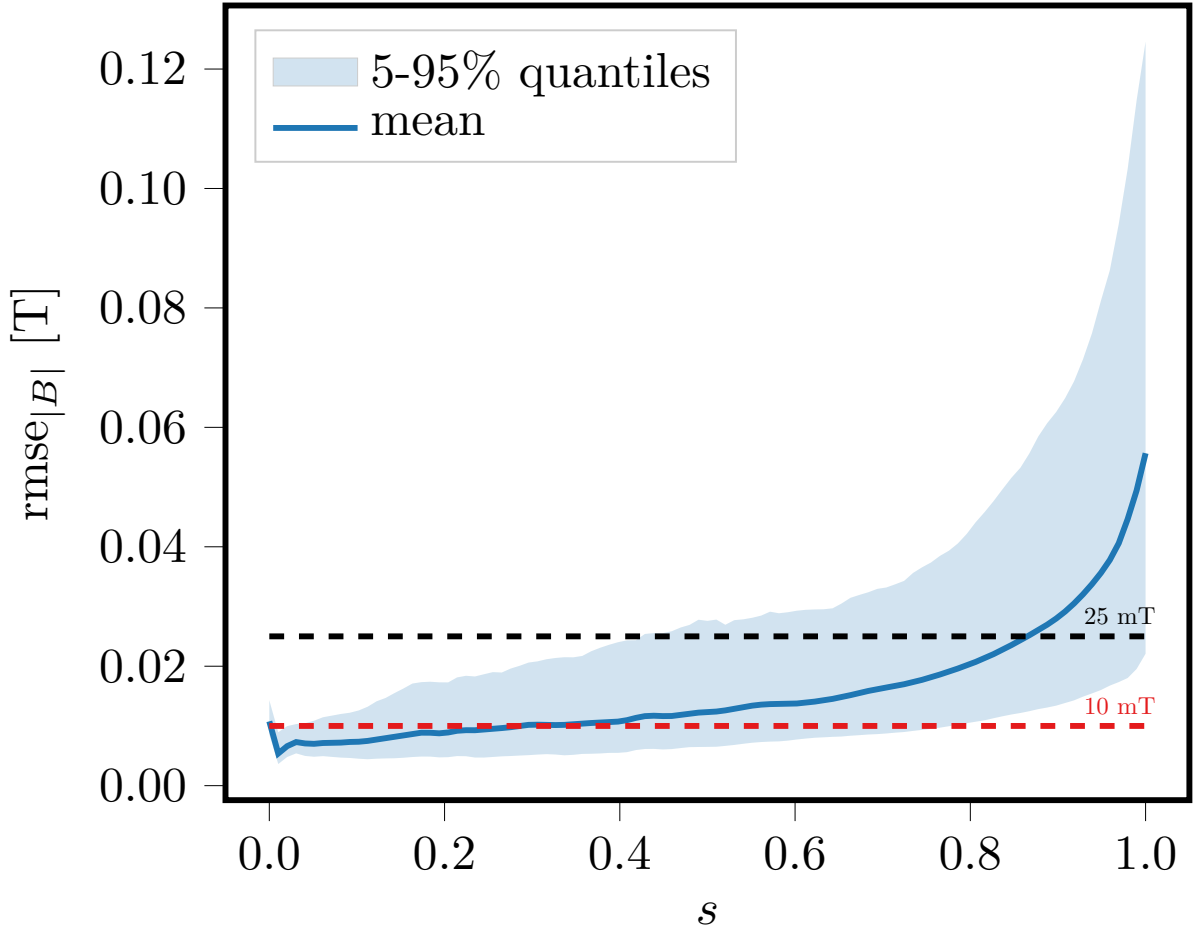

  \centering
  \igraph[]{content_figures_b_rmse.pgf}%
  \caption{
    The error on the magnetic field strength as a function of the radial location,
    averaged across all grid points on the flux surface.
    The solid line represents the mean across all equilibria in the test set,
    and the shaded regions represent the \SI{95}{\percent} and \SI{5}{\percent} quantiles.
    To guide the eye,
    two reference values are shown:
    \SI{25}{\milli\tesla} (dashed black) represents a \SI{1}{\percent} error
    (\gls{W7X} has a field of \SI{2.5}{\tesla} on average on axis),
    and \SI{10}{\milli\tesla} (dashed red) represents the expected accuracy needed on the magnetic field strength for equilibrium reconstruction routines~\cite{HoefelPC2021}.
    Numerical noise in the finite difference scheme invalidates the values on the magnetic axis (see~\cref{sec:results}).
  }%
  \label{fig:b-flux-surface-averaged-error}
\end{figure}

The model correctly reconstructs the magnetic field structure (\cref{fig:b-flux-surface-averaged-error}).
The model achieves a $\text{rmse}_B=\SI{23.6}{\milli\tesla}$ on all equilibria in the test set.
When compared with an average \gls{W7X} field of $B_0=\SI{2.5}{\tesla}$,
this value represents a relative error below \SI{1}{\percent}.

However,
the magnetic field strength error differs along the radial profile (\cref{fig:b-flux-surface-averaged-error}).
On average,
the field strength error is below \SI{10}{\milli\tesla} for a considerable fraction of the plasma volume,
up to $s\simeq\SI{0.3}{}$ (\ie, $\rho\simeq\SI{0.55}{}$).
The accuracy of the magnetic field strength in the core region is sufficient to navigate the magnetic configuration space of \gls{W7X} to obtain fast particle optimized configurations (see~\cref{sec:optim-fast-particle}).
Moreover,
an accuracy of \SI{10}{\milli\tesla} is expected to be enough to perform equilibrium reconstruction routines~\cite{HoefelPC2021}.

\begin{figure}[h!]
  \centering
  \igraph[]{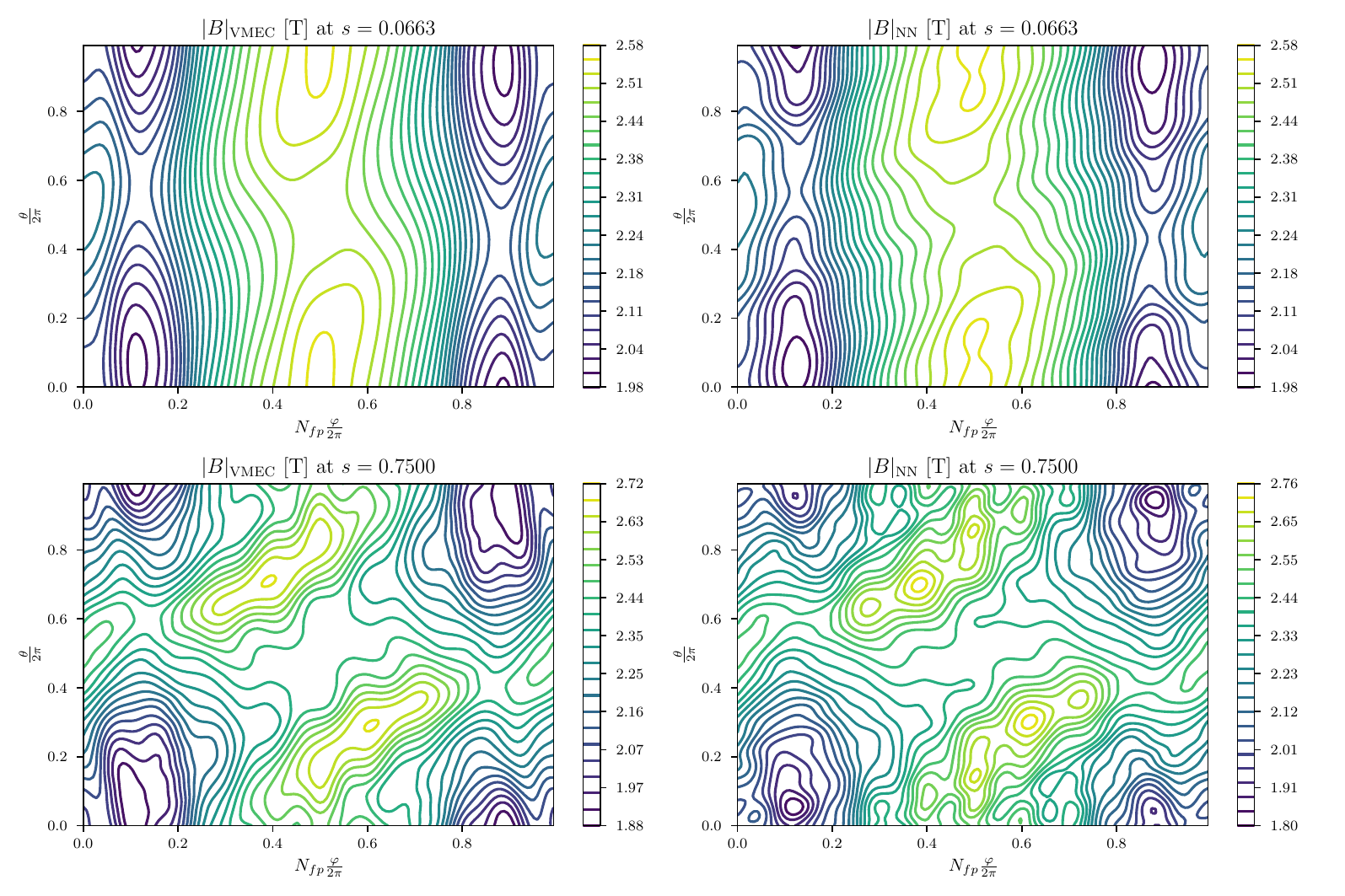}%
  \caption{
    True (\replaced[id=AM]{left}{right}) and predicted (\replaced[id=AM]{right}{left}) $B$ magnetic field strength structure at two radial locations:
    $s=\num{0.06}$ (top, close to the magnetic axis) and $s=\num{0.75}$ (bottom, close to the plasma boundary).
    The median predicted equilibrium in the test set is shown.
  }%
  \label{fig:b-median}
\end{figure}

The model introduces artificial field ripples in the equilibrium solution (\cref{fig:b-median}):
at the plasma edge,
and also to a less extent close to the magnetic axis,
the predicted $B$ field is not as smooth as the ground truth solution.
The flux surface geometry Jacobian $\sqrt{g}$ and the metric tensor elements $g_{ij}$ mainly define the magnetic field $B$ (\cref{sec:mhd}).
These quantities are highly sensitive to the radial derivative of the flux surface coordinates $(R, \lambda, Z)$.
High frequency components in the learned trunk network \textit{basis} functions might be the cause of such field ripples.

\subsection{Ideal-\gls{MHD} loss}\label{sec:mhd-loss}

A high condition number,
which has several negative implications,
characterizes the ideal-\gls{MHD} equilibrium problem (see~\cref{sec:physics-loss}).

Firstly,
a high condition number limits the rate of convergence.
During the physics regularization training stage,
the weights of the \gls{NN} model are adjusted to minimize the \gls{MHD} force residual for the equilibria in the training set.
The condition number sets the convergence rate of this gradient descent method.
To accelerate the descent,
non-linear \gls{MHD} \replaced[id=AM]{codes}{solvers} adopt a preconditioning matrix, which decreases the condition number of the problem~\cite{Hirshman1991}.
In this work,
this technique is not available:
in general,
preconditioning algorithms require an approximation of the Hessian matrix,
which is impractical due to the large number of free parameters of the \gls{NN} model.

Secondly,
a high condition number implies that the \gls{MHD} force residual is highly sensitive to the equilibrium solution.
For example,
if we define $\vec{x}_0$ to be that solution that satisfies the \gls{MHD} force residual
$F_{\text{MHD}}(\vec{x}_0)=0$
and $\kappa$ as the condition number,
then \mbox{$F_{\text{MHD}}(\vec{x}_0 + \vec{\epsilon}) \simeq \kappa \twoNorm{\vec{\epsilon}} \gg \num{1}$},
where $\vec{\epsilon}$ is a small displacement to the equilibrium solution.
Even if the model provides a good approximation of the equilibrium solution,
the \gls{MHD} force residual is not guaranteed to be null.

To quantify how well the equilibria satisfy the \gls{MHD} equations,
two measures are adopted.
The proxy and full \gls{MHD} force residuals:

\begin{gather}
  \varepsilon_{f_*} = \frac{\mu_0}{(2 \pi)^2} \langle F_s \vert_{F_{\beta}=0} \rangle_{\text{Vol}} = \int_{s_{\text{min}}}^1 f_* ds,\\
  \varepsilon_F = \frac{\mu_0}{(2 \pi)^2} \langle \twoNorm{\vec{F}} \rangle_{\text{Vol}} = \frac{\mu_0}{(2 \pi)^2} \int_{s_{\text{min}}}^1 \int_0^{2\pi} \int_0^{2\pi} \twoNorm{\vec{F}} \sqrt{g} ds d\theta d\varphi.
\end{gather}

And their normalized versions:

\begin{gather}
  \eta_{f_*} = \frac{\int_{s_{\text{min}}}^1 f_* ds}{\int_{s_{\text{min}}}^1 \int_0^{2\pi}\int_0^{2\pi} \twoNorm{\vec{\nabla} p} \sqrt{g} ds d\theta d\varphi},\\
  \eta_F = \frac{\int_{s_{\text{min}}}^1 \int \int \twoNorm{\vec{F}} \sqrt{g} ds d\theta d\varphi}{\int_{s_{\text{min}}}^1 \int_0^{2\pi} \int_0^{2\pi} \twoNorm{\vec{\nabla} p} \sqrt{g} ds d\theta d\varphi}.
\end{gather}

The model trained with the physics regularization term better satisfies the ideal-\gls{MHD} force residual compared with a model that has not been regularized (\cref{tab:mhd-loss}).
Both models show similar flux surface and iota errors,
however,
the physics regularized model shows a reduced ideal-\gls{MHD} force proxy residual.
However,
this number  is still three orders of magnitude higher than the value of ground truth equilibria from \gls{VMEC}.

\begin{table}[h!]
  \caption{
  Comparison between the not-regularized and regularized model in minimizing the \gls{MHD} force residual proxy across all equilibria in the test set.
  NN denotes a model that has not be regularized with the \gls{MHD} force residual
  (\ie, trained only in the data and gradient stages),
  $\text{NN}_{\text{regularized}}$ denotes the model that has been regularized
  (\ie, trained also in the physics regularization stage),
  and \gls{VMEC} denotes the value for the ground truth equilibria.
  }%
  \label{tab:mhd-loss}
  \centering
  \begin{tabular}{lccc}
    \toprule
                                                       & NN            & $\text{NN}_{\text{regularized}}$ & \gls{VMEC}    \\%
    \midrule
    $\text{rmse}_{\text{Fs}}$ [\si{\milli\meter}]      & \num{6.08e-1} & \num{6.16e-1}                    & -             \\%
    $\text{mape}_{\gls{ibar}}$                         & \num{9.24e-4} & \num{9.31e-4}                    & -             \\%
    $\varepsilon_{f_*}$ [\si{\newton\henry\per\meter}] & \num{7.27e-2} & \num{1.97e-2}                    & \num{2.43e-5} \\%
    \bottomrule
  \end{tabular}
\end{table}

\begin{figure}[h!]
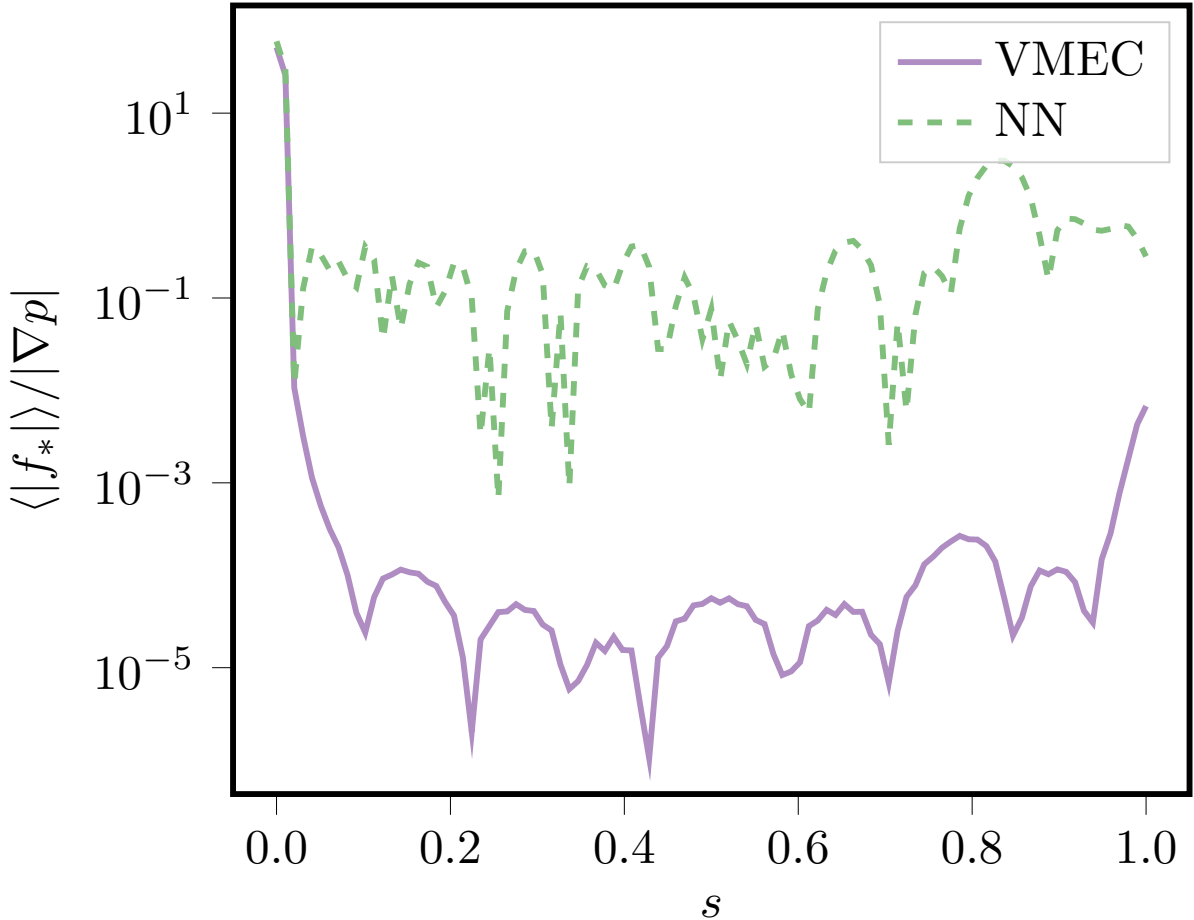

  \igraph[]{content_figures_normf.pgf}%
  \caption{
    True (solid pink) and predicted (dashed green) normalized \gls{MHD} force proxy residual for the median regressed equilibrium.
    Numerical noise in the finite difference scheme invalidates the values at $s \in [0, 0.01]$ (see~\cref{sec:results}).
  }%
  \label{fig:normf-median}
\end{figure}

Even though the model provides the solution of the ideal-\gls{MHD} equilibrium problem
(the geometry of the flux surfaces and the iota profile)
with high accuracy,
the ideal-\gls{MHD} force residual is poorly satisfied.
In case of the median predicted equilibrium (at $\averagePlasmaBeta=\SI{1.39}{\percent}$),
the normalized \gls{MHD} force proxy residual is $\eta_{f_*} = \SI{23.3}{\percent}$ (\cref{fig:normf-median}).
For comparison,
the ground truth equilibrium has a normalized residual of $\SI{3.04e-2}{\percent}$.

\subsubsection{Fine-tuning at inference time}\label{sec:fine-tuning}

Let us call
a \textit{single-pass} equilibrium an equilibrium solution provided by the model with a single forward pass.

\textit{Single-pass} equilibria poorly satisfy the ideal-\gls{MHD} equations (\cref{sec:mhd-loss}).
However,
without access to the ground truth solution,
equilibria can be improved at inference time.
The original free-boundary equilibrium problem is cast into an equivalent fixed-boundary one:
the plasma boundary of the \textit{single-pass} free-boundary equilibrium defines the boundary condition on $(R, Z)$ for the equivalent fixed-boundary equilibrium.
The error on $(R, Z)$ of the \textit{single-pass} equilibrium is below \SI{1}{\milli\meter} (\Cref{sec:geo}),
therefore,
the plasma boundary can be safely assumed to be close to the correct.

The \added[id=AM]{solution of the }equivalent fixed-boundary equilibrium problem is \replaced[id=AM]{approximated}{solved} by minimizing the \gls{MHD} force residual at inference time.
The loss function has three terms:
the full \gls{MHD} force residual,
a distance between the model and the fixed boundary $(R_b, Z_b)$ derived from the \textit{single-pass} equilibrium,
and the satisfaction of the required toroidal current profile $\Itor(s)$.
The loss function is then:

\begin{gather}%
  \label{eq:fine-tuning-loss}
  L = \alpha_{\text{MHD}} L_{\text{MHD}} + \alpha_b L_b + \alpha_I L_I ,\\
  L_{\text{MHD}} = \frac{\mu_0^2}{(2 \pi)^4} \int_{s_{\text{min}}}^1 \langle \twoNorm{\vec{F}} \rangle_{\text{Fs}}^2 ds ,\\
  L_b = \langle (\Delta R_i^b)^2 \rangle_i + \langle (\Delta Z_i^b)^2 \rangle_i ,\\
  L_I = \mu_0^2 \int_{s_{\text{min}}}^1 (\tilde{I}_{\text{tor}} - \Itor)^2 ds ,
\end{gather}

where $\alpha_{\text{MHD}}=\num{1e-3}$,
$\alpha_b =\num{0.2495}$,
$\alpha_I=\num{0.5}$,
and $s_{\text{min}}=\num{0.02}$.
The \textit{single-pass} equilibrium is used as initial guess for the minimization problem.
The AdamW algorithm minimizes the loss function.
The initial learning rate is set to \num{e-4},
and it is decreased to \num{e-6} at \num{e4} steps with an exponential decay.
The training is halted after \num{2e4} steps.
The branch network,
apart from the bias of the last layer,
is frozen.
For a complete description of the \glspl{HP},
see~\Cref{sec:appendix-hp}.
We call this optimization at the inference step \textit{fine-tuning}.

In terms of the full \gls{MHD} force residual,
the model can provide better-than-ground-truth equilibria (\cref{tab:mhd-loss-fine-tuning,fig:fine-tuning-normf,fig:fine-tuning-normF}).
As an example,
we consider a \gls{W7X} standard configuration at $\averagePlasmaBeta=\SI{2}{\percent}$, which is not included in the training data.
Both the \textit{single-pass} and fine-tuned equilibrium show low flux surface errors,
however,
the fine-tuned equilibrium satisfies the ideal-\gls{MHD} equations much better than the \textit{single-pass} equilibrium:
the normalized error is only $\eta_{f_*}\simeq\SI{1}{\percent}$.
Surprisingly,
the fine-tuned equilibrium minimizes the full \gls{MHD} force residual better than the ground truth equilibrium computed by \gls{VMEC}.

\begin{table}[h!]
  \caption{
  Comparison between the fine-tuned, not-regularized and regularized (\textit{single-pass}) model in minimizing the \gls{MHD} force residual in case of a \gls{W7X} standard configuration at $\averagePlasmaBeta=\SI{2}{\percent}$,
  which is not part of the training data.
  $\varepsilon_{f_*}$ and $\varepsilon_F$ denote the proxy and full \gls{MHD} force residuals,
  $\eta_{f_*}$ and $\eta_F$ denote the proxy and full normalized \gls{MHD} force residuals.
  Please refer to~\cref{sec:mhd-loss} for the definition of these quantities.
  NN denotes a model that has not be regularized with the \gls{MHD} force residual
  (\ie, trained only in the data and gradient stages),
  $\text{NN}_{\text{regularized}}$ denotes the model that has been regularized with \gls{MHD} force residual
  (\ie, the \textit{single-pass} model),
  and $\text{NN}_{\text{fine-tuned}}$ denotes the fine-tuned model.
  \gls{VMEC} denotes the value for the ground truth equilibrium.
  For each quantity,
  the best value is highlighted in bold.
  }%
  \label{tab:mhd-loss-fine-tuning}
  \centering
  \begin{tabular}{lcccc}
    \toprule
                                                       & NN                             & $\text{NN}_{\text{regularized}}$ & $\text{NN}_{\text{fine-tuned}}$ & \gls{VMEC}                     \\
    \midrule
    $\text{rmse}_{\text{Fs}}$ [\si{\milli\meter}]      & \num[math-rm=\mathbf]{3.11e-4} & \num{3.21e-4}                    & \num{3.63e-3}                   & -                              \\
    $\varepsilon_{f_*}$ [\si{\newton\henry\per\meter}] & \num{4.90e-2}                  & \num{1.18e-2}                    & \num{8.11e-4}                   & \num[math-rm=\mathbf]{9.86e-6} \\
    $\varepsilon_F$ [\si{\newton\henry\per\meter}]     & \num{8.83e-1}                  & \num{8.28e-1}                    & \num[math-rm=\mathbf]{1.24e-2}  & \num{5.43e-2}                  \\
    $\eta_{f_*}$                                       & \num{7.17e-1}                  & \num{1.73e-1}                    & \num{1.19e-2}                   & \num[math-rm=\mathbf]{1.44e-4} \\
    $\eta_F$                                           & \num{3.51e+0}                  & \num{3.29e+0}                    & \num[math-rm=\mathbf]{4.92e-2}  & \num{2.16e-1}                  \\
    \bottomrule
  \end{tabular}
\end{table}

\begin{figure}[h!]
  \centering
  \multigraph[labels={fig:fine-tuning-normf}{fig:fine-tuning-normF}]{content_figures_normf_fine_tuned.pgf;content_figures_normF_fine_tuned.pgf}%
  {%
    {}{}%
  }%
  \caption{
    The comparison between the \textit{single-pass} and fine-tuned model in minimizing the proxy and full \gls{MHD} force residuals.
    \subref{fig:fine-tuning-normf} The normalized \gls{MHD} force residual proxy for the \textit{single-pass} (dashed green),
    fine-tuned (solid green),
    and ground truth equilibrium (solid pink).
    A \gls{W7X} standard configuration at $\averagePlasmaBeta=\SI{2}{\percent}$ equilibrium,
    which was not part of the training data,
    is shown here.
    Numerical noise in the finite difference scheme invalidates the values at $s \in [0, 0.01]$ (see~\cref{sec:results}).
    \subref{fig:fine-tuning-normF} The normalized full \gls{MHD} force residual for the \textit{single-pass} (dashed green),
    fine-tuned (solid green),
    and ground truth equilibrium (solid pink).
    A \gls{W7X} standard configuration at $\averagePlasmaBeta=\SI{2}{\percent}$ equilibrium,
    which was not part of the training data,
    is shown here.
    Numerical noise in the finite difference scheme invalidates the values at $s \in [0, 0.01]$ (see~\cref{sec:results}).
  }
\end{figure}

\begin{figure}[h!]
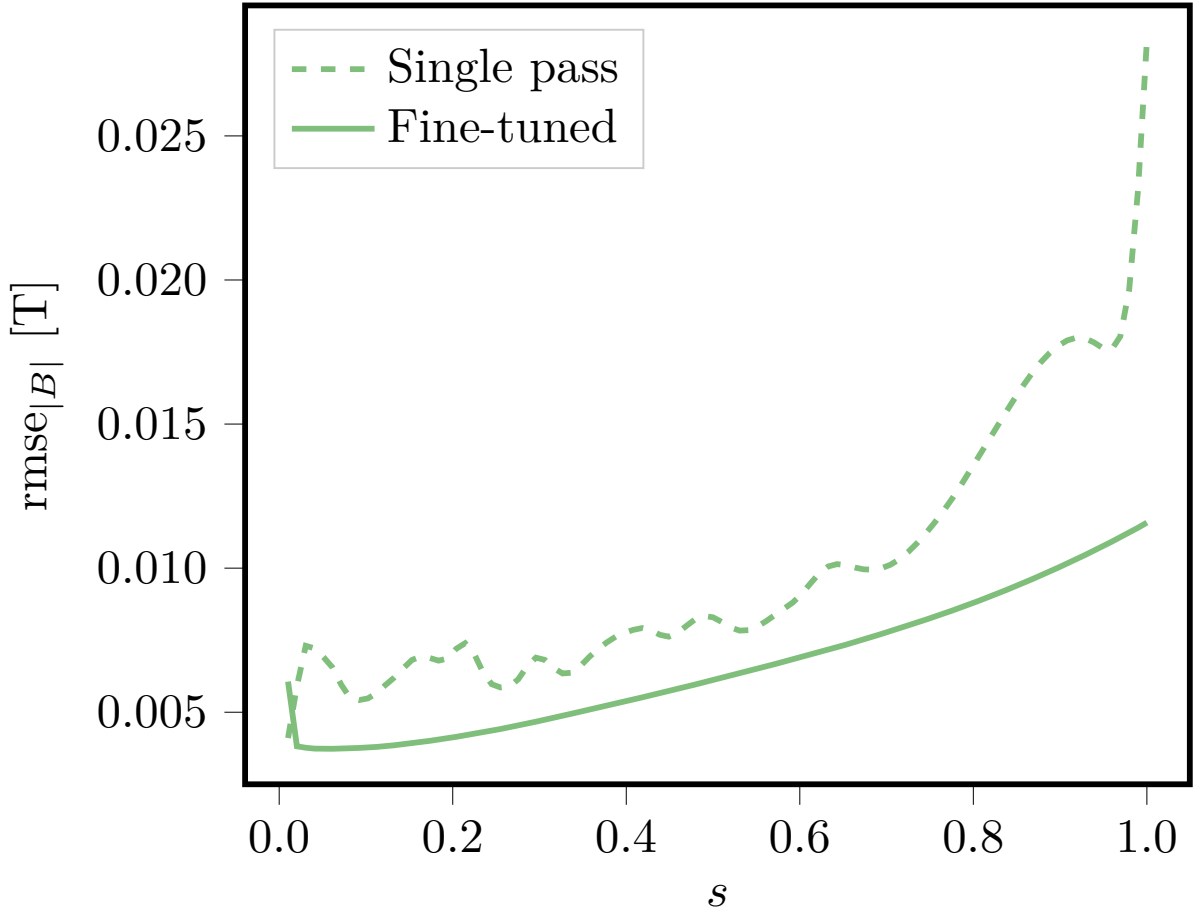

  \igraph[]{content_figures_b_rmse_fine_tuned.pgf}%
  \caption{
    The magnetic field strength error as a function of the radial profile,
    averaged across a flux surface,
    in case of a \gls{W7X} standard configuration at $\averagePlasmaBeta=\SI{2}{\percent}$ equilibrium,
    which was not part of the training data.
    The \textit{single-pass} (dashed green),
    and fine-tuned (solid green) equilibria are shown.
    Numerical noise in the finite difference scheme invalidates the values on the magnetic axis (see~\cref{sec:results}).
  }%
  \label{fig:fine-tuning-rmse-b}
\end{figure}

In addition,
the fine-tuning procedure also improves the reconstruction of the magnetic field strength:
the average error is below \SI{10}{\milli\tesla} for the entire plasma volume (\cref{fig:fine-tuning-rmse-b}).

The fine-tuning procedure is currently computationally expensive:
on a single \gls{GPU},
the fine-tuning procedure reaches the \gls{MHD} force residual of the \gls{VMEC} ground truth equilibrium after \num{4807} iterations (\ie, gradient update step) and \SI{3610}{\second},
whereas \gls{VMEC} converges after \num{13783} iterations and \SI{127}{\second} (on \num{20} \gls{CPU} cores).
Both the fine-tuning procedure and \gls{VMEC} use a gradient descent method,
however,
even if the model requires fewer iterations to achieve the same \gls{MHD} force residual,
the model runtime is longer:
the network and loss function are described in python,
while \gls{VMEC} is written in FORTRAN (thus resulting in compiled code).

The fine-tuning procedure simply shows how the single-pass model predictions can be improved in a self-supervised fashion
(\ie, without the use of a ground truth equilibrium).
The reduction of the fine-tuning runtime
(\eg, just-in-time compilation of both the network and loss function)
and the improvement of the fine-tuning convergence
(\eg, higher learning rate) are left for future investigations.

\subsection{Ideal-\gls{MHD} stability}\label{sec:mhd-stability}

The magnetic well $W$~\cite{Greene1997,Freidberg2014} represents a fast proxy for ideal-\gls{MHD} stability~\cite{Drevlak2018,Landreman2021b}.
The configuration is taken to be stable if~\cite{Bauer1984}:

\begin{gather}%
  \label{eq:pressure-well}
  p^{\prime} W < 0
\end{gather}

Therefore,
for a standard pressure profile with $p^{\prime} < 0$,
a positive $W$ is favourable for stability.

In the literature,
multiple definitions can be found.
When considering vacuum configurations,
the magnetic well can be expressed as the second radial derivative of the plasma volume:

\begin{gather}%
  \label{eq:well-vacuum}
  W_{\text{vacuum}} (s) = \frac{d^2 V_p}{d \psi^2} = \frac{d^2 V_p}{d s^2} \frac{d^2 s}{d \psi^2} = \frac{1}{\psiedge^2} \frac{d^2 V_p}{d s^2} ,
\end{gather}

where $V_p(s)$ is the plasma volume within the flux surface labelled $s$:

\begin{gather}%
  \label{eq:plasma-volume}
  V_p(s) = \int_0^s \int_0^{2 \pi} \int_0^{2 \pi} | \sqrt{g} | d s^{\prime} d \theta d \varphi .
\end{gather}

When a finite plasma pressure is introduced,
the expression is modified into~\cite{Greene1997}:

\begin{gather}%
  \label{eq:well}
  W (s) = \frac{V_p}{\langle B^2 \rangle} \frac{d}{d V_p} (2 \mu_0 p + \langle B^2 \rangle) .
\end{gather}

Moreover,
the magnetic well term as defined in the Mercier stability criterion is~\cite{Bauer1984,Carreras1988}:

\begin{gather}%
  \label{eq:dwell-median}
  D_W (s) =  \frac{dp}{d s} \left( \frac{d^2 V_p}{d s^2} - \frac{dp}{d s} \int_0^{2 \pi} \int_0^{2 \pi} \frac{|\sqrt{g}| d\theta d\varphi}{B^2} \right) \int_0^{2 \pi} \int_0^{2 \pi} |\sqrt{g}| d\theta d\varphi \frac{B^2}{|\nabla s|^2} ,
\end{gather}

where the $\frac{s^2}{\ibar^2 \pi^2}$ prefactor common to all Mercier terms has been omitted,
as employed in \gls{VMEC}.

The model only qualitatively reconstructs the local magnetic well (~\cref{fig:well-vacuum-median,fig:well-median,fig:dwell-median});
it correctly provides the trend of the profile,
however,
it introduces artificial wiggles:
because the magnetic well depends on the solution's second radial derivatives,
the magnetic well is particularly sensitive to the solution's radial dependency.
The relative error for all three expression of the magnetic well is high (see~\cref{tab:results}).

\begin{figure}[h!]
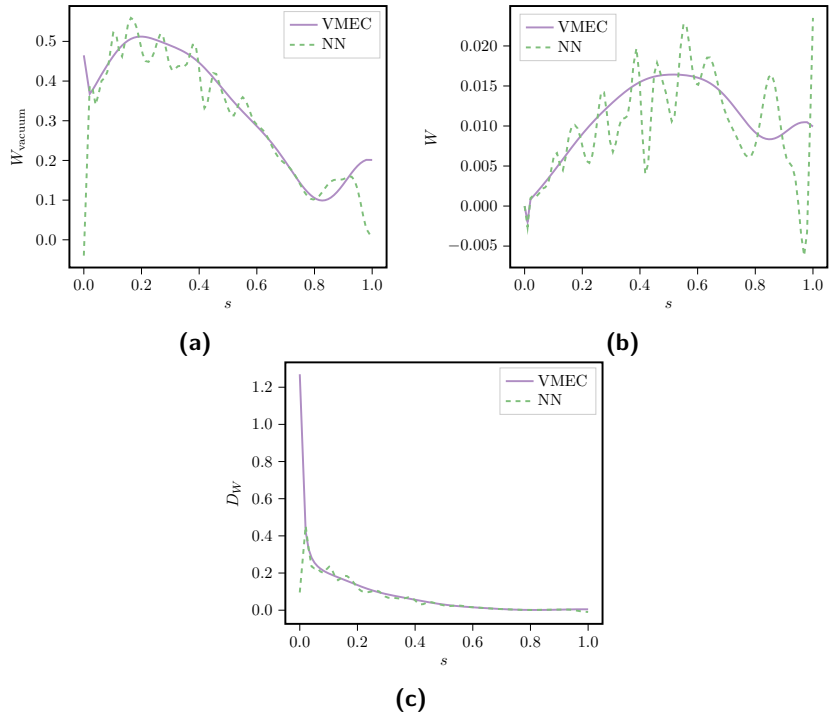

  \centering
  \multigraph[labels={fig:well-vacuum-median}{fig:well-median}{fig:dwell-median}]{content_figures_well.pgf;content_figures_well_.pgf;content_figures_dwell.pgf}%
  {%
    {}{}{}%
  }%
  \caption{
    True (solid pink) and predicted (dashed green) values of the vacuum magnetic well~\subref{fig:well-vacuum-median},
    the finite pressure magnetic well~\subref{fig:well-median},
    and the magnetic well term in the Mercier criterion~\subref{dwell-median},
    in case of the median regressed equilibrium.
    Numerical noise in the finite difference scheme invalidates the values at $s \in [0, 0.01]$ (see~\cref{sec:results}).
  }
\end{figure}

\begin{figure}[h!]
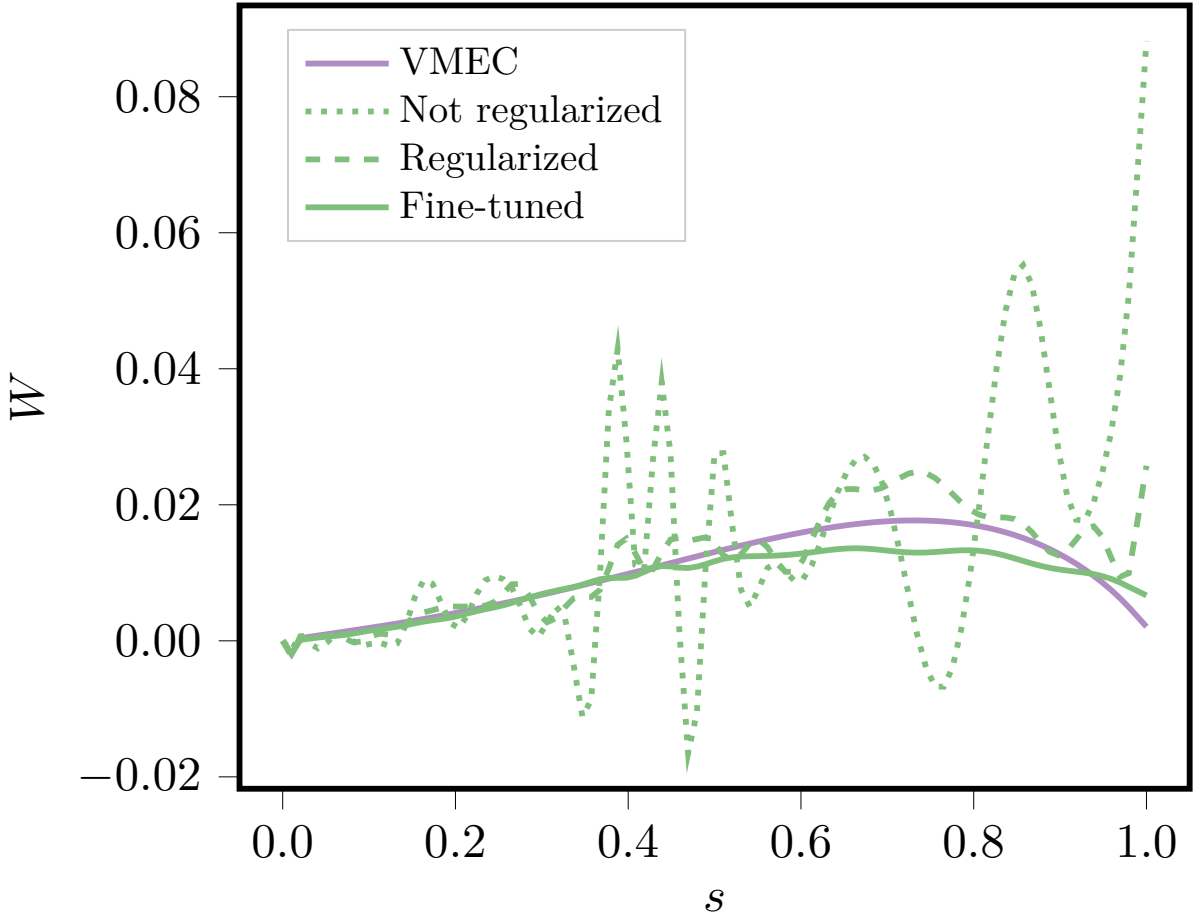

  \igraph[]{content_figures_finite_well_comparison.pgf}%
  \caption{
    Comparison on how the not-regularized,
    the regularized (\ie, single-pass),
    and the fine-tuned model reconstructs the local magnetic well $W$.
    Numerical noise in the finite difference scheme invalidates the values at $s \in [0, 0.01]$ (see~\cref{sec:results})
  }%
  \label{fig:well-comparison}
\end{figure}

The \gls{MHD} force residual regularization aids the model to faithfully reconstruct equilibrium properties.
As an example,
\cref{fig:well-comparison} shows the reconstructed local magnetic well by the not-regularized,
the regularized (\ie, single-pass),
and the fine-tuned models.
We again consider the \gls{W7X} standard configuration equilibrium at $\averagePlasmaBeta=\SI{2}{\percent}$ previously examined (see~\cref{sec:fine-tuning}).
The amplitude of the radial wiggle decreases from the not-regularized to the regularized models,
and it almost vanishes in the fine-tuned model.
Since the \gls{MHD} force residual depends on the equilibrium solution's second radial derivatives,
the \gls{MHD} force residual regularization also implicitly regularizes the equilibrium solution's radial derivatives.

To assess the \gls{MHD} stability of an equilibrium,
a globally defined magnetic well is usually considered.
For example,
stellarator optimization frameworks often use the magnetic well depth as a fast proxy for plasma stability~\cite{Andreeva2002,Landreman2022a}:

\begin{gather}%
  \label{eq:vacuum-well-depth}
  \average{W}_{\text{vacuum}} = \frac{V_p^{\prime}(s) |_{s=0} - V_p^{\prime}(s) |_{s=1} }{V_p^{\prime}(s) |_{s=0}} ,
\end{gather}

where $\average{W}_{\text{vacuum}}$ does not consider finite-\averagePlasmaBeta effects.
A magnetic well depth that considers also a finite pressure can be defined as:

\begin{gather}%
  \label{eq:well-depth}
  \average{W} = - \frac{p^*(s) |_{s=0} - p^*(s) |_{s=1} }{p^*(s) |_{s=0}} ,
\end{gather}

where $p^* = 2 \mu_0 p + \langle B^2 \rangle$ is the sum of the fluid and the magnetic pressure.

Despite only a qualitative agreement of the local magnetic well,
the model faithfully reconstructs the magnetic well depth (~\cref{fig:vacuum-magnetic-well-depth-scatter,fig:magnetic-well-depth-scatter}).
In case of the vacuum magnetic well depth,
the relative error is remarkably low:
$\text{mape}_{\average{W}_{\text{vacuum}}} = \SI{3.43}{\percent}$.
The accuracy on the magnetic well depth is sufficient to effectively navigate the magnetic configuration space of \gls{W7X} in finding precise,
negative-well configurations (see~\cref{sec:optim-negative-well}).

\begin{figure}[h!]
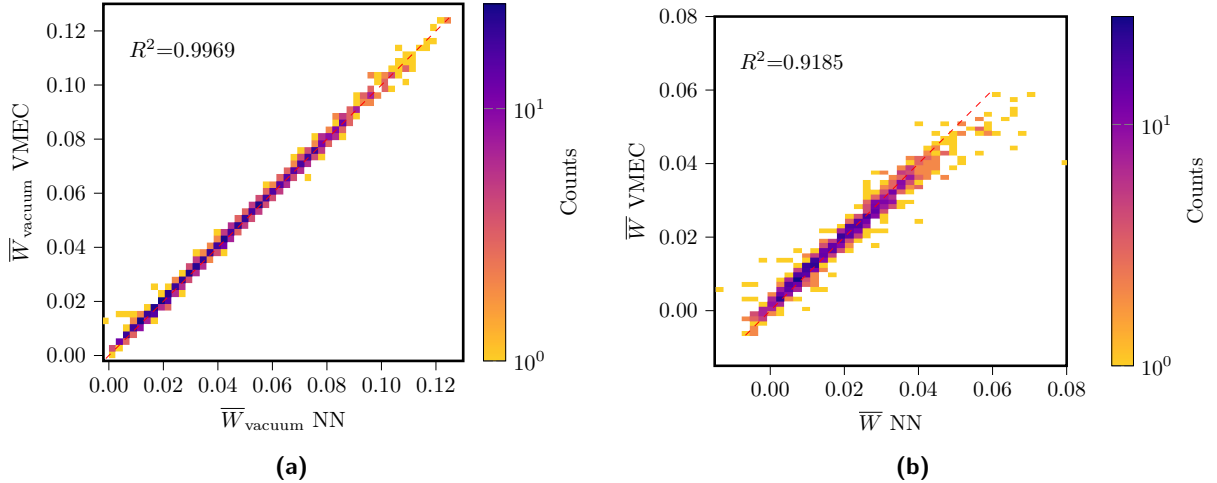

  \centering
  \multigraph[labels={fig:vacuum-magnetic-well-depth-scatter}{fig:magnetic-well-depth-scatter}]{content_figures_scatter_vacuum_well.pgf;content_figures_scatter_finite_well.pgf}%
  {%
    {}{}%
  }%
  \caption{
    Reconstruction of the vacuum~\subref{fig:vacuum-magnetic-well-depth-scatter} and finite pressure~\subref{fig:magnetic-well-depth-scatter} magnetic well depth across all equilibria in the test set.
    The color-bar indicates the counts in each histogram bin.
    To guide the eye,
    a red dashed line indicates the case of perfect reconstruction.
    The magnetic well depth is a measure of an average magnetic well value~\cite{Andreeva2002}.
  }
\end{figure}

\subsection{Neoclassical transport}\label{sec:neo-transport}

In a stellarator,
one critical transport regime is the so called $1/\nu$ regime,
where the neoclassical transport increases with decreasing collision frequency.
A measure of such transport is the effective ripple coefficient,
\epseff~\cite{Nemov1999}.
In this work,
the drift-kinetic code NEO~\mbox{\cite{Belli2008,Belli2011}} evaluates the \epseff.

No acceptable agreement is present between the predicted and ground truth \epseff values (\cref{fig:eps-eff-median}).
The effective ripple is greatly affected by the local structure of the magnetic field,
which the model struggles to smoothly reconstruct (\cref{fig:b-median}).
It seems coherent that the artificial field ripples in the model equilibrium solution affect the accuracy on the effective ripple.

A poor,
qualitative agreement between the ground truth and predicted \epseff is present only in the core region (\cref{fig:eps-eff-scatter}).
In such region,
namely for $s \leq \num{0.33}$ ($\rho \leq \num{0.57}$),
the magnetic field strength is reconstructed with an accuracy below \SI{10}{\milli\tesla}.
However,
the \epseff relative error is still high,
$\text{mape}_{\epseff}=\SI{54.6}{\percent}$.

\begin{figure}[h!]
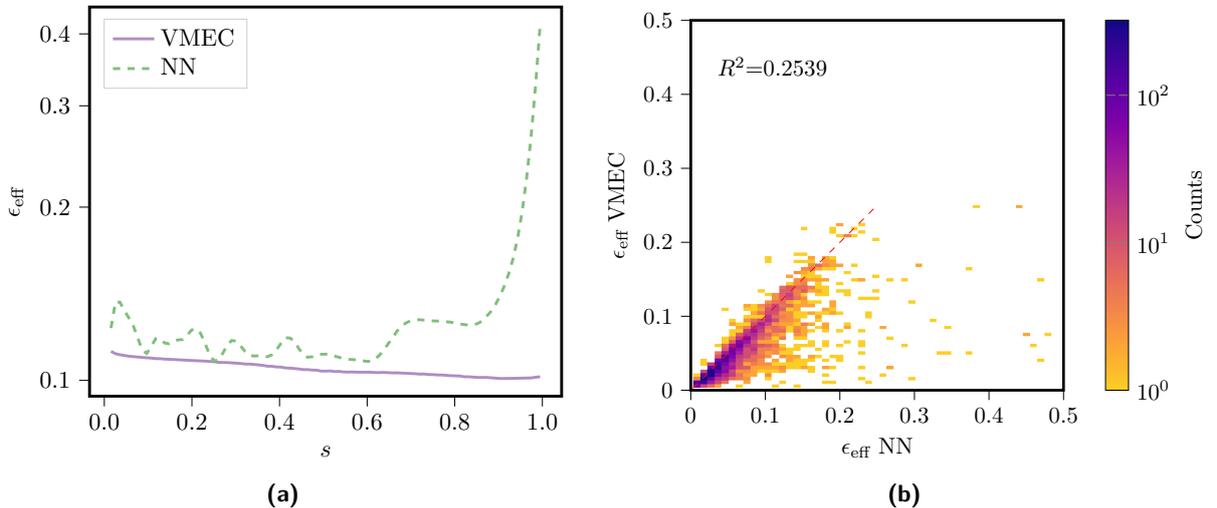

  \centering
  \multigraph[labels={fig:eps-eff-median}{fig:eps-eff-scatter}]{content_figures_eps_eff_.pgf;content_figures_scatter_eps_eff_.pgf}%
  {%
    {}{}%
  }%
  \caption{
    \subref{fig:eps-eff-median}
    True (solid pink) and predicted (dashed green) values of the effective ripple \epseff for the median regressed equilibrium.
    \subref{fig:eps-eff-scatter}
    Reconstruction of the effective ripple \epseff across all equilibria in the test set.
    The \epseff is computed on \num{5} equally spaced flux surfaces in the plasma region for $s \leq \num{0.33}$ ($\rho \leq \num{0.57}$).
    The color-bar indicates the counts in each histogram bin.
    To guide the eye,
    a red dashed line indicates the case of perfect reconstruction.
  }
\end{figure}

\begin{figure}[h!]
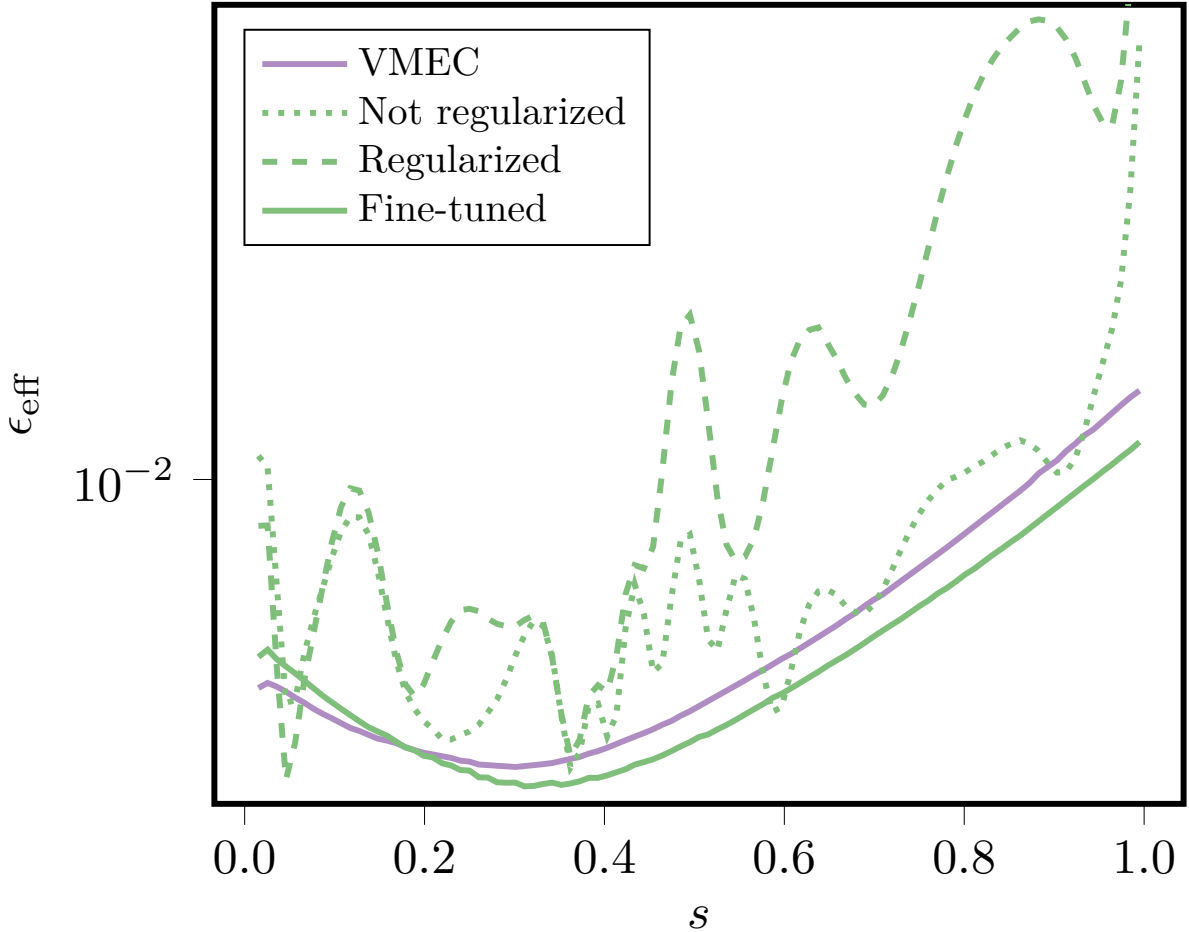

  \igraph[]{content_figures_eps_eff_comparison.pgf}%
  \caption{
    Comparison on how the not-regularized,
    the regularized (\ie, single-pass),
    and the fine-tuned model reconstructs \epseff.
    Numerical noise in the finite difference scheme invalidates the values at $s \in [0, 0.01]$ (see~\cref{sec:results})
  }%
  \label{fig:eps-eff-comparison}
\end{figure}

\added[id=AM]{
  The fine-tuned model can faithfully resolve \epseff.
  As an example,
  \cref{fig:eps-eff-comparison} shows the reconstructed \epseff by the not-regularized,
  the regularized,
  and the fine-tuned models in case of the \gls{W7X} standard configuration equilibrium at $\averagePlasmaBeta=\SI{2}{\percent}$ previously examined (see~\cref{sec:fine-tuning}).
  Both the not-regularized and regularized model fail to smoothly reconstruct \epseff.
}

Simplified proxies of the neoclassical transport do exist.
For example,
in case of \gls{W7X} magnetic configurations with a low mirror term ($b_{01} \simeq 0$),
the effective ripple can be approximated as~\cite{Geiger2014}:

\begin{gather}%
  \label{eq:eps-eff-proxy}
  \epseff \simeq \epseffproxy = - b_{11} \kappa^{4/3} ,
\end{gather}

where $b_{mn} = B_{mn} / B_0$,
$B_{mn}$ are the Fourier coefficients of the magnetic field strength $B$ in Boozer coordinates,
$B_0$ is a reference magnetic field value
(\eg, $B_{00}(s=1)$),
$\kappa = -b_{10} \frac{R}{r_{\text{eff}}}$ is the toroidal curvature term,
$R$ is the major radius,
and $r_{\text{eff}} = \rho a$ is the effective radius.

\begin{figure}[h!]
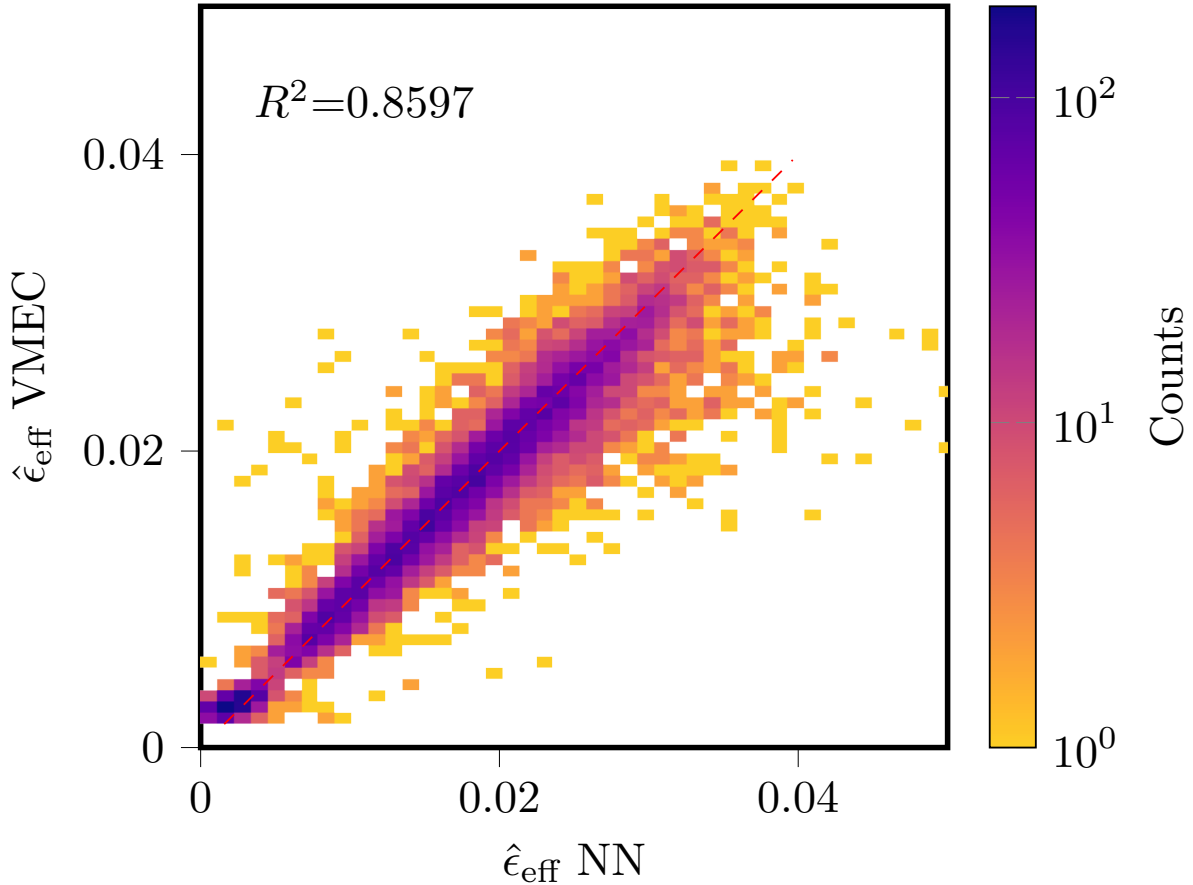

  \igraph[]{content_figures_scatter_eps_eff_proxy.pgf}%
  \caption{
    Reconstruction of the effective ripple proxy \epseffproxy across all equilibria in the test set.
    The color-bar indicates the counts in each histogram bin.
    To guide the eye,
    a red dashed line indicates the case of perfect reconstruction.
  }%
  \label{fig:eps-eff-proxy-scatter}
\end{figure}

The model well reconstructs the epsilon effective proxy \epseffproxy (\cref{fig:eps-eff-proxy-scatter}).
\epseffproxy relies on few geometrical quantities and on the leading Fourier components of the magnetic field strength $B$, which the model faithfully reconstructs (see~\mbox{\cref{sec:geo,sec:mag}}).
In case of the epsilon effective proxy \epseffproxy,
the model shows a relative error of $\text{mape}_{\epseffproxy} = \SI{13.3}{\percent}$,
lower than compared with  \epseff.

\subsection{Fast particle confinement}\label{sec:fast-ions}

Another critical measure of transport is the confinement of fast $\alpha$-particles (energetic ions),
which are the products of fusion reactions.
Any fusion facility should confine fast particles so that they can heat the bulk plasma during the plasma burn,
and to avoid damage to the first wall as a result of their losses.

Codes like the \gls{SIMPLE} compute the loss fraction of test particles,
for both trapped and passing particles,
when injected with an initial velocity in the plasma volume.
The collisionless guiding centers of particle motion are followed in time,
till a stopping condition is met or the particle leaves the plasma volume.

However,
such computations are expensive.
Stellarator optimization frameworks instead use explicit symmetries (\eg, quasi-symmetry),
equilibrium properties (\eg, omnigenity)
or simpler proxies that correlate with the confinement of fast particles (\eg, $\Gamma_c$)~\cite{Nemov2008,Drevlak2014,Drevlak2018,Bader2019,Bader2020a,Bader2021}.

For example,
\gls{W7X} is a \gls{QI} stellarator:
a \gls{QO} stellarator with magnetic field strength contours that close poloidally.
Omnigeneity means that the bounce-averaged radial drift of locally trapped particles vanishes~\cite{Hall1975}.
For a confining field $B$,
the conditions of omnigineity are~\cite{Cary1997}:

\begin{enumerate}
  \item{
        the contours of \Bmax are straight lines in Boozer coordinates;
        }
  \item{
        the magnetic field strength maxima and minima are the same across all field lines;
        }
  \item{
        equal Boozer angular separation $\delta$ along a field line between contours of constant magnetic field strength $B$ (\ie, bounce points) across all field lines;
        }
\end{enumerate}

Does the model reconstruct these equilibrium properties?

To assess property (\num{1}),
the standard deviation of the toroidal location of the contours of \Bmax,
$\sigma_{\varphi_{\Bmax}}$,
is considered.
In a \gls{QO} configuration,
the \Bmax contours are straight lines in Boozer coordinates.
Namely,
$\sigma_{\varphi_{\Bmax}} = 0$.

\begin{figure}[h!]
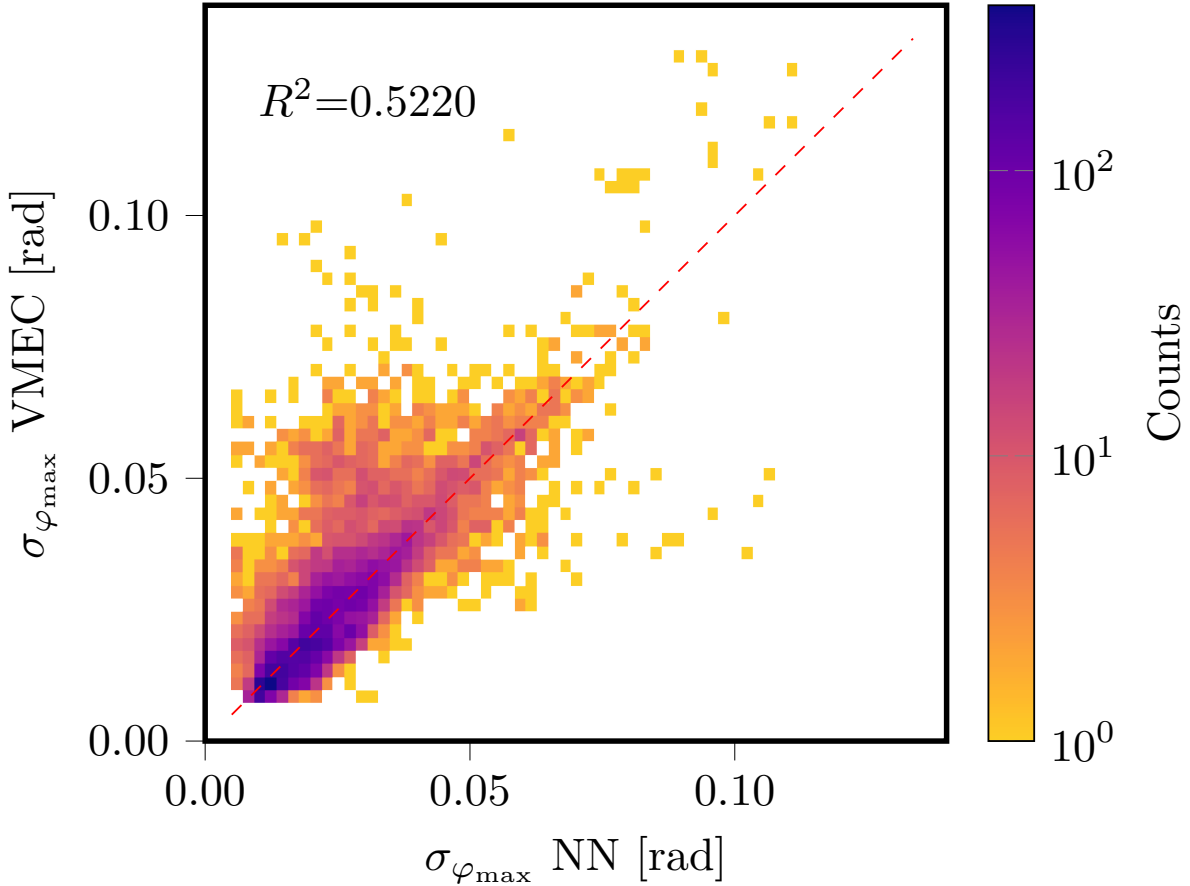

  \igraph[]{content_figures_scatter_vpmax.pgf}%
  \caption{
    Reconstruction of the standard deviation of the toroidal location of the contours of \Bmax,
    $\sigma_{\varphi_{\Bmax}}$.
    The color-bar indicates the counts in each histogram bin.
    To guide the eye,
    a red dashed line indicates the case of perfect reconstruction.
  }%
  \label{fig:vpmax-scatter}
\end{figure}

The model only qualitatively reconstructs the toroidal variance of the contours of \Bmax (\cref{fig:vpmax-scatter}).
The contours of \Bmax is a local equilibrium property that is affected by how well the model reconstructs the local magnetic field strength.

To assess property (\num{2}),
the standard deviation of the extrema of $B$ on a flux surface across all field lines are assessed,
$\sigma_{\Bmax}$ and $\sigma_{\Bmin}$.
In a \gls{QO} configuration,
$\sigma_{\Bmax} (s) = \sigma_{\Bmin} (s) = 0$ across the whole plasma volume.

The model equilibria correctly reproduce the standard deviation of the extrema of $B$ (\cref{fig:vbmax-scatter,fig:vbmin-scatter}):
$\text{rmse}_{\sigma_{\Bmax}}=\SI{9.43}{\milli\tesla}$,
and $\text{rmse}_{\sigma_{\Bmin}}=\SI{6.57}{\milli\tesla}$.

\begin{figure}[h!]
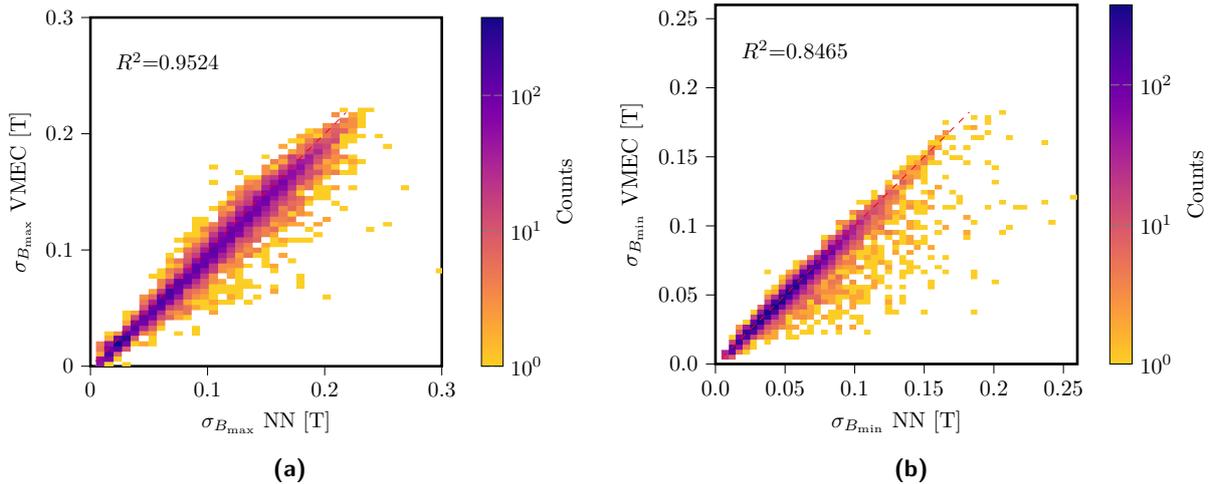

  \centering
  \multigraph[labels={fig:vbmax-scatter}{fig:vbmin-scatter}]{content_figures_scatter_vbmax.pgf;content_figures_scatter_vbmin.pgf}%
  {%
    {}{}%
  }%
  \caption{
    Reconstruction of the standard deviation of the \subref{fig:vbmax-scatter} maxima $\sigma_{\Bmax}$ and \subref{fig:vbmin-scatter} minima $\sigma_{\Bmin}$ of the magnetic field strength across all field lines on a flux surface.
    The standard deviation evaluated on ten equally spaced flux surfaces is considered in these figures.
    The color-bar indicates the counts in each histogram bin.
    To guide the eye,
    a red dashed line indicates the case of perfect reconstruction.
  }
\end{figure}

The Boozer angular separation $\delta(s, B, \alpha)$ is the angular distance along the field line $\alpha$ between the $B$ contours (see~\cref{fig:b-boozer-field-line}).
In a \gls{QO} configuration,
the angular separation $\delta(s, B, \alpha)$ should be independent of the field line label $\alpha$,
namely,
\mbox{$\frac{\partial \delta(s, B,\alpha)}{\partial \alpha} \vert_{s,B} = 0$}.
Numerically,
such condition can be verified by observing $\sigma_{\delta} (s, B) = 0$.

The Boozer angular separation $\delta$ relates to the bounce-averaged radial drift that trapped particles experience.
Therefore,
it makes sense to assess it only if a clear $B$ field well,
in which particles can be trapped,
exists.
Not all \gls{W7X} magnetic configurations exhibit such a $B$ field well:
the median regressed equilibrium in the test set is not guaranteed to possess it.
Therefore,
only to assess property (\num{3}),
the model is evaluated on a small (\num{58} equilibria) out-of-sample set of high-mirror \gls{W7X} configurations.
The high-mirror \gls{W7X} configuration has been constructed to feature a magnetic-mirror like $B$ field structure,
and it is the closest configuration to \gls{QI} out of the nine \gls{W7X} reference configurations.
This test set has been constructed as the main data set (see~\cref{sec:data}),
however,
the coil current ratios have been fixed to generate only high-mirror \gls{W7X} configurations.
Still,
finite-$\beta$ and toroidal current effects are present in the data set.
For completeness,
\cref{sec:fast-ions-test-set} presents the $\delta$ reconstruction in case of the median regressed equilibrium in the larger test set.

\begin{figure}[h!]
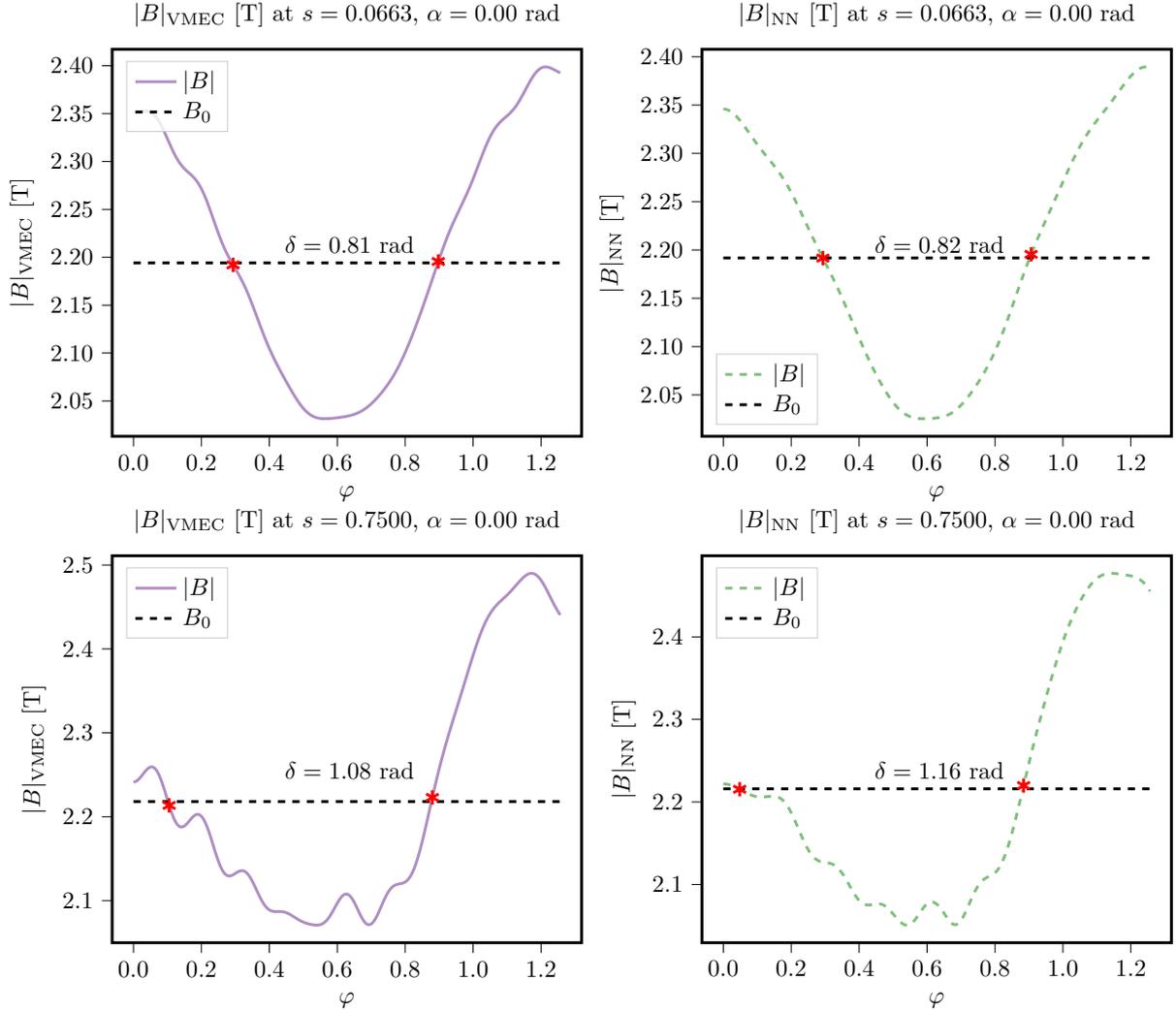

  \igraph[]{content_figures_b_fieldline_boozer_bounce_points_high_mirror.pgf}%
  \caption{
    True (left) and predicted (right) values for the magnetic field strength $B$ in Boozer coordinates along a field line in case of the median predicted equilibrium in the high-mirror out-of-sample test set.
    The $\alpha=\num{0}$ field line at the $s=\num{0.06}$ (top) and at the $s=\num{0.75}$ flux surface (bottom) are depicted.
    The dashed black horizontal line indicates the average magnetic field strength on the flux surface $B_0$.
    The red crosses highlight the intersections of the field line with the $B_0$ contours,
    and the Boozer angular separation $\delta(s, B_0, \alpha)$ between them is shown just beneath.
  }%
  \label{fig:b-boozer-field-line-high-mirror}
\end{figure}

\begin{figure}[h!]
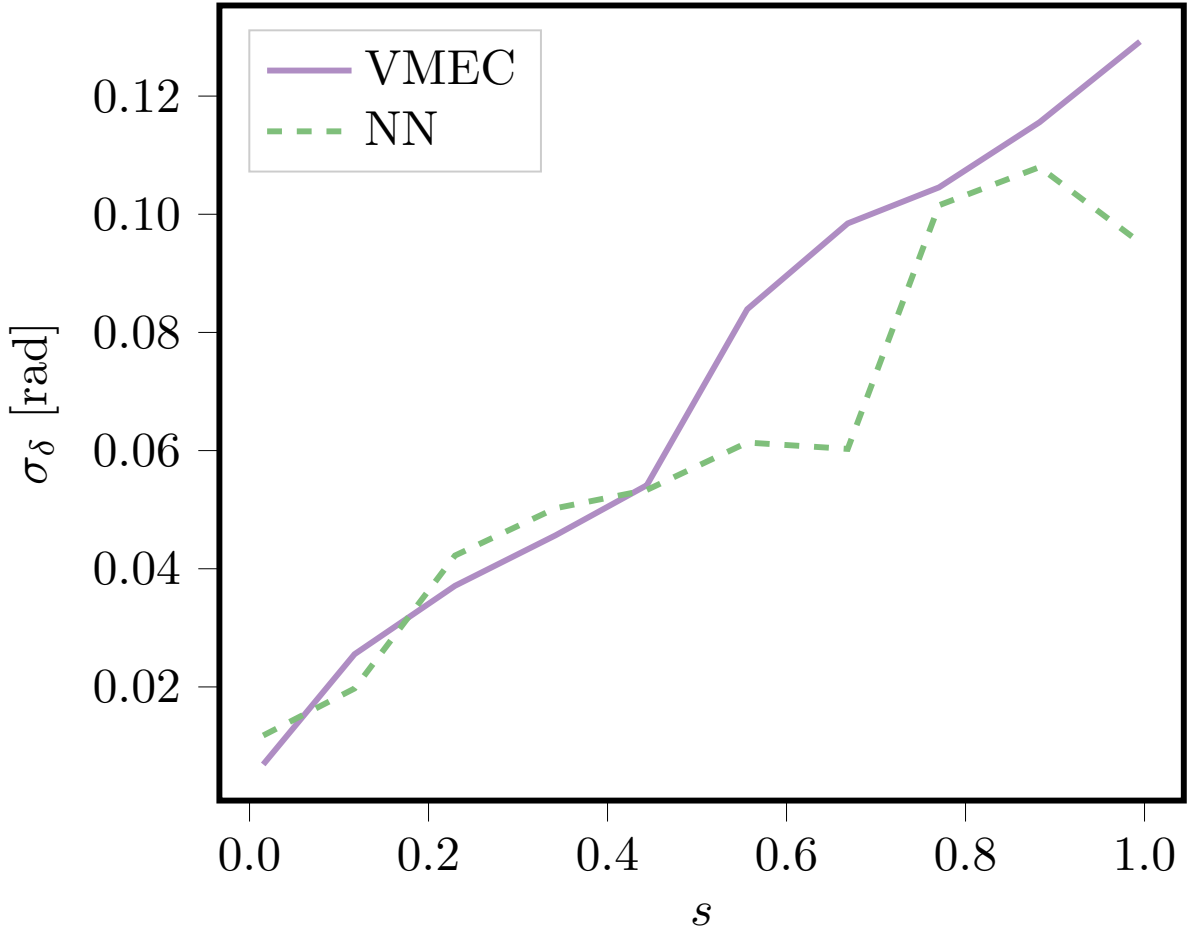

  \igraph[]{content_figures_delta_high_mirror.pgf}%
  \caption{
    True (solid pink) and predicted (dashed green) values of the standard deviation of Boozer separation angle $\sigma_{\delta}(s, B_0) = \sqrt{\langle \delta(s, B_0,\alpha)^2 \rangle_{\alpha}}$ for the median regressed equilibrium in the high-mirror out-of-sample test set.
    Without loss of generality,
    the contour of the average magnetic field strength $B_0$ on each flux surface is considered.
  }%
  \label{fig:delta-high-mirror-median}
\end{figure}

Even if the predicted magnetic field structure does not perfectly overlap with the ground truth field (\cref{fig:b-median}),
the model preserves the Boozer angular separation (\cref{fig:b-boozer-field-line-high-mirror}).
However,
the reconstruction accuracy is better in the core region (\cref{fig:delta-high-mirror-median}).

\subsection{Optimization in the configuration space of \gls{W7X}}\label{sec:optimization}

The exploration and optimization of the \gls{W7X} magnetic configuration space serve to further investigate how faithfully the \gls{NN} model reproduces equilibrium properties,
and to validate its use in downstream tasks.
For example,
\cite{Mikhailov2011,Drevlak2014,Wechsung2022a} proposed the a posteriori optimization of coil currents to further improve the equilibrium properties,
and to recover inevitable errors that occur during the coils manufacturing and placement.
Moreover,
especially in case of an island divertor that relies on a robust edge magnetic topology,
it is desirable to include finite-$\beta$ effects in the optimization.

A fast,
free-boundary ideal-\gls{MHD} equilibrium \replaced[id=AM]{model}{solver} benefits such optimization.
In this work,
\gls{W7X} configurations are targeted due to the limitations of the training data.
However,
the proposed technique can be applied to any present or future device once a training data set is generated.
The optimizations carried out in this work represent toy examples of real-world applications.
Indeed,
the proposal of new,
optimized \gls{W7X} configurations is beyond the scope of this paper.

\replaced[id=AM]{The equilibrium solution provided by the \gls{NN} is differentiable.}{The \gls{NN} model is a fully differentiable \gls{MHD} equilibrium solver.}
The analytic gradient of equilibrium properties
(\eg, the magnetic well)
with respect to the magnetic configuration and plasma profiles is available at no additional cost via \gls{AD}.
In a gradient-based optimization,
a single equilibrium evaluation is sufficient in each optimization step
(regardless of the number of free parameters).
In contrast,
when finite-differences approximate the objective function gradient,
the number of equilibrium evaluations each step scales linearly with the number of free parameters.

Both magnetic configuration and finite-$\beta$ effects are included in the optimization.
The search space is a subset of the \gls{NN} model input parameters:
the \gls{W7X} non-planar $i_{[2,...,5]}$ and planar $i_{[A,B]}$ coil current ratios,
the toroidal magnetic flux at the edge \phiedge,
and the pressure on axis $p_0$.
The pressure profile shape is fixed,
and it is assumed to be of the form $p(s)=p_0(1 - s)^2$.
Moreover,
we assume the toroidal plasma current to vanish
(\ie, $\Itor (s) = 0$).

To avoid extrapolation,
the search space is bounded by the distribution of the input parameters as seen during training.
Denoting $\vec{p} = [p_1, \hdots, p_{d}] \in \real^d$ the optimization vector,
each input parameter $p_i$ is bounded by $[ \mu_i - c \sigma_i, \mu_i + c \sigma_i ]$,
where $\mu_i$ and $\sigma_i$ are the mean and the standard variation of the distribution of $p_i$,
respectively.
$c=\sqrt{3}$ for the parameters that have been uniformly sampled in the training data,
and $c=\num{2}$ for all the others.
The configuration with $p_i = \mu_i \ \forall i$ serves as initial guess
(\ie, the \textit{average} configuration seen during training).

The objective function is the weighted sum of the squared residual between the proposed and target equilibrium properties:

\begin{gather}\label{eq:optim-obj-function}
  f(\vec{p}) = \sum_i w_i (f_i(\vec{p}) - f_i^{*})^2 ,
\end{gather}

where $f_i^{*}$ is a target equilibrium property,
$f_i(\vec{p})$ is the proposed equilibrium property,
and $w_i$ is the weight of each residual term.
Such form of the objective function is usually employed in stellarator optimization frameworks~\cite{Hirshman1998,Spong1998,Drevlak2018}.

A combination of the global \gls{TPE} algorithm from \textit{hyperopt}~\cite{Bergstra2013} and the local \gls{LBFGS}~\cite{Nocedal1980} algorithm from \textit{scipy}~\cite{Virtanen2020} drives the optimization.
Whenever possible,
\gls{AD} computes the analytical Jacobian of the objective function.
On the other hand,
if the analytical Jacobian is not available,
(\eg, an objective function term requires an external software package that breaks the model computational graph),
finite differences approximate the Jacobian.

\subsubsection{Negative well}\label{sec:optim-negative-well}

\begin{figure}[h!]
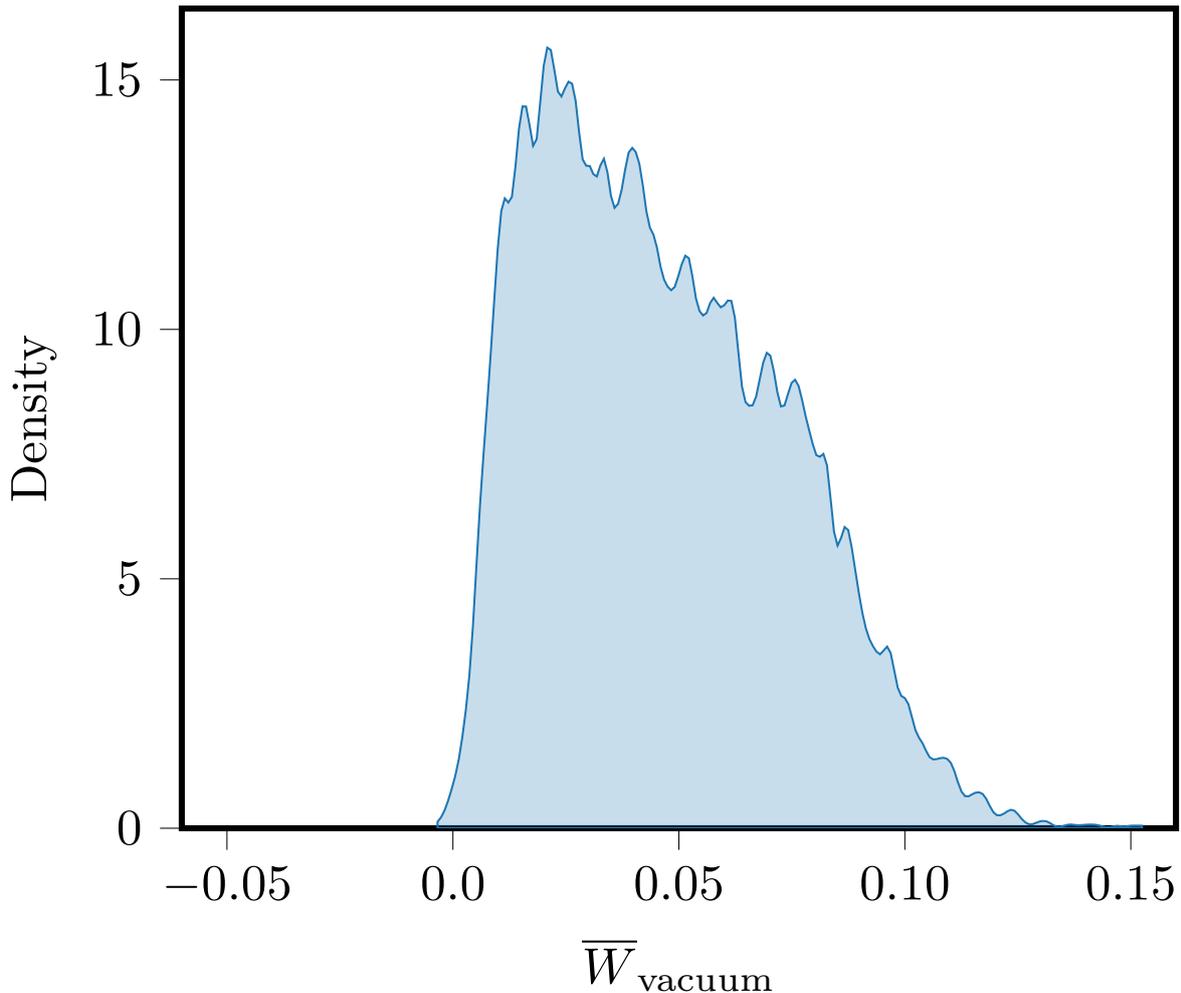

  \igraph{content_figures_hist_vacuum_well.pgf}%
  \caption{
  The distribution of the vacuum magnetic well depth $\average{W}_{\text{vacuum}}$ for all equilibria in the training set.
  }%
  \label{fig:well-dist}
\end{figure}

Almost all \gls{W7X} configurations in the training set feature a positive magnetic well (\cref{fig:well-dist}).
\gls{W7X} was optimized with respect to good \gls{MHD} stability~\cite{Grieger1993},
therefore,
in general,
a positive magnetic well is to be expected.

Can the model generalize to negative well configurations?
To search for a negative well configuration,
a small,
but negative magnetic well depth is targeted:
$\average{W}_{\text{vacuum}}^* = \num{-0.01}$ and $w_{\average{W}_{\text{vacuum}}} = \num{e4}$.
Finite-$\beta$ effects deepen the magnetic well,
therefore,
the search is performed in the limit of vacuum configurations.
After \num{500} iterations of \gls{TPE},
\gls{LBFGS} refines the best configuration.

\begin{figure}[h!]
  \igraph[]{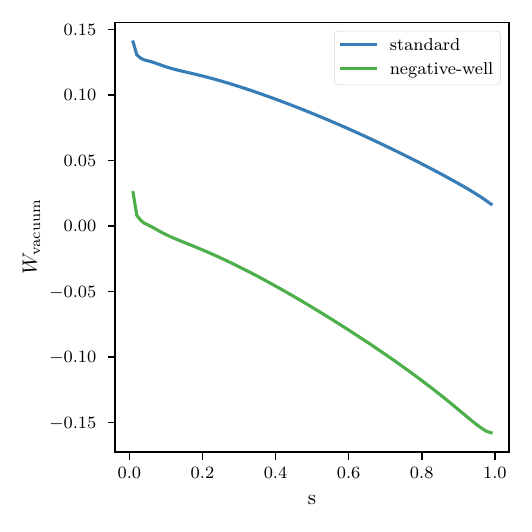}%
  \caption{
    The magnetic well of the obtained \gls{W7X} negative well configuration (green) as evaluated by \gls{VMEC}.
    For comparison,
    the magnetic well of the \gls{W7X} standard reference configuration (blue) is also shown (the \gls{W7X} standard configuration has a positive magnetic well).
    Numerical noise in the finite difference scheme invalidates the values at $s \in [0, 0.01]$ (see~\cref{sec:results}).
  }%
  \label{fig:negative-well}
\end{figure}

Although the \gls{NN} model has rarely seen a negative well configuration during training,
it can reliably predict equilibria with a negative well (\cref{tab:optim-negative-well}).
The optimization converged to a configuration for which the model predicts a magnetic well depth of \num{-0.01}.
The configuration has been evaluated also with \gls{VMEC},
which reports a magnetic well depth of \num{-0.009}.
Indeed,
the vacuum magnetic well ($W(\psi) = \frac{d^2 V_p(\psi)}{d \psi^2}$) is negative (\cref{fig:negative-well}).

\begin{table}[h!]
  \caption{
  Relative magnitude of the non-planar and planar \gls{W7X} coil currents for the negative well configuration.
  $\average{W}_{\text{vacuum}}^*=\num{-0.01}$ is the small,
  negative magnetic well depth targeted in the optimization.
  In case of the obtained configuration,
  the \gls{NN} model predicts $\average{W}_{\text{vacuum}}=\num{-0.010}$,
  and \gls{VMEC} computes $\average{W}_{\text{vacuum}}=\num{-0.009}$.
  }%
  \label{tab:optim-negative-well}
  \centering
  \begin{tabular}{lc}
    \toprule
    Quantity                           & Value        \\
    \midrule
    $i_1$                              & \num{1.000}  \\
    $i_2$                              & \num{1.027}  \\
    $i_3$                              & \num{0.689}  \\
    $i_4$                              & \num{0.614}  \\
    $i_5$                              & \num{0.614}  \\
    $i_A$                              & \num{0.205}  \\ 
    $i_B$                              & \num{0.263}  \\ 
    $\average{W}_{\text{vacuum}}^*$    & \num{-0.010} \\
    $\average{W}_{\text{vacuum}}$ NN   & \num{-0.010} \\
    $\average{W}_{\text{vacuum}}$ VMEC & \num{-0.009} \\
    \bottomrule
  \end{tabular}
\end{table}

\subsubsection{Improved fast particle confinement}\label{sec:optim-fast-particle}

A classical stellarator does not confine fast particles as good as a tokamak,
therefore,
the magnetic geometry has to be optimized to do so.
\gls{W7X} standard and high-mirror configurations have been optimized to improve fast particle confinement at high-\plasmaBeta.
Moreover,
\cite{Mikhailov2011,Drevlak2014} have also found improved configurations.

Can we use the model to search for a \gls{W7X} fast particle optimized configuration?
To address this question,
we employ a similar objective function to the function used for the \textit{min-max} configuration in~\cite{Drevlak2014},
and we investigate whether the optimized configuration shows the same performance in terms of fast particle confinement.

In a perfect \gls{QI} configuration,
the variance of the extrema of the magnetic field strength across all field lines vanishes.
As in the case of the \textit{min-max} configuration,
the variance of $B$ at the bean-shaped cross-section (\ie, $\varphi=0$ or the $B_{\text{max}}$-contour)
and at the triangular cross-section (\ie, $\varphi=\pi/\gls{Nfp}$ or the $B_{\text{min}}$-contour)
is minimized.
The flux surface at $s=\num{0.015}$ is targeted (roughly \SI{6}{\centi\meter} from the magnetic axis).
The optimization is performed at a fixed $\phiedge$ value.
No toroidal current is assumed in the plasma.

The objective function is:

\begin{gather}\label{eq:optim-fast-particle-obj-function}
  f = w_{B} ( \text{Var}_{\Bmax} + \text{Var}_{\Bmin} ) + w_{\beta} (\averagePlasmaBeta - \averagePlasmaBeta^*)^2 + w_{mr} (mr - mr^*)^2 ,
\end{gather}

where $\text{Var}_{\Bmax}$ and $\text{Var}_{\Bmin}$ are the variances of the magnetic field strength at the bean-shaped and triangular cross-section,
respectively,
and

\begin{gather}\label{eq:mirror-factor}
  mr=\langle \frac{B(\theta,\varphi=0) - B(\theta,\varphi=\pi/\gls{Nfp})}{B(\theta,\varphi=0) + B(\theta,\varphi=\pi/\gls{Nfp})} \rangle_{\theta} ,
\end{gather}

is a mirror factor to ensure that the extrema of the magnetic field strength are located at the correct angular position
(maxima at $\varphi=0$ and minima at $\varphi=\pi/\gls{Nfp}$).
$\averagePlasmaBeta^*=\SI{2}{\percent}$ and $mr^*=\SI{10}{\percent}$ (like in the \gls{W7X} \textit{high-mirror} configuration).
Finally,
$w_B=\num{e2}$,
$w_{\averagePlasmaBeta}=\num{e3}$,
and $w_{mr}=\num{e2}$.
The optimization is carried out for \num{500} iterations of the \gls{TPE} algorithm.

\gls{SIMPLE}~\cite{Albert2020,Albert2020a} assess the fraction of lost particles from the plasma volume.
\num{e4} particles at \SI{60}{\keV} are launched from $s=\num{0.06}$,
and their trajectories are followed for \SI{0.1}{\second},
or till they are lost from the computational domain.
The optimized configuration is compared against the \gls{W7X} \textit{standard}, \textit{high-mirror} and \textit{min-max} configurations.
All configurations have been scaled to have the same magnetic field strength on axis $B(\varphi=0)\simeq\SI{2.52}{\tesla}$,
the same plasma beta $\averagePlasmaBeta\simeq\SI{2}{\percent}$,
and the same plasma volume $V_p\simeq\SI{30}{\cubic\meter}$.

\begin{figure}[h!]
  \igraph[]{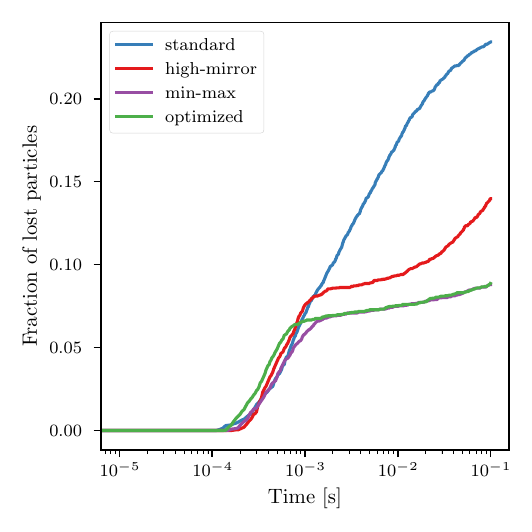}%
  \caption{
    The fraction of lost particles for the \gls{W7X} standard,
    high-mirror,
    \textit{min-max},
    and optimized configuration at \averagePlasmaBeta=\SI{2}{\percent}.
  }%
  \label{fig:optim-fast-ions-simple}
\end{figure}

The optimized configuration confines fast particles better than the \textit{standard} and \textit{high-mirror} configurations,
and as well as the \textit{min-max} configuration (\cref{fig:optim-fast-ions-simple}):
the total fraction of lost particles after \SI{0.1}{\second} is only \SI{8.86}{\percent} (\cref{tab:optim-fast-ions}).

\begin{table}[h!]
  \caption{
    Relative magnitude of the non-planar and planar \gls{W7X} coil currents for the $\averagePlasmaBeta=\SI{2}{\percent}$ fast particle confinement optimized configuration.
    The total fraction of lost \SI{60}{\keV} fast ions initialized at $s=\num{0.06}$ is reported.
  }%
  \label{tab:optim-fast-ions}
  \centering
  \begin{tabular}{lc}
    \toprule
    Quantity      & Value               \\
    \midrule
    $i_1$         & \num{1.000}         \\
    $i_2$         & \num{0.737}         \\
    $i_3$         & \num{0.889}         \\
    $i_4$         & \num{0.776}         \\
    $i_5$         & \num{0.899}         \\
    $i_A$         & \num{0.011}         \\
    $i_B$         & \num{-0.149}        \\
    Lost fraction & \SI{8.86}{\percent} \\ 
    \bottomrule
  \end{tabular}
\end{table}

\section{Discussion and conclusion}\label{sec:conclusion}

We propose a \gls{NN} model that quickly approximates the ideal-\gls{MHD} solution operator in a stellarator geometry.
The model correctly predicts the equilibrium solution,
and it faithfully reconstructs global equilibrium properties and proxy functions that are commonly used in stellarator optimization frameworks.
Compared with previous data-driven approaches,
the model fulfils equilibrium symmetries by construction.
In addition,
the \gls{MHD} force residual regularizes the model to better satisfy the ideal-\gls{MHD} equations.

The error between the model and the ground truth equilibrium solution is less than \SI{1}{\milli\meter} in case of the flux surface locations.
However,
mainly due to the high condition number of the equilibrium problem,
the model equilibria only poorly minimize the \gls{MHD} force residual.
Yet,
the model predictions can be improved at inference time to yield better than ground truth equilibria
(\ie, equilibria that minimize the \gls{MHD} force residual better than the ground truths).

Equilibrium properties that are highly sensitive to the magnetic field structure cannot be precisely reconstructed.
The magnetic field depends on first-order derivatives of the equilibrium solution,
therefore,
even small errors in the equilibrium geometry might lead to large discrepancies in the computed magnetic field.
Nevertheless,
the magnetic field strength is smoothly reconstructed in the core region,
but the model introduces artificial field ripples at the edge.
As a consequence,
the model fails to meet the required magnetic field accuracy to faithfully compute some quantities of interest
(\eg, the effective ripple, which is a measure of neoclassical transport).
Still,
the accuracy of the model is sufficient to explore and optimize \gls{W7X} magnetic configurations,
in terms of ideal-\gls{MHD} stability and fast particle confinement.
This result suggests that \gls{NN}-based surrogate models can be used to find optimized configurations for current and future stellarator devices.

The model has still many limitations.
The \gls{NN} is trained in a supervised learning fashion,
therefore,
the training data limit the applicability of the model
(the model can only approximate \gls{W7X} equilibria).
Moreover,
noise-free plasma profiles are assumed.
Finally,
the model accuracy on the magnetic field limits its use in tasks that are highly sensitive to it.

This work can be improved along multiple dimensions.
The robustness of the model with respect to experimental noise can be investigated and enhanced with established robustness and regularization techniques:
injecting noise to the input plasma profiles and training the convolutional layers with dropout.
The artificial field ripples introduced by the model can be ameliorated by adding a regularization term on the second-order radial derivative
(\eg, $\frac{\partial^2 R}{\partial s^2}$),
or,
by replacing the learned trunk network functions with an orthonormal set of basis functions
(\eg, Legendre polynomials).
In addition,
the model can be trained on an extended space of stellarator equilibria,
relaxing the constraint of \gls{W7X} configurations.
Finally,
equilibrium reconstructions in a stellarator geometry must be demonstrated.

Contrary to a tokamak,
the computation of \gls{MHD} equilibria \replaced[id=AM]{is a limiting factor in}{limits} stellarator research and design.
Fast,
differentiable and accurate three-dimensional \gls{MHD} \replaced[id=AM]{equilibrium solutions}{solvers} would reduce the gap between tokamaks and stellarators.
In particular,
they would enable:
real-time equilibrium reconstructions~\cite{Ferron1998,Yue2013},
first-principles flight simulators~\cite{Fable2021},
data-driven plasma control~\cite{Degrave2022},
and extensive understanding and exploration of the stellarator optimization space.
\section{Author Statement}\label{sec:author-statement}

The contributions to this paper are described using the CRediT taxonomy~\cite{Brand2015}:

\begin{description}
    \item[Andrea Merlo] Conceptualization, Data Curation, Formal Analysis, Investigation, Methodology, Software, Visualization, Writing - original draft, Writing - review \& editing.
    \item[Daniel Böckenhoff] Methodology, Software, Supervision, Validation, Writing - review \& editing.
    \item[Jonathan Schilling] Formal Analysis, Methodology.
    \item[Samuel Aaron Lazerson] Supervision, Writing – review \& editing.
    \item[Thomas Sunn Pedersen] Conceptualization, Funding acquisition, Supervision.
\end{description}

\section{Acknowledgement}\label{sec:acknowledgement}



We wish to acknowledge the helpful discussions on the limitations of this work with Ralf Schneider.
Furthermore,
we are indebted to the communities behind the multiple open-source software packages on which this work depends:
hydra~\cite{Yadan2019},
hyperopt~\cite{Bergstra2013},
matplotlib~\cite{Hunter2007},
numpy~\cite{Harris2020},
pytorch~\cite{Paszke2019},
pytorch lightning~\cite{Falcon2019},
scipy~\cite{Virtanen2020}.

The data sets were generated on the \gls{MPCDF} cluster ``COBRA'',
Germany.
Financial support by the European Social Fund (ID: ESF/14-BM-A55-0007/19) and the Ministry of Education, Science and Culture of Mecklenburg-Vorpommern,
Germany via the project ``NEISS'' is gratefully acknowledged.
This work has been carried out within the framework of the EUROfusion Consortium,
funded by the European Union via the Euratom Research and Training Programme (Grant Agreement No 101052200 — EUROfusion).
Views and opinions expressed are however those of the author(s) only and do not necessarily reflect those of the European Union or the European Commission.
Neither the European Union nor the European Commission can be held responsible for them.


\ifthenelseproperty{compilation}{appendix}{%
    \section{Appendix}\label{sec:appendix}

\subsection{Equilibria Fourier scaling}\label{sec:fourier-scaling}

A trade-off must be struck between model capacity and computational complexity.
Given the poloidal and toroidal expansions of the equilibrium solution in Fourier series,
the number of Fourier modes utilized to represent the solution limits its accuracy:
a large number of Fourier modes enhances the model accuracy.
On the other hand,
it increases the computational complexity of the model,
as well as the training and inference times.
How many Fourier modes are adequate?

To examine such trade-off,
the approximation error on magnetic field strength is investigated.
One of the objectives of this study is to investigate the precision with which the model reconstructs equilibrium properties.
The magnetic field strength is one of the critical equilibrium properties.
Therefore,
the Fourier resolution of the equilibrium solution should not be the largest source of error in representing the field strength. 

To represent the magnetic field strength with an error of less than \SI{1}{\percent},
up to $m=8$ poloidal and $|n|=11$ toroidal models are required.
\Cref{fig:b-fourier-scaling} shows the magnetic field approximation error when the Fourier resolution is varied,
using VMEC equilibria with up to $m=11$ and $|n|=12$ Fourier modes as \quotes{high-fidelity} references.

\begin{figure}[h!]
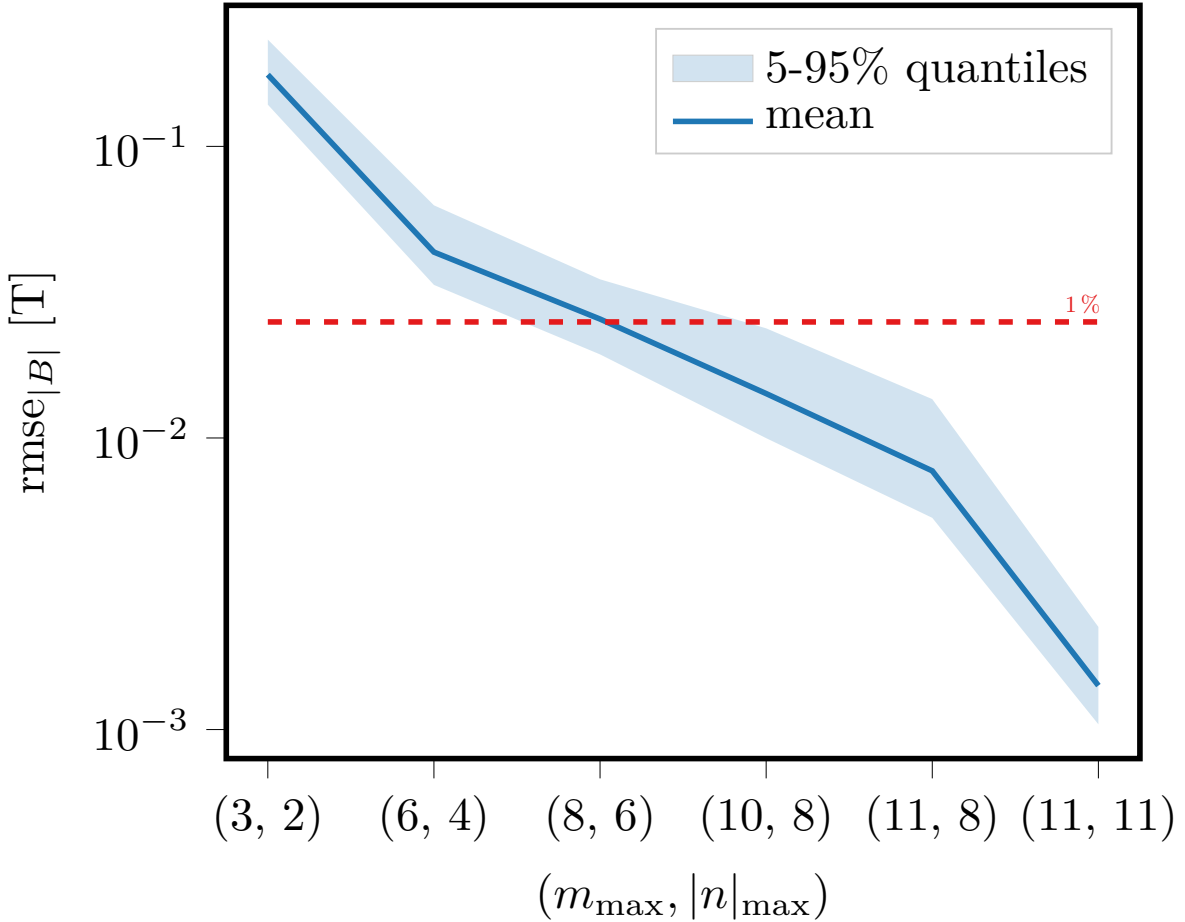

  \igraph[]{content_figures_b_fourier_scaling.pgf}%
  \caption{
    The approximation error on the magnetic field strength when representing the equilibrium solution $\{R, \lambda, Z\}$ with a reduced number of Fourier modes.
    The equilibrium with up to $m=11$ poloidal and $|n|=12$ toroidal mode is taken as a reference.
    The solid line represents the average error across \num{128} randomly sampled equilibria in the data set,
    and the shaded region represents the \SI{95}{\percent} and \SI{5}{\percent} quantiles.
    \gls{W7X} has a mean field $\simeq\SI{2.5}{\tesla}$,
    therefore,
    a field error of \SI{25}{\milli\tesla} represents a \SI{1}{\percent} error.
  }%
  \label{fig:b-fourier-scaling}
\end{figure}

\subsection{Model architecture and \glspl{HP}}\label{sec:appendix-hp}

The trunk network maps the radial dimension $s$ into a set of non-linear basis functions,
in which the $X_{mn}^j$ Fourier coefficients and \gls{ibar} profile are radially expanded (see~\cref{sec:architecture}).
Each Fourier coefficient and \gls{ibar} has its own set of basis functions.
The input dimension is \mbox{$N_{\text{ti}} = \num{1}$},
and the output dimension is \mbox{$N_{\text{to}} = (N_{X_{mn}^j} + N_{\gls{ibar}}) N_{\text{tb}} = \num{4696}$},
where \mbox{$N_{X_{mn}^j} = 3 [ (m_{\text{pol}} + 1) (2 n_{\text{tor}} + 1) - n_{\text{tor}} ] -2 = \num{586}$} is the number of Fourier modes used to represent $(R, \lambda, Z)$,
$m_{\text{pol}} = 8$ is the highest Fourier poloidal mode,
$n_{\text{tor}} = 11$ is the highest Fourier toroidal mode,
$N_{\gls{ibar}} = 1$ is the single output needed to represent the iota profile,
and $N_{\text{tb}} = \num{8}$ is the number of trunk basis functions per output.

The trunk network is a \gls{MLP} network with \gls{SiLU} as non-linear activation functions (\cref{tab:trunk-hp}).
Each layer is initialized accordingly to~\cite{Glorot2010}.

\begin{table}
  \caption{
    \glspl{HP} of the trunk network.
    See~\cref{sec:architecture} for details on the model macro-architecture.
  }%
  \label{tab:trunk-hp}
  \centering
  \begin{tabular}{lc}
    \toprule
    \gls{HP} & value \\
    \midrule
    depth & \num{4} \\
    width & \num{2048} \\
    trunk basis & \num{8} \\
    activation function & \gls{SiLU} \\
    \bottomrule
  \end{tabular}
\end{table}

The branch network maps the \gls{MHD} parameters and input plasma profiles into the weighting coefficients of the trunk basis functions (see~\cref{sec:architecture}).
A profile head extracts high-level features from the input profiles into a latent representation.
A set of blocks made of one \num{1}D convolutional and pooling layers composes the profile head.
The branch body then processes the concatenation of the profile latent representation with the input scalar parameters.
The output dimension is \mbox{$N_{\text{bo}} = (N_{X_{mn}^j} + N_{\gls{ibar}}) (N_{\text{tb}} + 2) = \num{5870}$},
where the \quotes{+\num{2}} is for the constant and identity functions.

The branch body is a \gls{MLP} network with \gls{SiLU} as non-linear activation functions (\cref{tab:trunk-hp}).
Each layer is initialized accordingly to~\cite{Glorot2010}.
The last layer is initialized at the median value of the model outputs evaluated on \SI{20}{\percent} of the training data set.

\begin{table}
  \caption{
    \glspl{HP} of the branch network.
    See~\cref{sec:architecture} for details on the model macro-architecture.
  }%
  \label{tab:branch-hp}
  \centering
  \begin{tabular}{lc}
    \toprule
    \gls{HP} & value \\
    \midrule
    \multicolumn{2}{c}{Profile Head} \\
    \midrule
    depth & \num{2} \\
    convolution kernel size & \num{5} \\
    convolution stride & \num{2} \\
    convolution padding & \num{0} \\
    filters & \num{16} \\
    pooling kernel size & \num{3} \\
    pooling stride & \num{2} \\
    pooling padding & \num{0} \\
    \midrule
    \multicolumn{2}{c}{Body} \\
    \midrule
    depth & \num{5} \\
    width & \num{2048} \\
    activation function & \gls{SiLU} \\
    \bottomrule
  \end{tabular}
\end{table}

The overall training process is divided in three different stages (see~\cref{sec:training-stages}):
data, gradient, and physics regularization stages.
During the data stage,
the learning rate is decreased accordingly to an exponential decay schedule with $\gamma$ as decay constant.
During the gradient and physics regularization stages,
the learning is kept constant.
Only after the first step of the curriculum learning procedure in the physics regularization stage,
the learning rate is decreased by a factor of \num{10}.
In addition,
in every training stage,
the norm of the model gradients is clipped.
\Cref{tab:data-stage-hp,tab:gradient-stage-hp,tab:physics-stage-hp} list the \glspl{HP} in each training stage.
\Cref{tab:fine-tuning-hp} lists the \glspl{HP} for the fine-tuning procedure.

\begin{table}
  \caption{
    \glspl{HP} for the data training stage.
  }%
  \label{tab:data-stage-hp}
  \centering
  \begin{tabular}{lc}
    \toprule
    \gls{HP} & value \\
    \midrule
    batch size & \num{128} \\
    initial learning rate & \num{8.0e-4} \\
    $\gamma$ & \num{9.9e-1} \\
    $L^2$ regularization & \num{6.0e-2} \\
    $\alpha_R$ & \num{2.5e-1} \\
    $\alpha_Z$ & \num{2.5e-1} \\
    $\alpha_{\lambda}$ & \num{2.5e-1} \\
    $\alpha_{\gls{ibar}}$ & \num{2.5e-1} \\
    patience epochs & \num{150} \\
    gradient clip & \num{1.0e-2} \\
    \bottomrule
  \end{tabular}
\end{table}

\begin{table}
  \caption{
    \glspl{HP} for the gradient training stage.
  }%
  \label{tab:gradient-stage-hp}
  \centering
  \begin{tabular}{lc}
    \toprule
    \gls{HP} & value \\
    \midrule
    batch size & \num{32} \\
    initial learning rate & \num{1.0e-5} \\
    $\gamma$ & \num{1} \\
    $L^2$ regularization & \num{6.0e-2} \\
    $\alpha_R$ & \num{2.4e-1} \\
    $\alpha_Z$ & \num{2.4e-1} \\
    $\alpha_{\lambda}$ & \num{2.4e-1} \\
    $\alpha_{\gls{ibar}}$ & \num{2.4e-1} \\
    $\alpha_{R_s}$ & \num{1.0e-2} \\
    $\alpha_{Z_s}$ & \num{1.0e-2} \\
    $\alpha_{{\lambda_s}}$ & \num{1.0e-2} \\
    $\alpha_{\gls{ibar}^{\prime}}$ & \num{1.0e-2} \\
    patience epochs & \num{150} \\
    gradient clip & \num{1.0e-2} \\
    \bottomrule
  \end{tabular}
\end{table}

\begin{table}
  \caption{
    \glspl{HP} for the physics regularization training stage.
  }%
  \label{tab:physics-stage-hp}
  \centering
  \begin{tabular}{lc}
    \toprule
    \gls{HP} & value \\
    \midrule
    batch size & \num{32} \\
    initial learning rate & \num{1.0e-6} \\
    $\gamma$ & \num{1} \\
    $L^2$ regularization & \num{6.0e-2} \\
    $\alpha_R$ & \num{2.4e-1} \\
    $\alpha_Z$ & \num{2.4e-1} \\
    $\alpha_{\lambda}$ & \num{2.4e-1} \\
    $\alpha_{\gls{ibar}}$ & \num{2.399e-1} \\
    $\alpha_{R_s}$ & \num{1.0e-2} \\
    $\alpha_{Z_s}$ & \num{1.0e-2} \\
    $\alpha_{{\lambda_s}}$ & \num{1.0e-2} \\
    $\alpha_{\gls{ibar}^{\prime}}$ & \num{1.0e-2} \\
    $\alpha_{\text{MHD}}$ & \num{1.0e-4} \\
    patience epochs & \num{50} \\
    gradient clip & \num{1.0e-2} \\
    \bottomrule
  \end{tabular}
\end{table}

\begin{table}
  \caption{
    \glspl{HP} for the fine-tuning procedure.
  }%
  \label{tab:fine-tuning-hp}
  \centering
  \begin{tabular}{lc}
    \toprule
    \gls{HP} & value \\
    \midrule
    initial learning rate & \num{1.0e-4} \\
    $\gamma$ & \num{9.996e-1} \\
    $L^2$ regularization & \num{1.0e-1} \\
    $\alpha_b$ & \num{2.495e-1} \\
    $\alpha_I$ & \num{5.0e-1} \\
    $\alpha_{\text{MHD}}$ & \num{1.0e-3} \\
    gradient clip & \num{1.0e-2} \\
    \bottomrule
  \end{tabular}
\end{table}

\subsection{Derivation of \gls{MHD} force residual}\label{sec:mhd-residual}

This section shows how the \gls{MHD} force residual can be derived once the mapping $\vec{x} = (R, \lambda, Z)$,
as well as the $p(s)$ and $\ibar(s)$ profiles,
are known.
As an example,
$F_{\beta}$ is derived;
$F_s$ can be derived in a similar manner.

Let us recall that:

\begin{gather}
  F_{\beta} = J^s = \frac{1}{\mu_0 \sqrt{g}} (\frac{\partial B_{\varphi}}{\partial \theta} - \frac{\partial B_{\theta}}{\partial \varphi}) ,
\end{gather}

where $\sqrt{g}$ is the Jacobian of the \num{3}D coordinate transformation $f : (s, \theta, \varphi) \rightarrow (R, \phi, Z)$,
and it is expressed as $\sqrt{g} = (\vec{e}_s \cdot \vec{e}_{\theta} \times \vec{e}_{\varphi})$.
Recalling the definition of the covariant basis vectors (see~\cref{sec:mhd}),
the choice of $\varphi = \phi$ effectively yields a \num{2}D Jacobian:

\begin{gather}
  \sqrt{g} = R (\frac{\partial R}{\partial \theta} \frac{\partial Z}{\partial s} - \frac{\partial R}{\partial s} \frac{\partial Z}{\partial \theta}) .
\end{gather}

The covariant magnetic field components (see~\cref{eq:b-theta-down,eq:b-zeta-down}) are connected to the contravariant magnetic field components (see~\cref{eq:b-theta-up,eq:b-zeta-up}) through the covariant metric tensor elements:

\begin{gather}
  g_{ij} = \vec{e}_i \cdot \vec{e}_j = \frac{\partial R}{\partial \alpha_i} \frac{\partial R}{\partial \alpha_j} + R^2 \frac{\partial \phi}{\partial \alpha_i} \frac{\partial \phi}{\partial \alpha_j} + \frac{\partial Z}{\partial \alpha_i} \frac{\partial R}{\partial \alpha_j}.
\end{gather}

Finally,
the expression for $F_{\beta}$ reads:

\begin{gather}
  F_{\beta} = \frac{\Phi^{\prime}}{\mu_0 \sqrt{g}} \{ \frac{\partial}{\partial \theta} [ \frac{1}{\sqrt{g}} [ (\gls{ibar} - \frac{\partial \lambda}{\partial \varphi}) ( \frac{\partial R}{\partial \theta} \frac{\partial R}{\partial \varphi} + \frac{\partial Z}{\partial \theta} \frac{\partial R}{\partial \varphi} ) + (1 + \frac{\partial \lambda}{\partial \theta}) ( \frac{\partial R}{\partial \varphi} \frac{\partial R}{\partial \varphi} + R^2 + \frac{\partial Z}{\partial \varphi} \frac{\partial R}{\partial \varphi} ) ] ] + \\ - \frac{\partial}{\partial \varphi} [ \frac{1}{\sqrt{g}} [ (\gls{ibar} - \frac{\partial \lambda}{\partial \varphi}) ( \frac{\partial R}{\partial \theta} \frac{\partial R}{\partial \theta} + \frac{\partial Z}{\partial \theta} \frac{\partial R}{\partial \theta} ) + (1 + \frac{\partial \lambda}{\partial \theta} ( \frac{\partial R}{\partial \varphi} \frac{\partial R}{\partial \theta} + \frac{\partial Z}{\partial \varphi} \frac{\partial R}{\partial \theta} ) ] ] \} .
\end{gather}

$F_{\beta}$ depends only on $\vec{x}$,
its first and second derivatives,
and \ibar.

\subsection{Physics regularization}\label{sec:f-star}

The proxy of the \gls{MHD} force residual $f_*$ is derived as follows.
\deleted[id=AM,comment=The whole section of red colored text has been removed.]{Let review the defining equations of the ideal-\gls{MHD} model:}

\begin{gather}
  \vec{F} = - \vec{J} \times \vec{B} + \vec{\nabla} p ,\\
  \vec{\nabla} \times \vec{B} = \mu_0 \vec{J} ,\\
  \vec{\nabla} \cdot \vec{B} = 0 .
\end{gather}

The magnetic field can be written in contravariant form as:

\begin{gather}
  \vec{B} = B^{\theta} \vec{e}_{\theta} + B^{\varphi} \vec{e}_{\varphi} ,\\
  B^{\theta} = \frac{1}{\sqrt{g}}(\chi^{\prime} - \Phi^{\prime} \frac{\partial \lambda}{\partial \varphi}) ,\\
  B^{\varphi} = \frac{\Phi^{\prime}}{\sqrt{g}}(1 + \frac{\partial \lambda}{\partial \theta}).
\end{gather}

By substituting the expression of $\vec{B}$ into $\vec{F}$,
the covariant form of the \gls{MHD} force is:

\begin{gather}
  \vec{F} = F_s \vec{\nabla}s + F_{\beta} \vec{\beta} , \\
  F_s = \sqrt{g}(J^{\varphi} B^{\theta} - J^{\theta} B^{\varphi}) + p^{\prime} ,\\
  F_{\beta} = J^s ,
\end{gather}

where \mbox{$\vec{\beta} = \sqrt{g}(B^{\varphi} \vec{\nabla} \theta - B^{\theta} \vec{\nabla} \varphi)$}.

The contravariant components of the current density,
\mbox{$J^i = \vec{J} \cdot \vec{\nabla} \alpha_i = \frac{1}{\mu_0} (\vec{\nabla} \times \vec{B}) \cdot \vec{\nabla} \alpha_i$},
are:

\begin{gather}
  J^s = \frac{1}{\mu_0 \sqrt{g}} (\frac{\partial B_{\varphi}}{\partial \theta} - \frac{\partial B_{\theta}}{\partial \varphi})\\
  J^{\theta} = \frac{1}{\mu_0 \sqrt{g}} (\frac{\partial B_s}{\partial \varphi} - \frac{\partial B_{\varphi}}{\partial s})\\
  J^{\varphi} = \frac{1}{\mu_0 \sqrt{g}} (\frac{\partial B_{\theta}}{\partial s} - \frac{\partial B_s}{\partial \theta})
\end{gather}

By substitute the contravariant current density components in the radial force component,
it follows that:

\begin{gather}
  F_s = \frac{1}{\mu_0} (B^{\theta} \frac{\partial B_{\theta}}{\partial s} - B^{\theta} \frac{\partial B_s}{\partial {\theta}} - B^{\varphi} \frac{\partial B_s}{\partial {\varphi}} + B^{\varphi} \frac{\partial B_{\varphi}}{\partial s}) + p^{\prime}
\end{gather}

Let us define the proxy of the \gls{MHD} residual as the flux surface averaged force residual norm when $F_{\beta}=0$:

\begin{gather}\label{eq:f-star-appendix}
  f_* = \frac{1}{(2 \pi)^2} \langle \mu_0 \sqrt{g} \twoNorm{F} \vert_{F_{\beta}=0} \rangle = \frac{1}{(2 \pi)^2} \langle \mu_0 \sqrt{g} F_s \rangle
\end{gather}

where \mbox{$\langle A \rangle = \int_0^{2 \pi} \int_0^{2 \pi} A d \theta d \varphi$} denotes a flux surface average operator (note that the $\sqrt{g}$ factor is not included in the average operator here).

Substituting the contravariant magnetic field components \added[id=AM]{(see~\cref{eq:b-theta-up,eq:b-zeta-up}) }in~\cref{eq:f-star-appendix},
the following terms result:

\begin{gather}
  \langle \sqrt{g} B^{\theta} \frac{\partial B_{\theta}}{\partial s} \rangle = \langle (\chi^{\prime} - \Phi^{\prime} \frac{\partial \lambda}{\partial \varphi}) \frac{\partial B_{\theta}}{\partial s} \rangle = \chi^{\prime} \frac{\partial \langle B_{\theta} \rangle}{\partial s} + \Phi^{\prime} \langle \lambda \frac{\partial B_{\theta}}{\partial s \partial \varphi} \rangle ,\label{eq:fs-1}\\
  \langle \sqrt{g} B^{\theta} \frac{\partial B_s}{\partial \theta} \rangle = \langle (\chi^{\prime} - \Phi^{\prime} \frac{\partial \lambda}{\partial \varphi}) \frac{\partial B_s}{\partial {\theta}} \rangle = \chi^{\prime} \langle \frac{\partial B_s}{\partial \theta} \rangle + \Phi^{\prime} \langle \lambda \frac{\partial B_s}{\partial \theta \partial \varphi} \rangle = \Phi^{\prime} \langle \lambda \frac{\partial B_s}{\partial \theta \partial \varphi} \rangle ,\label{eq:fs-2}\\
  \langle \sqrt{g} B^{\varphi} \frac{\partial B_{\varphi}}{\partial s} \rangle = \langle \Phi^{\prime} (1 + \frac{\partial \lambda}{\partial \theta}) \frac{\partial B_{\varphi}}{\partial s} \rangle = \Phi^{\prime} \frac{\partial \langle B_{\varphi} \rangle}{\partial s} - \Phi^{\prime} \langle \lambda \frac{\partial B_{\varphi}}{\partial s \partial \theta} \rangle ,\label{eq:fs-3}\\
  \langle \sqrt{g} B^{\varphi} \frac{\partial B_s}{\partial \varphi} \rangle = \langle \Phi^{\prime} (1 + \frac{\partial \lambda}{\partial \theta}) \frac{\partial B_s}{\partial \varphi} \rangle = \Phi^{\prime} \langle \frac{\partial B_s}{\partial \varphi} \rangle - \Phi^{\prime} \langle \lambda \frac{\partial B_s}{\partial \varphi \partial {\theta}} \rangle = - \Phi^{\prime} \langle \lambda \frac{\partial B_s}{\partial \varphi \partial \theta} \rangle .\label{eq:fs-4}\\
\end{gather}

The \mbox{$\frac{\partial \lambda}{\partial \varphi}$} and \mbox{$\frac{\partial \lambda}{\partial \theta}$} terms have been integrated by parts exploiting the fact that $\lambda$ and the covariant component of the B field are periodic functions in the poloidal and toroidal directions.
The second and fourth terms cancel each other.

In addition,
when $F_{\beta} = 0$ and $\sqrt{g} \neq 0$,
it follows that:

\begin{gather}
  F_{\beta} = J^s = \frac{1}{\mu_0 \sqrt{g}} (\frac{\partial B_{\varphi}}{\partial \theta} - \frac{\partial B_{\theta}}{\partial \varphi}) = 0 \rightarrow \frac{\partial B_{\varphi}}{\partial \theta} = \frac{\partial B_{\theta}}{\partial \varphi}
\end{gather}

Therefore,
also the second addenda of the first and third terms cancel each other.
Then,
the \gls{MHD} force residual proxy is:

\begin{gather}\label{eq:f-star-residual}
  (2 \pi)^2 f_* = \langle \mu_0 \sqrt{g} F_s \rangle = \chi^{\prime} \frac{\partial \langle B_{\theta} \rangle}{\partial s} + \Phi^{\prime} \frac{\partial \langle B_{\varphi} \rangle}{\partial s} + \mu_0 p^{\prime} \langle \sqrt{g} \rangle = 0
\end{gather}

\subsection{Boozer angular separation $\delta$ on test set}\label{sec:fast-ions-test-set}

The evaluation of the reconstruction of the Boozer angular separation $\delta$ on the test set is shown here.
The median regressed equilibrium in the test set does not have a minimum of the magnetic well strength for $\varphi = \frac{\pi}{\gls{Nfp}}$,
therefore,
the Boozer angular separation $\delta$ depicted in~\cref{fig:b-boozer-field-line} does not represent the angular separation between bounce points.
Nevertheless,
\cref{fig:b-boozer-field-line} shows how the model smoothly reconstructs the local magnetic field strength close to the magnetic axis,
and it introduces artificial field ripples in the edge region.

\begin{figure}[h!]
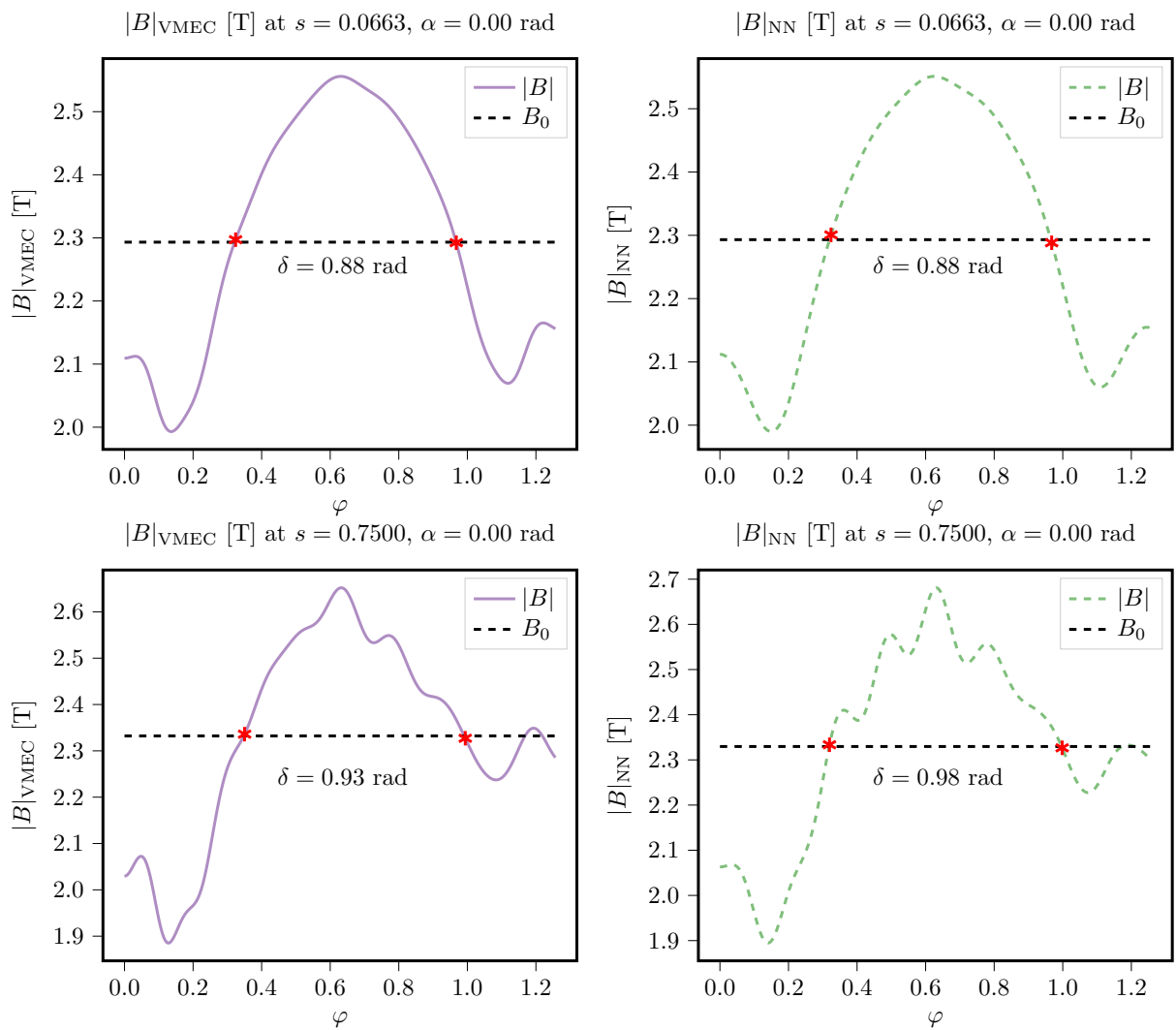

  \igraph[]{content_figures_b_fieldline_boozer_bounce_points.pgf}%
  \caption{
    True (left) and predicted (right) values for the magnetic field strength $B$ in Boozer coordinates along a field line in case of the median predicted equilibrium in the test set.
    The $\alpha=\num{0}$ field line at $s=\num{0.06}$ (top) and $s=\num{0.75}$ flux surface (bottom) are depicted.
    The dashed black horizontal line indicates the average magnetic field strength on the flux surface $B_0$.
    The red crosses highlight the $B_0$ contours,
    and the Boozer angular separation $\delta(s, B_0, \alpha)$ between them is shown just beneath.
  }%
  \label{fig:b-boozer-field-line}
\end{figure}

}{}

\ifthenelseproperty{compilation}{acknowledgement}{%
    \chapter{Acknowledgements}\label{sec:acknowledgement}
This work was not possible without the help of many people.
\TODO{Thanksgiving}
\par
\TODO{Maybe the following applies.}
This work has been carried out within the framework of the EUROfusion Consortium and has received funding from the Euratom research and training programme 2014-2018 and 2019-2020 under grant agreement No 633053. 
The views and opinions expressed herein do not necessarily reflect those of the European Commission.

}{}

\ifthenelseproperty{compilation}{affidavit}{%
    \thispagestyle{empty}
\ifthenelseproperty{compilation}{clsdefineschapter}{%
	\ifKOMA
		\addchap[Statutory declaration]{Statutory declaration}
	\else
    	\chapter[Statutory declaration]{Statutory declaration}
    \fi
}{%
	\ifKOMA
		\addsec[Statutory declaration]{Statutory declaration}
	\else
    	\section[Statutory declaration]{Statutory declaration}
    \fi
}
I hereby declare in accordance with the examination regulations that I myself have written this document, that no other sources as those indicated were used and all direct and indirect citations are properly designated, that the document handed in was neither fully nor partly subject to another examination procedure or published and that the content of the electronic exemplar is identical to the printing copy.

\Signature{\getproperty{document}{location}}{\textsc{%
    \ifluatex
        \IfSubStr{\getproperty{author}{firstname}}{TODO}{%
            \getproperty{author}{firstname}
        }{%
            \FirstWord{\getproperty{author}{firstname}}
        }
    \else
        \getproperty{author}{firstname}
    \fi
    \getproperty{author}{familyname}}}

}{}

\ifthenelseproperty{compilation}{lof}{%
    \disabledprotrusion{\listoffigures}
}{}

\ifthenelseproperty{compilation}{lot}{%
    \disabledprotrusion{\listoftables}
}{}

\ifthenelseproperty{compilation}{lol}{%
    \disabledprotrusion{\lstlistoflistings}
}{}

\ifthenelseproperty{compilation}{listofpublications}{%
    \begin{refsection}
	\newrefcontext[sorting=ndymdt]
	\nocite{*}
    \ifthenelseproperty{compilation}{clsdefineschapter}{%
		\ifKOMA
			\addchap[Publications as first author]{Publications as first author}\label{sec:publications_as_first_author}
		\else
	    	\chapter[Publications as first author]{Publications as first author}\label{sec:publications_as_first_author}
	    \fi
    }{%
		\ifKOMA
			\addsec[Publications as first author]{Publications as first author}\label{sec:publications_as_first_author}
		\else
	    	\section[Publications as first author]{Publications as first author}\label{sec:publications_as_first_author}
	    \fi
    }
	\printbibliography[
						keyword=firstAuthor,
						keyword=refereed,
						heading=subbibliography,
						title={Peer-reviewed articles},
						resetnumbers=true
					]
\end{refsection}
\begin{refsection}
	\newrefcontext[sorting=ndymdt]
	\nocite{*}
    \ifthenelseproperty{compilation}{clsdefineschapter}{%
		\ifKOMA
			\addchap[Publications as coauthor]{Publications as coauthor}\label{sec:publications_as_coauthor}
		\else
	    	\chapter[Publications as coauthor]{Publications as coauthor}\label{sec:publications_as_coauthor}
	    \fi
    }{%
		\ifKOMA
			\addsec[Publications as coauthor]{Publications as coauthor}\label{sec:publications_as_coauthor}
		\else
	    	\section[Publications as coauthor]{Publications as coauthor}\label{sec:publications_as_coauthor}
	    \fi
    }
	\setboolean{isCoauthorList}{true}
	\printbibliography[
						keyword=coAuthor,
						keyword=refereed,
						heading=subbibliography,
						title={Peer-reviewed articles},
						resetnumbers=true
					]
\end{refsection}

}{}

\ifthenelseproperty{compilation}{glossaries}{%
    \ifthenelseproperty{compilation}{acronyms}{%
	\printglossary[type=\acronymtype,style=mcoltree]%
}{}%
\ifthenelseproperty{compilation}{los}{%
	\setlength\extrarowheight{5pt}%
	\printglossary
			[
				title=List of Symbols,
				type=symbols,
				style=customListOfSymbols,
			]
	\setlength\extrarowheight{0pt}%
}{}%

}{}

\ifthenelseproperty{compilation}{bibliography}{%
    \printbibliography
}{}

    \typeout{----- END OF DOCUMENT -----}

\end{document}
\typeout{----- END OF MAIN -----}